\documentclass[fleqn,usenatbib]{mnras}

\usepackage{newtxtext,newtxmath}
\usepackage[T1]{fontenc}
\usepackage{ae,aecompl}

\usepackage{adjustbox}
\usepackage{amsmath}
\usepackage{amssymb}
\usepackage{epic}
\usepackage{float}
\usepackage{graphicx}
\usepackage[graphicx]{realboxes}
\usepackage{epstopdf}
\usepackage{multirow}
\usepackage{natbib}
\usepackage{overpic}
\usepackage{pdflscape}
\usepackage[dvipsnames]{xcolor}
\usepackage{morefloats}
\usepackage{rotating}
\usepackage{afterpage}
\AtBeginShipout{%
  \ifnum\value{page}>1 %
    \typeout{* Additional boxing of page `\thepage'}%
    \setbox\AtBeginShipoutBox=\hbox{\copy\AtBeginShipoutBox}%
  \fi
}

\title[Environment and radio source asymmetry]{Radio Galaxy Zoo: Observational evidence for environment as the cause of radio source asymmetry}

\author[Rodman et al.]{
P. E. Rodman$^{1}$\thanks{E-mail: prodman@utas.edu.au}, R. J. Turner$^{1,2}$, S. S. Shabala$^{1}$, J. K. Banfield${^3}$, O. I. Wong${^{4,5}}$,\newauthor H. Andernach${^6}$, A. F. Garon${^7}$, A. D. Kapi\'nska${^8}$, R. P. Norris${^9}$, L. Rudnick${^7}$\\
	$^{1}$ School of Natural Sciences, Private Bag 37, University of Tasmania, Hobart, TAS 7001, Australia\\
	${^2}$ CSIRO Australia Telescope National Facility, PO Box 76, Epping, NSW, 1710, Australia\\
	${^3}$ Research School of Astronomy and Astrophysics, Australian National University, Canberra ACT, 2611, Australia\\
	${^4}$ International Centre for Radio Astronomy Research-M468, The University of Western Australia, 35 Stirling Hwy, Crawley, WA 6009, Australia\\
	${^5}$ ARC Centre of Excellence for All-sky Astrophysics (CAASTRO), Australia\\
	${^6}$ Departamento de Astronom\'ia, DCNE, Universidad de Guanajuato, Apdo. Postal 144, CP 36000, Guanajuato, Gto., Mexico\\
	${^7}$ Minnesota Institute for Astrophysics, University of Minnesota, 116 Church St. SE, Minneapolis, MN 55455\\
	${^8}$ National Radio Astronomy Observatory (NRAO), 1003 Lopezville Road, Socorro, NM 87801-0387, USA\\
	${^9}$ Western Sydney University, Locked Bag 1797, Penrith, NSW 2751, Australia}

\date{Accepted 2018 November 8. Received 2018 October 14; in original form 2018 July 24}

\pubyear{2018}

\begin{document}
\label{firstpage}
\pagerange{\pageref{firstpage}--\pageref{lastpage}}
\maketitle

\begin{abstract}
We investigate the role of environment on radio galaxy properties by constructing a sample of large ($\gtrsim100$~kpc), nearby ($z<0.3$) radio sources identified as part of the Radio Galaxy Zoo citizen science project. Our sample consists of 16 Fanaroff-Riley Type II (FR-II) sources, 6 FR-I sources, and one source with a hybrid morphology. FR-I sources appear to be hosted by more massive galaxies, consistent with previous studies. In the FR-II sample, we compare the degree of asymmetry in radio lobe properties to asymmetry in the radio source environment, quantified through optical galaxy clustering. We find that the length of radio lobes in FR-II sources is anti-correlated with both galaxy clustering and lobe luminosity. These results are in quantitative agreement with predictions from radio source dynamical models, and suggest that galaxy clustering provides a useful proxy for the ambient gas density distribution encountered by the radio lobes.
\end{abstract}


\section{Introduction}
\label{sec:intro}

In the local Universe, radio-loud active galactic nuclei (AGN) are predominantly hosted by massive elliptical galaxies \citep{SadlerEA89,Burns90}. Optical properties of host galaxies of these AGN show a striking dichotomy: radio AGN with strong emission lines (so-called high-excitation radio galaxies, or HERGs) are hosted by lower-mass galaxies than low-excitation radio galaxies \citep[LERGs;][]{BestHeckman12}. The radio AGN fraction in LERGs shows a strong mass dependence \citep{BestEA05}, with massive galaxies more likely to host an AGN. Even at fixed stellar mass, brightest cluster galaxies host a higher fraction of radio AGN \citep{BestEA07}. Dynamical state of the cluster is also important, with almost all rapidly cooling clusters hosting radio AGN, compared to fewer than half for clusters with cooling times in excess of 1 Gyr \citep{MittalEA09}. These findings can be explained in a framework where LERGs are triggered by gradual cooling of the gas out of the hot halo \citep{HardcastleEA07}, a fraction of this gas being eventually accreted by the black hole, possibly via chaotic cold accretion \citep{GaspariEA15}. Models in which LERGs operate as thermostats are quantitatively consistent with the observed scaling of the radio AGN duty cycle with mass \citep{ShabalaEA08,PopeEA12}. Conversely, the fraction of high-excitation radio AGN is enhanced in the presence of one-on-one interactions between galaxies \citep{SabaterEA13,EllisonEA15}, and their host galaxies often show disturbed morphologies and high dust masses \citep{KavirajEA12,ShabalaEA12,TadhunterEA14}, consistent with being triggered through galaxy interactions.

Once a jet is triggered, environment plays an important role in determining the final lobe morphology. At low redshift, edge-darkened, or Fanaroff-Riley \citep[FR; ][]{FanaroffRiley74} Type I, objects are mostly found in clusters, while edge-brightened, or FR-II, sources prefer poorer environments \citep{HillLilly91}. FR-IIs have higher radio luminosities than FR-Is, and the two FR classes show a difference in characteristic radio luminosity which increases with host galaxy optical luminosity \citep{OwenLedlow94}; however, both FR classes show a large scatter and overlap in radio luminosity \citep{Best09,MiraghaeiBest17}. This morphological distinction is an important factor in the radio source--environment interaction: while FR-Is provide gentle, quasi-continuous heating to cluster cores \citep[e.g.][]{ChurazovEA01,Fabian12}, FR-IIs drive powerful shocks capable of affecting satellite galaxies on scales of hundreds of kiloparsecs \citep{RawlingsJarvis04,ShabalaEA11}. There is a loose association between FR morphology and optical classification, in the sense that almost all FR-Is are LERGs, but FR-IIs can be hosted by both high and (less often) low-excitation galaxies. A growing consensus appears to be that the FR classification may be determined by jet-environment interaction on approximately kiloparsec scales: if the initially relativistic jet can be slowed down sufficiently via entrainment from the interstellar medium \citep{Bicknell95} or stellar winds \citep{Komissarov94, PeruchoEA14}, it will eventually be disrupted and form a FR-I; on the other hand, if the jet is not entrained appreciably, it will form a characteristic FR-II structure with lobes inflated via backflow of overpressured plasma from jet termination shocks, seen in radio images as hotspots. Alternatives to jet entrainment are jet stalling in a rising pressure profile \citep{MassagliaEA16}, or failed collimation of an initially conical jet by the environment \citep{Alexander06,KrauseEA12}; both are followed by eventual jet disruption and transition to an FR-I as a result of jet--environment interaction. A small fraction \citep[less than 1 per cent; ][]{GawronskiEA06} of extended radio sources show hybrid morphologies \citep{Gopal-KrishnaWiita00,KapinskaEA17}, with FR-I morphology on one side of the central engine, and FR-II morphology of the other. These objects are ideal laboratories for studying the interaction of the small-scale jets with their environment: the two jets are intrinsically identical, and the difference in the final morphology can be attributed to different kinds of interaction with the environment.

On larger scales (tens and hundreds of kpc), environment is a key factor in lobe (rather than jet) evolution. Dynamical radio source models \cite[e.g. ][]{KaiserAlexander97,BlundellRawlings00,TurnerShabala15,Hardcastle18} and numerical simulations \citep{HardcastleKrause13,HardcastleKrause14} predict that the temporal evolution of lobe size and luminosity should be strongly environment dependent; these predictions are consistent with X-ray observations \citep{ArnaudEA10}. Compact radio AGN are more prevalent in low-mass galaxies and poor environments \citep{Shabala18}, consistent with models in which extended emission may be below the surface brightness detection threshold of existing instruments \citep{ShabalaEA17,TurnerEA18b}; recent increased sensitivity observations of giant lobes in the archetypal FR-I source 3C31 \citep{HeesenEA18} confirms this picture. Estimates of AGN lifetimes and jet kinetic powers from radio continuum data are therefore environment-dependent, and hence so too are estimates of the energy budget available for AGN feedback; this feedback process is responsible for shaping the bright end of the galaxy luminosity function \citep{SilkRees98,BowerEA06,CrotonEA06}. Environment quantification through galaxy clustering provides a natural connection between galaxy formation and lobe evolution models. \citet{TurnerShabala15} used a semi-analytic galaxy formation model to quantify radio source environments, and showed that this approach can explain many properties of the observed radio galaxy populations. \citet{ShabalaAlexander09} and \citet{RaoufEA17} modelled radio AGN and galaxy properties self-consistently within a semi-analytic galaxy formation model, and showed that the requirement to match radio AGN and galaxy properties simultaneously places powerful constraints on feedback models.

In this paper, we present an analysis of a sample of Radio Galaxy Zoo \citep[hereafter RGZ;][]{BanfieldEA15} radio galaxies with large-scale environment information from galaxy clustering. The focus of the paper is on exploring the relationship between the asymmetry in radio lobe properties, and asymmetry in radio source environments quantified through galaxy clustering. We present our sample in Sections~\ref{sec:sample} and \ref{sec:results}, and discuss the role of environment on lobe evolution in Section~\ref{sec:results_by_type}. We conclude in Section~\ref{sec:conclusions}.

Throughout the paper, we assume a flat Universe with $H_0=68$\,km\,s$^{-1}$\,Mpc$^{-1}$, $\Omega_\Lambda=0.685$ and $\Omega_{\rm M}=0.315$ \citep{PlanckCollab14}.

\section{Sample selection}
\label{sec:sample}

The starting point for our sample is the set of multi-component radio sources identified by citizen scientists through the Radio Galaxy Zoo project \citep{BanfieldEA15}. RGZ enlists citizen scientists to classify 1.4 GHz radio continuum images from the Faint Images of the Radio Sky at Twenty Centimeters \citep[FIRST; ][]{BeckerEA95} and the Australian Telescope Large Area Survey \citep[ATLAS;][]{NorrisEA06,MiddelbergEA08} projects. RGZ offers to its volunteers a superposition of these radio images with mid-infrared images at 3.4\,$\mu$m from the Wide-field Infrared Survey Explorer \citep[WISE;][]{WrightEA10} and at 3.6\,$\mu$m from the \textit{Spitzer} Wide-Area Infrared Extragalactic Survey \citep[SWIRE; ][]{LonsdaleEA03}. Citizen scientists are asked to identify whether separate radio source components may belong to a single radio structure, and whether there is a corresponding infra-red host galaxy. Upon achieving a sufficient number of classifications for one image (20 for non-compact sources), the consensus level $C$ is evaluated for the radio and infra-red classifications of each source. The consensus on the position of the host galaxy is then determined through application of a kernel-density estimator (KDE) on the positions clicked by the volunteers since these positions may differ by a few pixels and yet be identifying the same source. We refer the interested reader to \citet{BanfieldEA15} for further details. 

The initial dataset contained 2679 candidate sources with two or three radio components from the subset of the RGZ catalogue investigated by \citet{BanfieldEA15}. The sample was reduced to those sources with a consensus level above 0.7 and redshift $z<0.3$, reducing the sample to 169 sources. Radio lobes were required to be approximately straight to allow for accurate quantification of lobe length and environment (within $\sim10$ degrees; see below), reducing the sample to 89. We required at least one of the radio lobes to be at least 100~kpc in length as measured along a straight line from the host to the emission region farthest from the host, to ensure sufficient image resolution. Finally, an integrated flux density threshold of 20~mJy was imposed to ensure sufficiently high signal-to-noise ratio. 

In this work we seek to probe the relationship between radio source morphology and environment; this is achieved using the Sloan Digital Sky Survey (SDSS) DR10 \citep{AhnEA2014} and the photometric redshifts of \citet{BeckEA16} to quantify galaxy clustering near our radio sources. To ensure robust quantification of the environment, neighbouring galaxies were associated to the host using photometric redshifts; galaxies with photometric redshifts consistent with the host spectroscopic or photometric redshift (within 3 sigma uncertainties) were admitted as neighbours. In addition, we required each source to have at least 20 such neighbour galaxies within a projected distance of 1 Mpc of the host galaxy.

The large tolerance in redshift is chosen to ensure that for host galaxies located in clusters, all cluster members are included; we note some nearby isolated galaxies may also be included, adding noise to our measurements. Imposing more stringent requirements of $2\sigma$ or $1\sigma$ would reduce the number of neighbours by $\sim17$ per cent and $\sim38$ per cent respectively. We repeated our analysis below using these more stringent cuts, and reproduced the same conclusions with lower statistical significance. The threshold galaxy counts requirement biases the sample against sources in poor environments, but makes our analysis more robust to outliers.

Our final sample consists of 23 extended RGZ sources seen in the FIRST survey at $z<0.3$ with environment information. Radio and optical images of these sources and their surrounding environment are presented in Appendices~\ref{sec:appendixa} and \ref{sec:appendixb}, with one example of each class (FR-II, FR-I and hybrid) shown in the main text.

\section{Results}
\label{sec:results}

The objective of this work is to investigate the relationship between asymmetries in radio source properties and those of its environment. Below, we briefly outline quantitative measures of asymmetry in radio source size, luminosity and morphology (Section~\ref{sec:Radio source properties}), as well as environment as described by galaxy clustering (Section~\ref{sec:clustering}).

\subsection{Radio source properties}
\label{sec:Radio source properties}

Following the approach of \citet{FanaroffRiley74}, radio lobes were separated into Fanaroff-Riley Type I (edge-darkened) or Type II (edge-brightened) morphological classes; radio sources with lobes of different morphology (i.e. FR-I on one side of the central engine, FR-II on the other) are classified as hybrids. Visual classifications were performed by three of the authors (PER, RJT, SSS), resulting in a sample consisting of 16 FR-IIs, 6 FR-Is and one hybrid. The Northern lobe of one further source, RGZ J102733.6+481718, has a borderline FR-I/II morphology and could not be reliably classified; this source was excluded from further analysis. Examples of an FR-I and FR-II source in our sample are shown in Figure 1; the one hybrid source is shown in Figure 2. The remaining FR-II sources are presented in Appendix~\ref{sec:appendixa}, and the remaining FR-I sources in Appendix~\ref{sec:appendixb}. Relevant properties of all sources in our sample are given in Table~\ref{tab:all_sources}.

\begin{figure*}
	\centering
	\includegraphics[width=0.8\columnwidth]{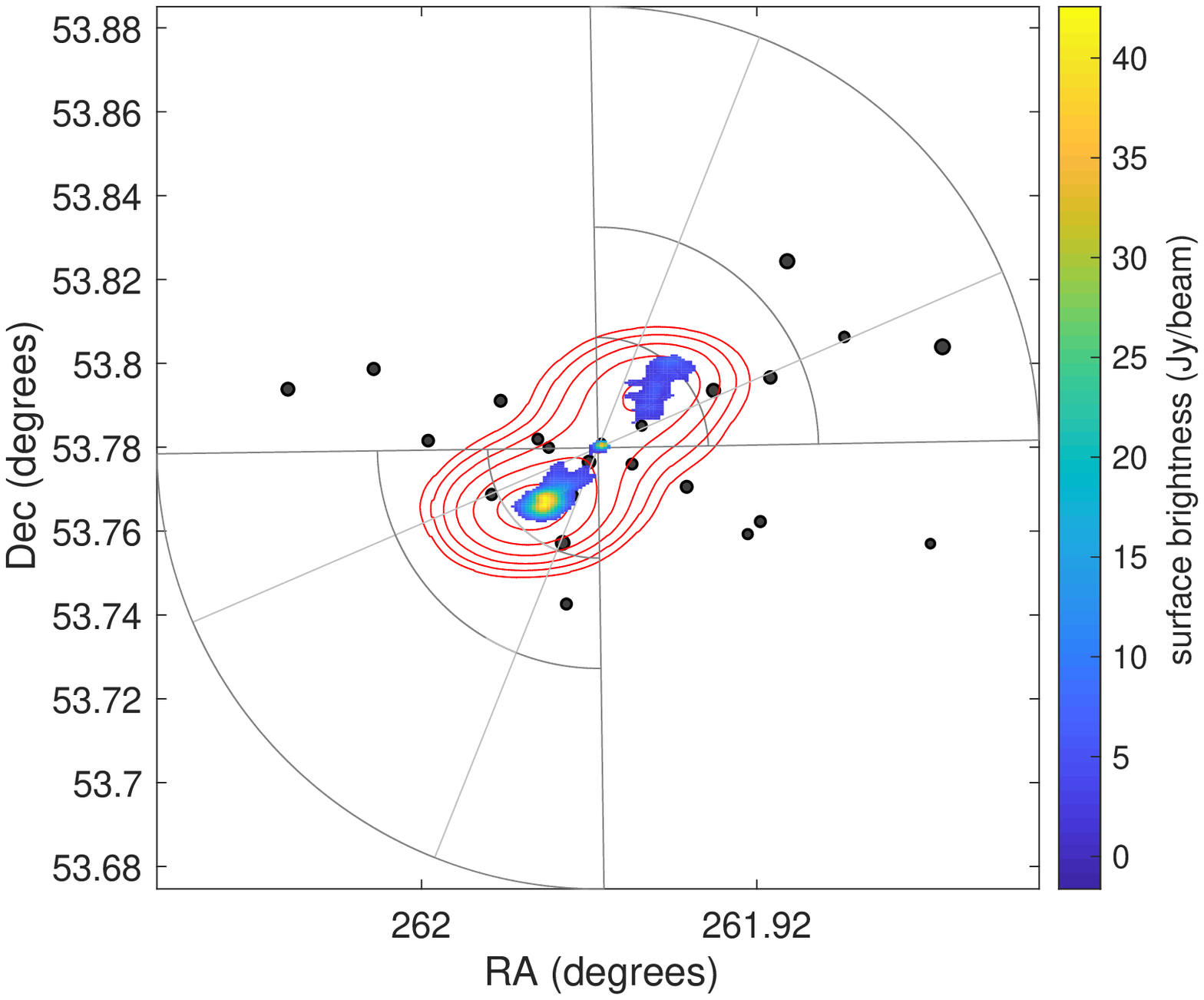} \hspace{0.15\textwidth}
	\includegraphics[width=0.8\columnwidth]{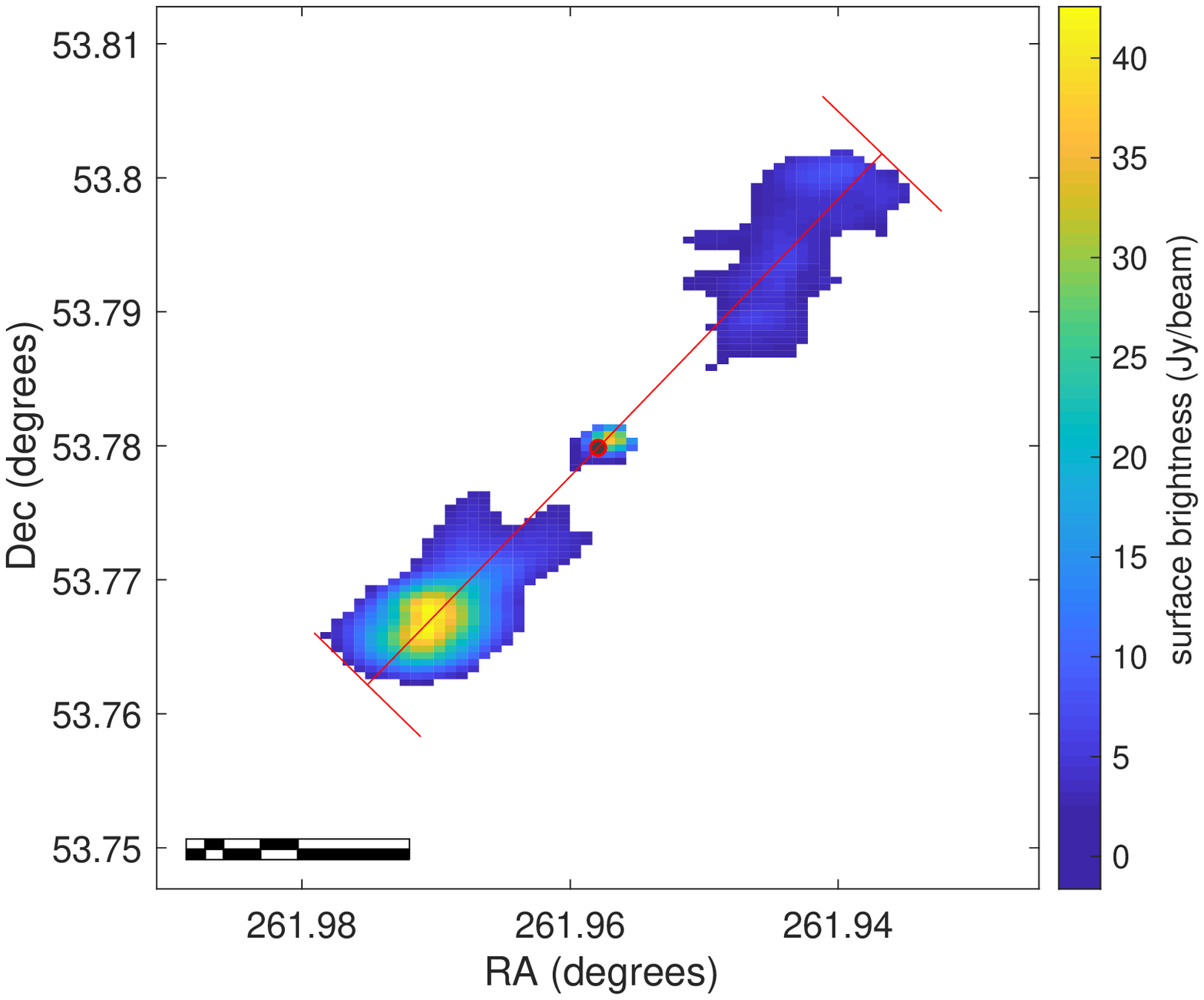}\\ \vspace{0.15cm}
	\includegraphics[width=0.8\columnwidth]{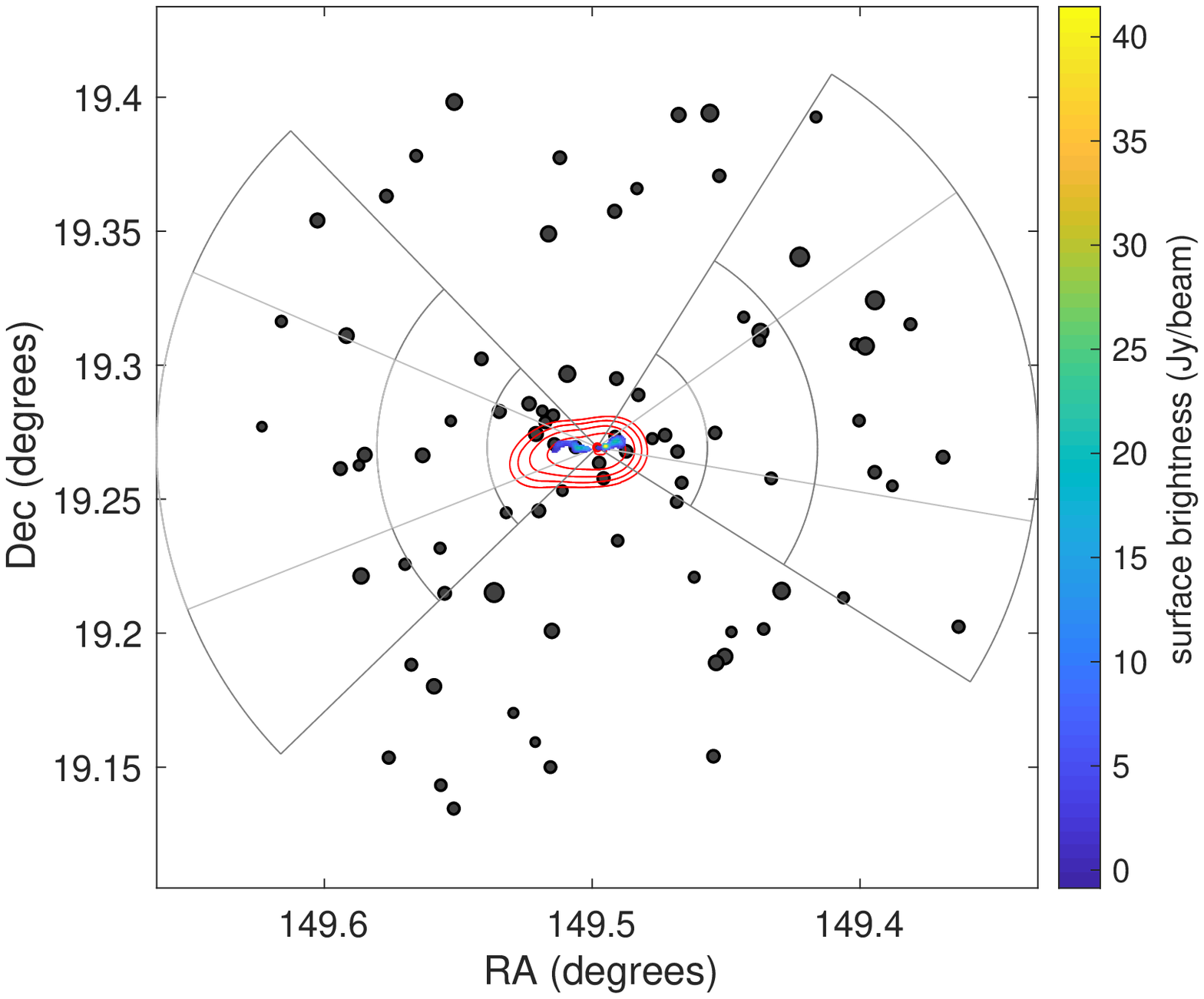}\hspace{0.15\textwidth}
	\includegraphics[width=0.8\columnwidth]{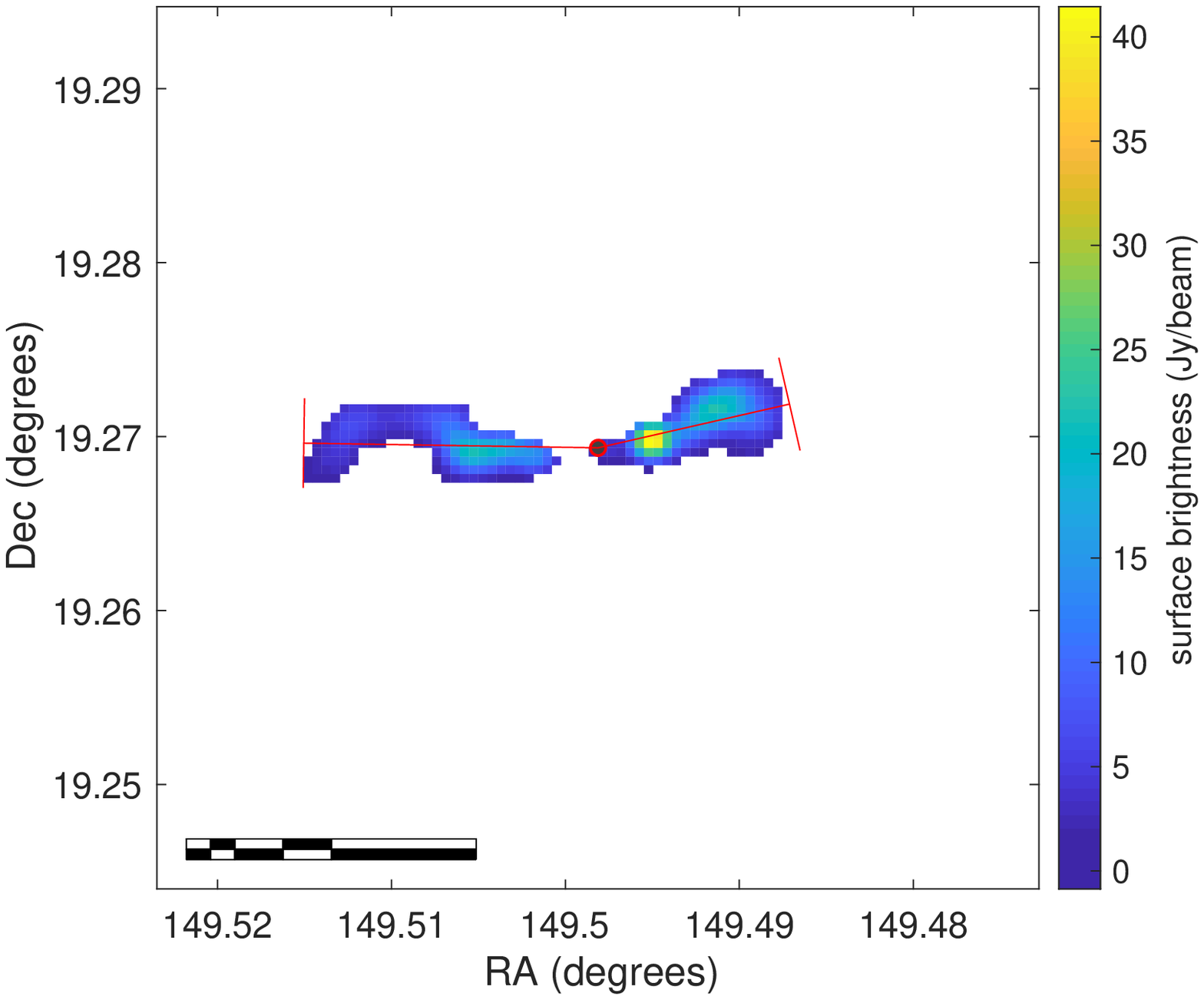}
	\caption{{\it Top left:} Radio-optical overlays for the FR-II source RGZ J172749.5+534647. Colour scale shows 1.4 GHz FIRST surface brightness, red contours are 1.4 GHz NVSS surface brightness at 5, 10, 20, 40, 80, and 160$\sigma$. Black circles are optical galaxies with redshifts consistent with the radio source host; symbol size is proportional to $z$-band magnitude of each galaxy. Wedges of 45 and 90 degrees centered on the lobe axis are shown, with arcs drawn at 250, 500 and 1000 kpc from the AGN host. {\it Top right:} Zoom-in on the radio emission; the scale bar is one arcminute in length, and subdivided into units of 5, 10, 20 and 30 arcseconds. Red lines are indicative of axis lengths and jet directions; cross-bars at the end are the width of the fitted box. 
	{\it Bottom: }FR-I source RGZ J095759.3+191609; symbols are as above. Radio-optical overlays for the remaining FR-II and FR-I sources can be found in Appendix~\ref{sec:appendixa} and Appendix~\ref{sec:appendixb} respectively.}
	\label{fig:FRII_alpha_11}
\end{figure*}

\begin{table*}
\vspace{-1.5cm}
	\rotatebox{90}{
		\centering
		\tiny
		\tabcolsep=0.1cm
		\begin{tabular}{ c c c c c c c p{0.1cm} | c c c c c c c c c c p{0.1cm} | c c c c c c c c c c }
			\hline\hline
			\multicolumn{7}{c|}{Galaxy and cluster environment}
			&& \multicolumn{10}{|c|}{Eastern lobe}
			&& \multicolumn{10}{|c}{Western lobe}\\
			\hline
			RGZ Name &  $z$ & $d_{\rm L}$ & $M_{\rm r}$ & $N_{\rm 1\,Mpc}$ & $d_5$ & $d_{10}$ & &  \multicolumn{2}{c}{$D$ [kpc]} & \multicolumn{2}{c}{$S_{1.4}$ [mJy]} &  Hotspot &  FR &  \multicolumn{2}{c}{$N_{\rm 500\,kpc}$} &  \multicolumn{2}{ c }{$N_{\rm 1\,Mpc}$} & & \multicolumn{2}{c}{$D$ [kpc]} & \multicolumn{2}{c}{$S_{1.4}$ [mJy]} &  Hotspot &  FR & \multicolumn{2}{c}{$N_{\rm 500\,kpc}$} & \multicolumn{2}{c}{$N_{\rm 1\,Mpc}$}\\
			&  & [Mpc] &  & ($m_z$>$20.5$) & [kpc] & [kpc] & & 5$\sigma$ & 3$\sigma$ & 5$\sigma$ & 3$\sigma$ & fraction &  & $45^\circ$ & $90^\circ$ & $45^\circ$ & $90^\circ$ & & 5$\sigma$ & 3$\sigma$ & 5$\sigma$ & 3$\sigma$ & fraction &  & $45^\circ$ & $90^\circ$ & $45^\circ$ & $90^\circ$\\
			(1)&(2)&(3)&(4)&(5)&(6)&(7)&&(8)&(9)&(10)&(11)&(12)&(13)&(14)&(15)&(16)&(17)&&(18)&(19)&(20)&(21)&(22)&(23)&(24)&(25)&(26)&(27)\\
			\hline\hline
			\multicolumn{29}{c}{\bf FR-IIs}\\
			\hline
			RGZ J090542.6+465809&0.195$^\dagger$&979&-21.19&16&446&1023&&141&142&76.4&83.1&0.23&2.17&0&2&1&6&&187&189&60.3&67.7&0.30&2.31&1&3&1&5\\
			RGZ J091445.5+413714&0.140&680&-23.86&14&154&240&&133&137&202.0&205.2&0.13&1.98&1&3&4&8&&106&106&164.0&167.3&0.16&2.09&6&8&10&14\\
			RGZ J093821.5+554333&0.221&1126&-22.53&26&187&370&&91&114&26.7&28.4&0.19&1.58&3&8&5&12&&148&156&19.3&21.8&0.14&2.01&0&1&3&6\\
			RGZ J100128.8+043437&0.222&1131&-23.29&37&351&484&&192&197&27.6&30.0&0.19&2.11&1&9&2&12&&147&150&34.7&37.7&0.22&2.15&3&8&6&15\\
			RGZ J102040.8+315509&0.286&1508&-23.15&17&213&441&&138&145&13.1&17.3&0.18&2.15&0&4&0&5&&175&181&17.2&19.9&0.17&2.17&2&4&2&4\\
			RGZ J110253.0+125904&0.140&680&-22.94&17&219&444&&211&216&11.9&17.3&0.19&2.41&0&1&5&9&&243&252&7.2&13.4&0.27&2.42&1&2&4&11\\
			RGZ J120118.2+124500&0.278$^\dagger$&1460&-22.22&18&295&444&&112&114&134.0&136.4&0.09&1.75&3&7&4&8&&167&170&85.5&89.0&0.08&2.09&1&1&2&2\\
			RGZ J125721.9+122820&0.208&1052&-22.43&13&196&431&&119&123&84.0&90.1&0.11&2.13&0&1&0&8&&131&136&170.0&176.4&0.31&2.17&0&1&1&5\\
			RGZ J125724.2+063114&0.175&868&-21.91&12&480&586&&123&126&19.9&25.8&0.18&1.80&1&3&7&10&&175&179&9.1&16.5&0.29&2.02&2&4&2&6\\
			RGZ J135110.8+072846&0.150&733&-22.67&9&270&380&&151&152&105.0&115.0&0.23&1.98&3&6&3&6&&130&135&89.7&97.5&0.24&2.15&5&6&5&9\\
			RGZ J153008.0+231616&0.090&423&-23.29&31&393&732&&265&272&50.4&62.4&0.10&2.25&5&10&5&12&&265&271&55.7&64.1&0.09&2.07&5&7&5&10\\
			RGZ J154936.7+361417&0.236&1212&-23.01&14&334&597&&72&78&5.2&8.5&0.29&2.25&2&4&2&4&&71&72&9.41&14.5&0.21&2.27&1&4&1&4\\
			RGZ J161037.5+060509&0.241&1241&-22.45&13&506&603&&177&189&18.1&23.9&0.16&2.08&1&4&2&7&&158&176&17.9&23.0&0.15&2.13&2&2&3&5\\
			RGZ J172749.5+534647&0.244$^\dagger$&1259&-22.17&21&165&321&&191&193&470.0&486.7&0.05&1.93&3&5&3&5&&237&239&166.0&188.2&0.03&2.24&3&3&5&6\\
			RGZ J172957.2+450623&0.255$^\dagger$&1323&-22.29&12&490&663&&299&305&142.0&155.8&0.08&2.20&2&3&2&3&&281&291&127.0&142.3&0.08&2.23&6&7&6&7\\
			RGZ J210030.5+100529$^\star$&0.156$^\dagger$&765&-22.46&12&294&466&&125&175&1.2&12.7&$\sim$1&2.32&0&1&0&1&&160&169&10.8&18.7&0.44&2.16&1&2&5&7\\
			\hline\hline
			\multicolumn{29}{c}{\bf Hybrids}\\
			\hline
			RGZ J082835.2+322825&0.280&1472&-22.93&19&446&1023&&140&149&7.7&12.7&0.48&1.25&1&3&2&4&&139&144&7.6&10.0&0.40&2.24&2&7&2&7\\
			\hline\hline
			\multicolumn{29}{c}{\bf FR-Is}\\
			\hline
			RGZ J085549.1+420420&0.238&1224&-22.19&13&446&1023&&246&253&22.6&35.8&0.05&1.56&1&1&3&4&&219&237&29.8&37.9&0.05&1.55&3&4&3&4\\
			RGZ J094443.2+024754&0.220$^\dagger$&1120&-23.49&37&447&1024&&280&293&72.6&79.8&0.16&1.66&4&6&7&11&&233&243&98.7&104.6&0.24&1.50&2&5&3&6\\
			RGZ J095759.3+032725&0.165&814&-23.57&32&448&1025&&169&176&256.2&259.9&0.41&0.97&4&9&9&15&&134&142&292.8&295.3&0.29&0.96&4&9&7&14\\
			RGZ J095759.3+191609&0.088$^\dagger$&413&-23.42&11&449&1026&&98&98&122.7&125.9&0.17&1.36&6&13&11&23&&65&66&149.6&153.1&0.26&0.96&6&9&16&22\\
			RGZ J145001.5+144747&0.300&1593&-23.28&13&450&1027&&87&89&28.1&29.9&0.28&1.48&0&3&0&3&&80&84&32.0&34.1&0.34&1.17&2&2&2&2\\
			RGZ J145039.8+441829&0.286&1508&-23.04&1&451&1028&&320&328&84.4&90.5&0.14&1.20&3&4&5&8&&255&264&110.6&114.1&0.36&1.03&2&5&5&10\\
			\hline\hline
		\end{tabular}
	}
	\caption{The samples of FR-IIs, hybrids and FR-Is. Columns are: (1) source name; (2) redshift, where a dagger denotes only photometric available; (3) luminosity distance; (4) $z$-band absolute magnitude; (5) number of galaxies within $1\rm\, Mpc$ above $m_z = 20.5$~mag; (6) distance to fifth nearest neighbour; and (7) distance to tenth nearest neighbour. For the eastern / western radio lobes the columns are: (8/18, 9/19) lobe lengths for 5$\sigma$ and 3$\sigma$ flux density cuts, respectively, measured along the jet axis; (10/20, 11/21) $1.4\rm\, GHz$ flux density from FIRST at the 5$\sigma$ and 3$\sigma$ cuts; (12/22) fraction of flux density in the hotspot; (13/23) Fanaroff-Riley morphology; (14/24, 15/25) number of galaxies within $500\rm\, kpc$ of the host and 45 or 90 degrees of the lobe axis; and (16/26, 17/27) same as the previous columns but for $1\rm\, Mpc$. $^\star$ Continuum emission in RGZ~J210030.5+100529 is hotspot-dominated, and this source is excluded from analysis.}
	\label{tab:all_sources}
\end{table*}

We define the spatial extent of each source by the outermost 5$\sigma$ contours ($\sigma \sim 0.15$~mJy/beam). Following \citet{TurnerEA18a}, the source is divided into two radio lobes by finding the two most extreme points and associating the other pixels of emission to the extreme point on the same side of the host optical galaxy. A rectangular box is then fitted to each lobe so that it intersects the extreme point and has the host optical galaxy bisect the opposite edge; the box is rotated so that the two edges parallel to the lobe major axis are tangent to the 5$\sigma$ contour (see Figure~\ref{fig:lobefits}). The length of each lobe is then taken to be the length of this rectangular box outwards from the central engine, and the flux density as the integrated surface brightness. The uncertainty in these parameters is quantified by comparing these estimates to those fitted assuming a lower 3$\sigma$ contour. For FR-II sources, the lobe length thus defined typically corresponds to the distance between the core and edge of the lobe. For FR-Is, lobe length defined in this way will necessarily be sensitivity limited \citep[e.g.][]{TurnerEA18a}; however, this does not bias our results as we are interested in the ratio of the lobe lengths of individual sources.

We use the location of the brightest point in each lobe to calculate the Fanaroff-Riley index, ${\rm FR}=2 x_{\rm bright} / x_{\rm length} +1/2$; in this formulation FR-Is have indices $0.5<{\rm FR}<1.5$ and FR-IIs have $1.5<{\rm FR}<2.5$ \citep{KrauseEA12}. The numerical values obtained are in excellent agreement with visual classifications. As shown in Table~\ref{tab:all_sources}, all visually-identified FR-II lobes have FR indices well in excess of 1.5, with the exception of the southern lobe of RGZ J093821.5+554333, where the small lobe size and source curvature are likely to render lobe length measurement unreliable. For FR-Is, only the eastern lobe of RGZ J094443.2+024754 has an FR index marginally in excess of 1.5; it also has an unusual double-knot structure in the jet. We note that our approach of measuring lobe lengths is an approximation, and cannot deal with curved sources. Below, we restrict our analysis to straight, FR-II sources for which our methodology is valid. There is one hybrid source, RGZ J082835.2+322825 (Figure~\ref{fig:hybrid}), with a clear FR-I morphology in the eastern lobe, and FR-II morphology in the western lobe. The source J125721.9+122820 shows diffuse emission almost perpendicular to the current lobe axis; such radio morphology could come about due to backflow in highly non-axysymmetric environments \citep{Hodges-KluckReynolds2011}, or be due to a previous jet episode \citep[e.g.][]{ChonEA2012}. For the purposes of the present analysis we note that the morphology of the bright FR-II lobes is well-defined, and we therefore retain this source in our sample. 

Finally, the lobe luminosities of some of the FR-IIs in our sample may be contaminated by hotspot emission (as seen in Figures~\ref{fig:FRII_alpha_1}-\ref{fig:FRII_beta_4}), which is unaccounted for in analytical radio source models. Hotspots are typically $\lesssim 1$ per cent of source size \citep{HardcastleEA98,GodfreyShabala13}, and hence sources with total length shorter than $\sim 200$ arcsec are likely to contain both lobe and hotspot emission at the peak brightness location. In Table~\ref{tab:all_sources}, we quote the fraction of the total lobe flux density contained in hotspots. The integrated flux density in the eastern lobe of RGZ J210030.5+100529 (Figure~\ref{fig:FRII_beta_4}) arises almost entirely from unresolved (hotspot) emission, whereas 41 per cent of the emission in the western lobe is attributed to the hotspot. We conclude that this source has substantial hotspot contamination, and exclude it from the analysis.

\subsection{Galaxy clustering}
\label{sec:clustering}

\begin{figure*}
\centering
{\includegraphics[width=0.65\columnwidth]{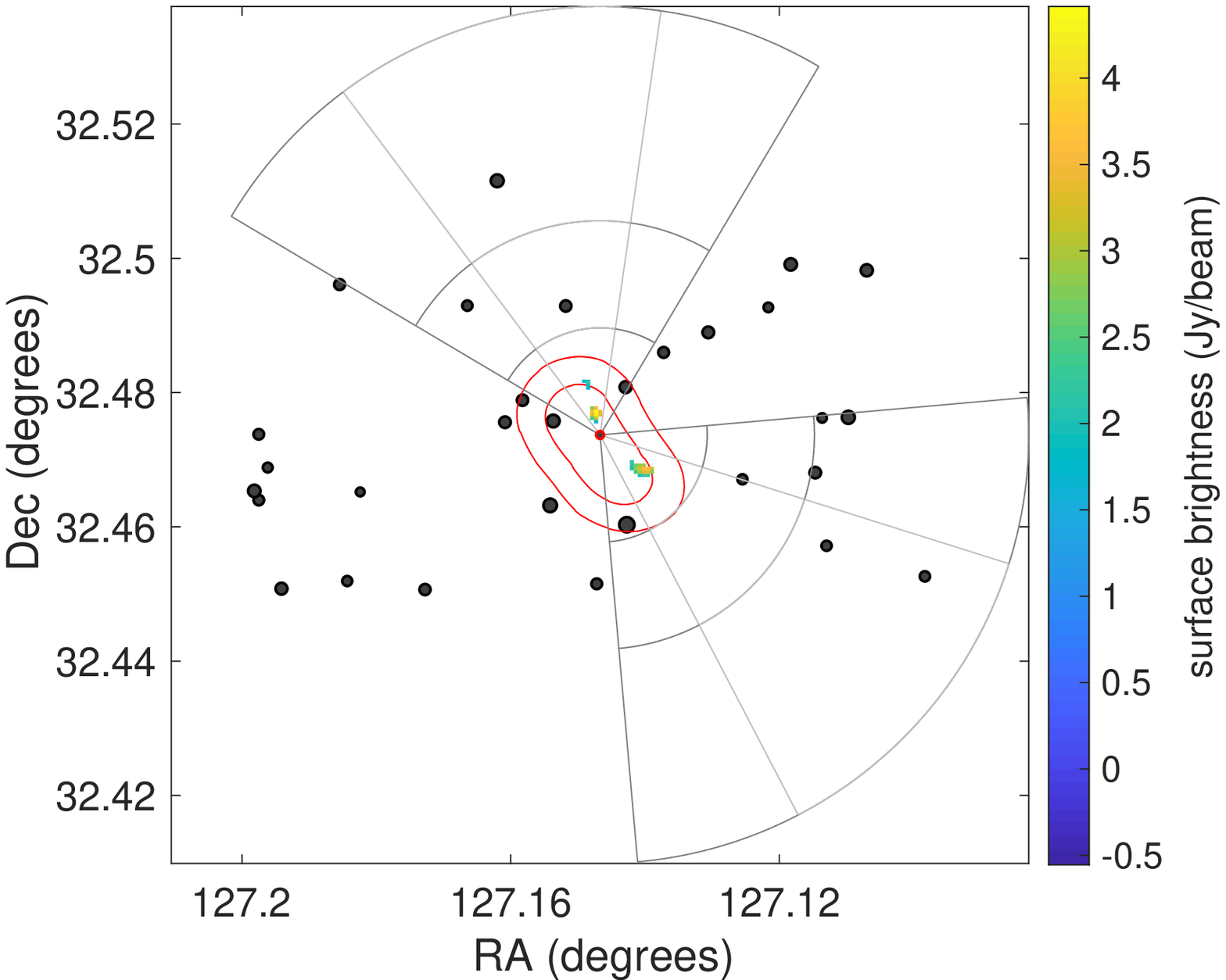} \;\:
\includegraphics[width=0.65\columnwidth]{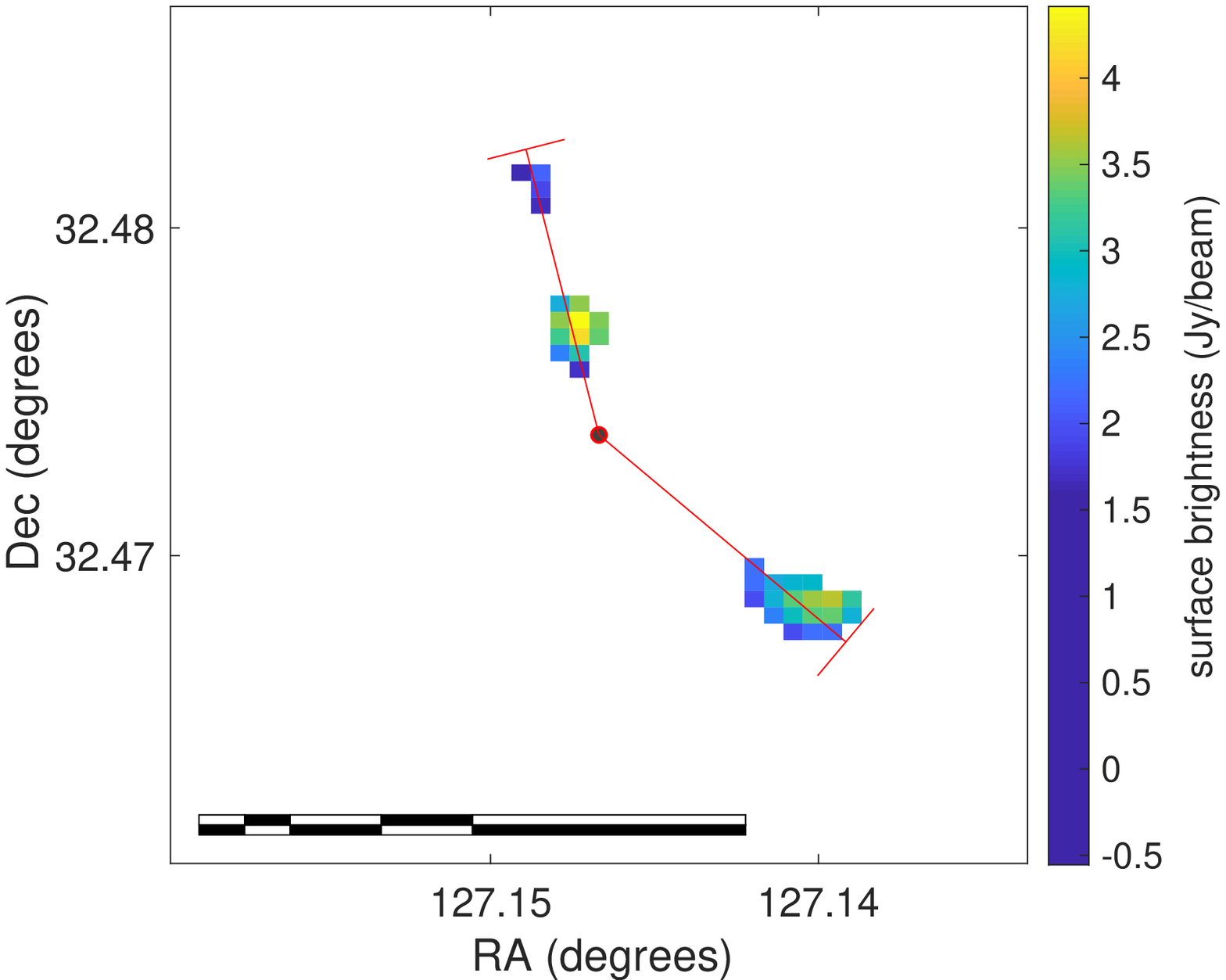} \;\:
\includegraphics[width=0.72\columnwidth]{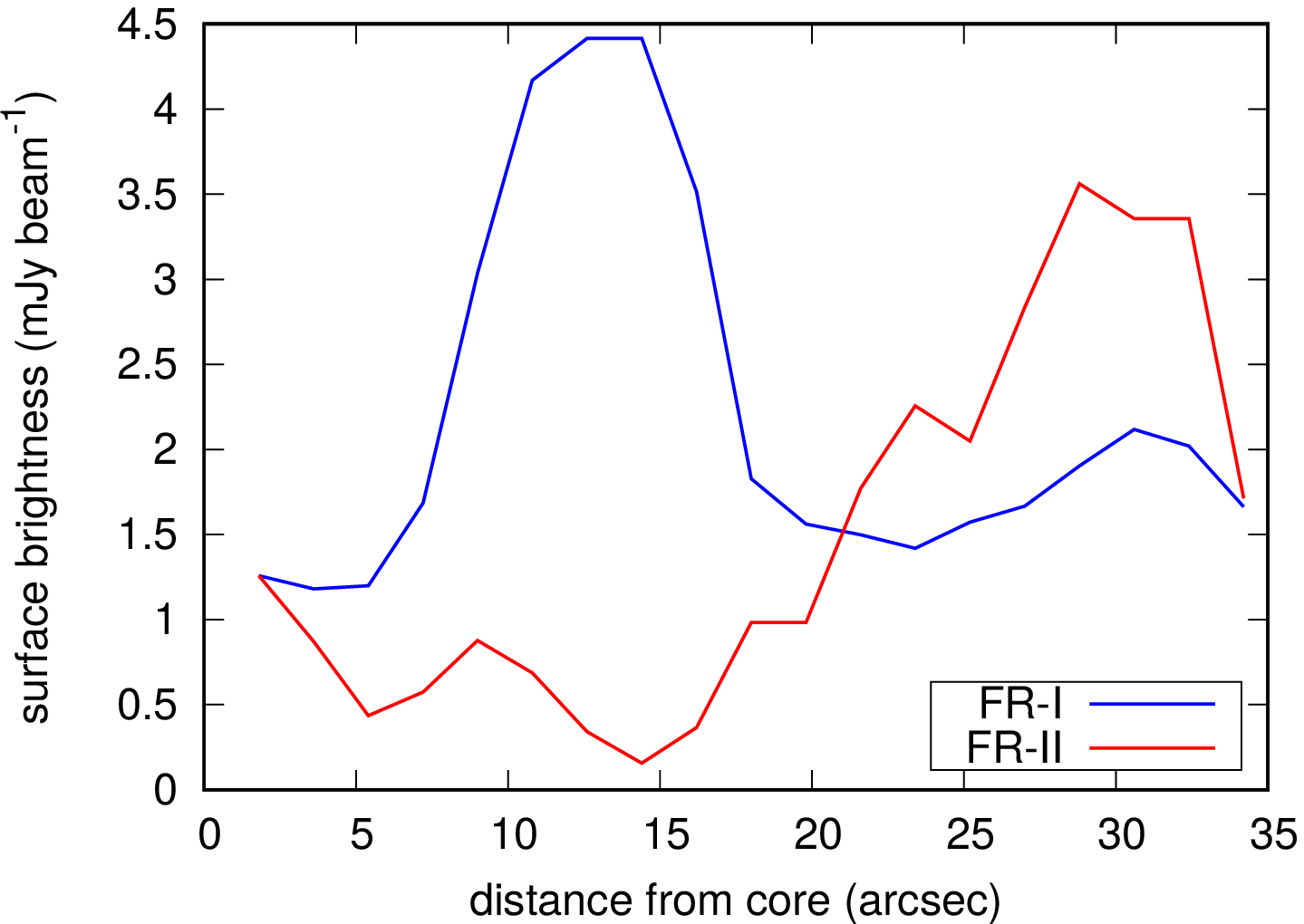}}
\caption{{\it Left and Middle: }Hybrid source RGZ J082835.2+322825; symbols are as in Figure~\ref{fig:FRII_alpha_11}. {\it Right: } The surface brightness profile along the lobe major axis of the hybrid source RGZ J082835.2+322825, measured from the active nucleus towards the extrema of each lobe. The profile of the eastern-most lobe (blue) is of characteristic FR-I type, whereas that of the western-most lobe (red) describes an FR-II.}
\label{fig:hybrid}
\end{figure*}

\begin{figure}
	\centering
	\includegraphics[width=0.8\columnwidth]{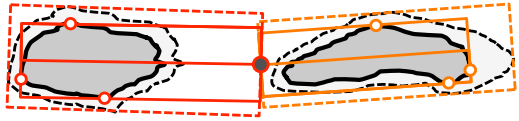}
	\caption{Definition of the spatial extent of the radio lobes using the outermost 5$\sigma$ (solid rectangles) for an FR-II type lobe ({\it left}) and an FR-I type lobe ({\it right}). Associated 3$\sigma$ contours (dashed rectangles) are used to estimate the uncertainty of the spatial extent. Bounding rectangles are chosen such that (i) the active nucleus lies at the midpoint on the base of the rectangle (filled circle), and (ii) the remaining sides are tangent to the extrema of the lobe (open circles). Radio lobes with lobe axes differing by more than approximately ten degrees, or whose axes are not confined within the horizontal extent of the lobe (i.e. are highly bent), are excluded from our analysis.}
	\label{fig:lobefits}
\end{figure}

In quantifying galaxy clustering associated with each lobe, we consider galaxies within a 1~Mpc radius and with redshifts consistent within 3$\sigma$ of the AGN host. The rectangular boxes fitted to the lobes, as described in Section \ref{sec:Radio source properties}, are used to estimate the direction of jet propagation are shown in Figure~\ref{fig:hybrid} by the red line extending outwards from the host galaxy. We then define `wedges' around each lobe, mirrored about the expansion axis and subtending angles of $45$ and $90$ degrees; these are shown by dashed and solid grey lines in Figure~\ref{fig:hybrid}. Neighbouring galaxies are assigned to a lobe if they are located within the appropriate wedge. By selecting for our sample only relatively straight pairs of lobes which are close to anti-parallel (to within approximately ten degrees), we ensure that all galaxies are associated with a maximum of only one lobe. The likelihood of including nearby isolated galaxies increases for lobes projected close to the line of sight; as sources with strong jet and hotspot emission have been excluded from this sample, projection effects will add further scatter but not cause any systematic effect.

\begin{figure}
\centering
	\includegraphics[width=1.05\columnwidth]{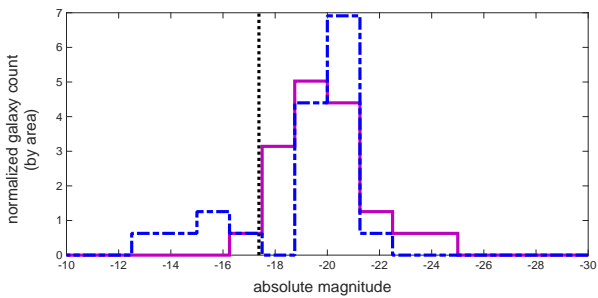}
	\caption{The distribution of absolute $z$-band magnitudes of galaxies surrounding FR-I source RGZ J095759.3+191609 (Figure~\ref{fig:FRI_3}), normalised to the area of the containing 90 degree wedge. The solid line corresponds to the western-most lobe, and the dot-dashed line to the eastern-most lobe. The dotted line is the 95 per cent completeness limit on the absolute $z$-band magnitude. Normalised galaxy counts are lower for galaxies less luminous than this limit, as expected.}
	\label{fig:zcompleteness}
\end{figure}

Our sample consists of both nearby and more distant ($z=0.3$) AGN. SDSS is 95 per cent complete at $m_z=20.5$~mag, and we use this cut-off, converted to absolute magnitude, when calculating clustering associated with each lobe. This procedure allows us to compare environments of radio sources with different morphologies (Section~\ref{sec:results_by_type}). In Section~\ref{sec:FRII_analysis}, we consider the differences between two lobes of the same radio source. We quantify environment asymmetry by taking the ratio of total galaxy counts associated with each lobe, above the completeness limit (Figure~\ref{fig:zcompleteness}). This is broadly equivalent to taking the ratio of the normalization in the stellar mass function of galaxies associated with each lobe: we repeated the analysis below using estimates of the total stellar mass (summed over all galaxies) associated with each wedge (calculated using $z$-band magnitudes), and found similar results.

\subsection{Sample statistics}

\begin{figure*}
\centering
\includegraphics[width=0.69\columnwidth]{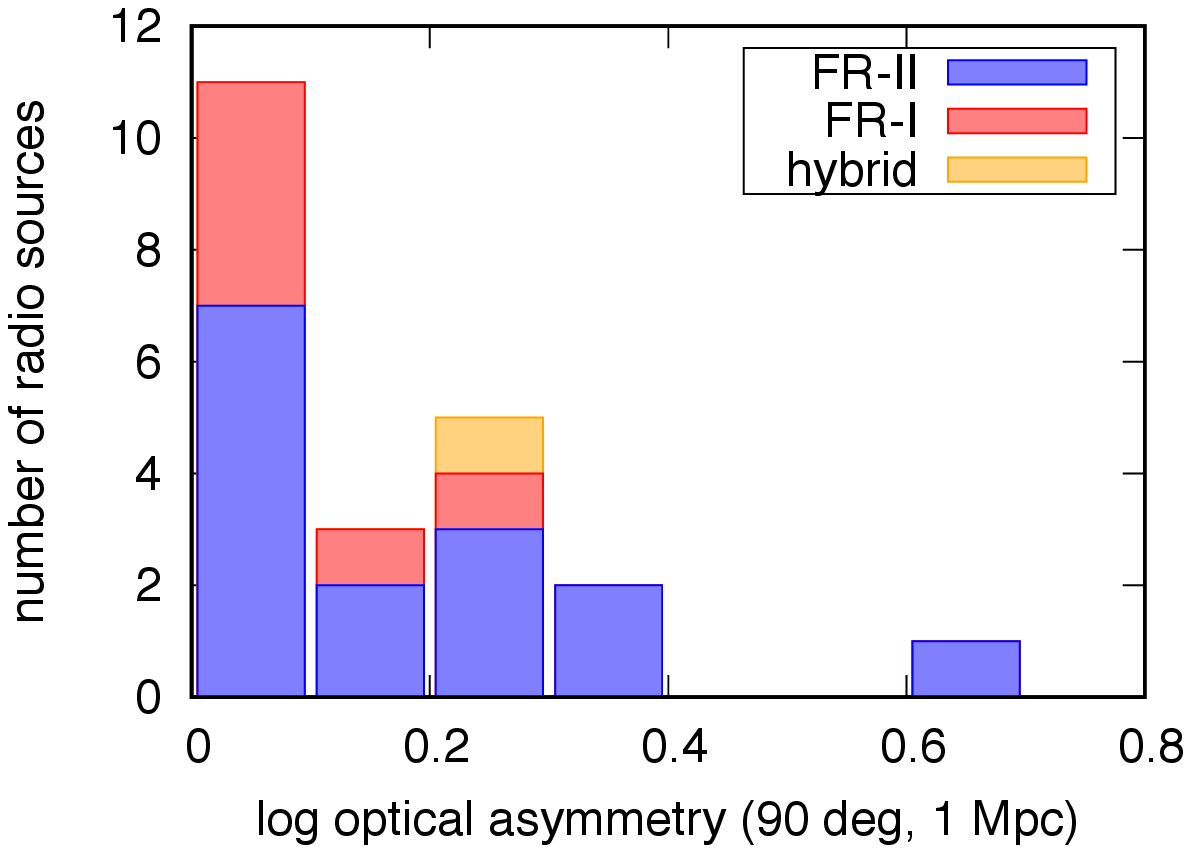}
\includegraphics[width=0.69\columnwidth]{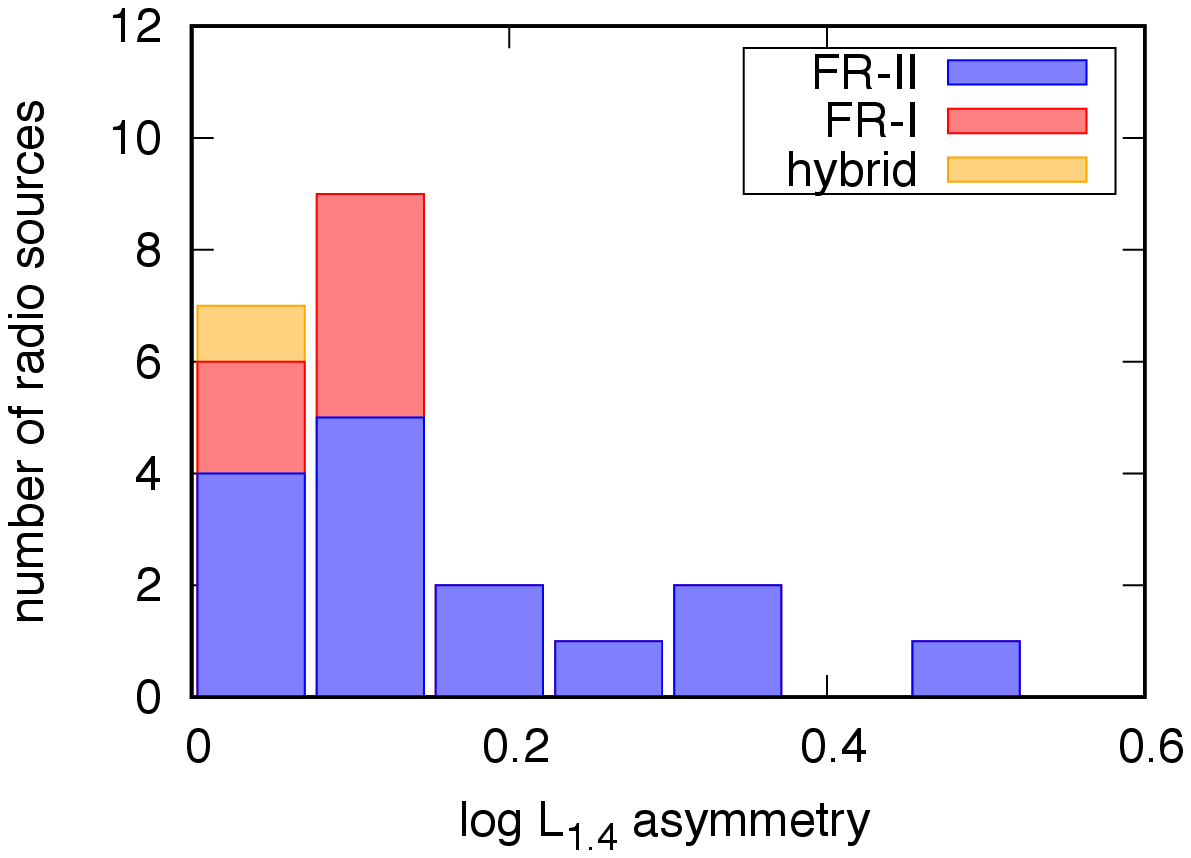}
\includegraphics[width=0.69\columnwidth]{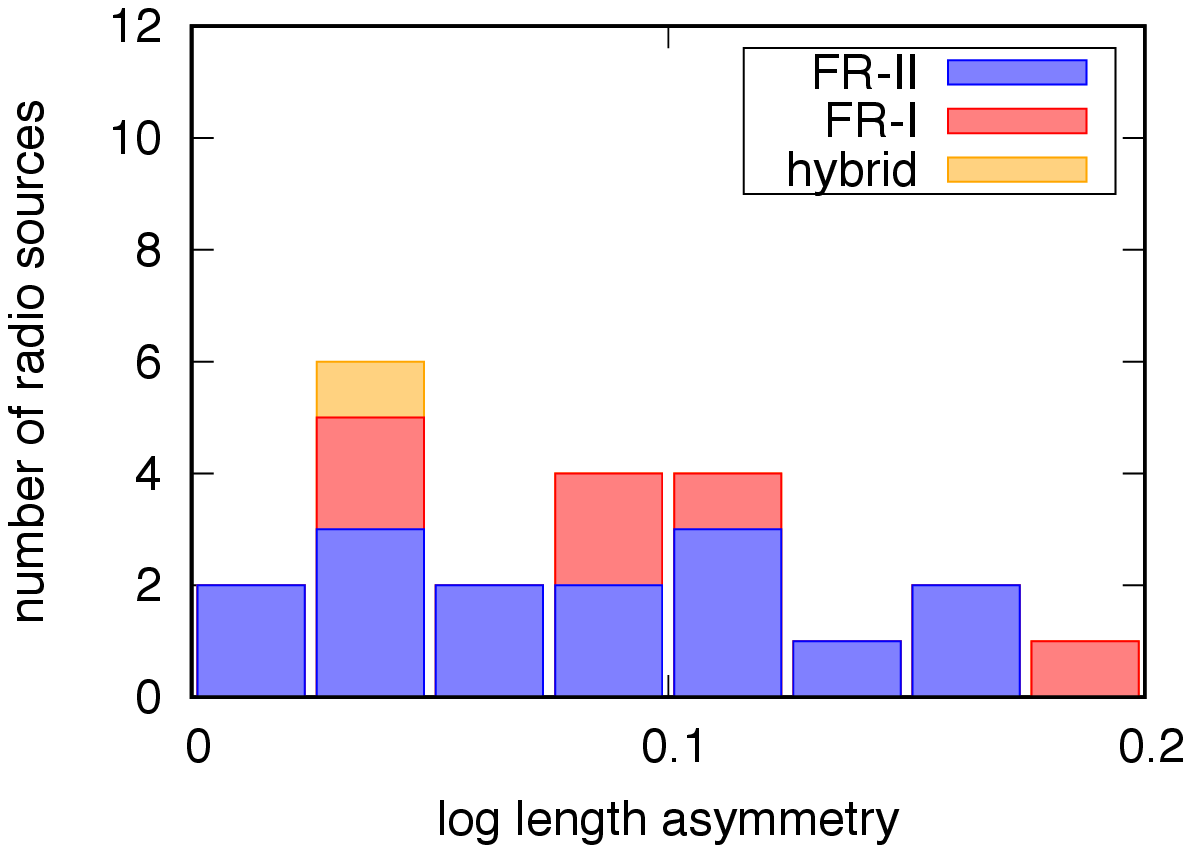}
\caption{Distribution of asymmetries in radio source properties for FR-II (blue), FR-I (red), and hybrid (yellow) sources, shown as stacked bars. {\it Left}: Distribution of asymmetry in galaxy clustering associated with each lobe, defined as the decimal log of the ratio of the number counts of galaxies brighter than 20.5~mag in SDSS $z$-band, located within a 90 degree wedge of 1~Mpc radius centred on the host galaxy. {\it Middle}: Asymmetry in 1.4 GHz FIRST radio luminosity, integrated out to 5$\sigma$ for each lobe. {\it Right}: Asymmetry in lobe lengths which are defined by the bounding rectangle described in Section~\ref{sec:Radio source properties} and Figure~\ref{fig:lobefits}.}
\label{fig:asymm_dist}
\end{figure*}

The aim of this paper is to investigate the relationship between asymmetries in radio lobe and environment properties. The left panel of Figure~\ref{fig:asymm_dist} shows the distribution of optical asymmetries, quantified as the logarithm of the ratio of galaxy counts above the limiting $z$-band magnitude of 20.5 within 1~Mpc in a 90-degree wedge associated with each lobe. Asymmetries in excess of 0.6~dex (i.e. a factor of 4) are observed for FR-II sources. The middle panel of Figure~\ref{fig:asymm_dist} shows a similar distribution for asymmetry in the radio continuum luminosity of each pair of lobes. Although FR-Is appear to show smaller asymmetries in both lobe continuum luminosity and galaxy counts, these differences are not statistically significant at the $p=0.05$ level using the Wilcoxon-Mann-Whitney test. We note that although FR-Is are known to preferentially reside in more massive environments \citep{HillLilly91,SabaterEA13} than FR-IIs, our selection bias against highly curved and asymmetric sources such as wide-angle tailed or head-tail radio galaxies will be an important contributing factor -- for example, overdense environments may preferentially produce bent sources, as found in Garon et al. (submitted). The right panel of Figure~\ref{fig:asymm_dist} shows the distribution of length asymmetries for the sources in our sample; again, we find no statistically significant difference at the $p=0.05$ level between the FR-I and FR-II samples.

In Section~\ref{sec:FRII_analysis} we investigate the relationship between radio and optical asymmetry diagnostics for the FR-II sources in our sample. We note again that length and luminosity measurements of FR-I sources are strongly affected by surface brightness sensitivity of observations \citep{ShabalaEA17,TurnerEA18a} and should be interpreted with caution; we do not consider the FR-I sample in detail further in this work.

\section{The role of environment}
\label{sec:results_by_type}

\subsection{The relationship between radio source and host galaxy properties}
\label{sec:The relationship between radio source and host galaxy properties}

\begin{figure}
\centering
\includegraphics[angle=-90,width=1\columnwidth]{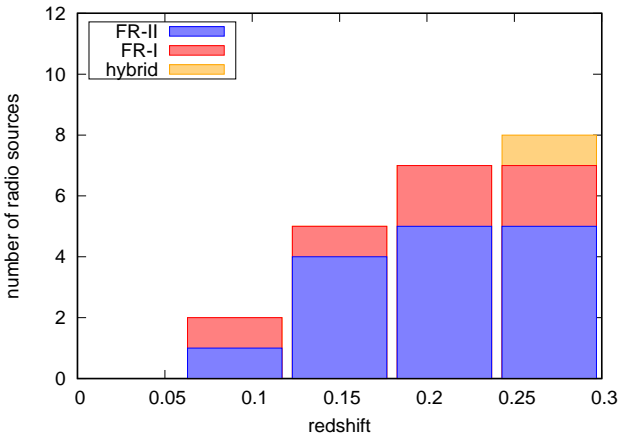}
\caption{The distribution in redshift for sources out to the redshift limit $z=0.3$, shown as stacked bars. Colours are as in Figure~\ref{fig:asymm_dist}.}
\label{fig:z_dist}
\end{figure}

\begin{figure}
\centering
\includegraphics[angle=-90,width=1\columnwidth]{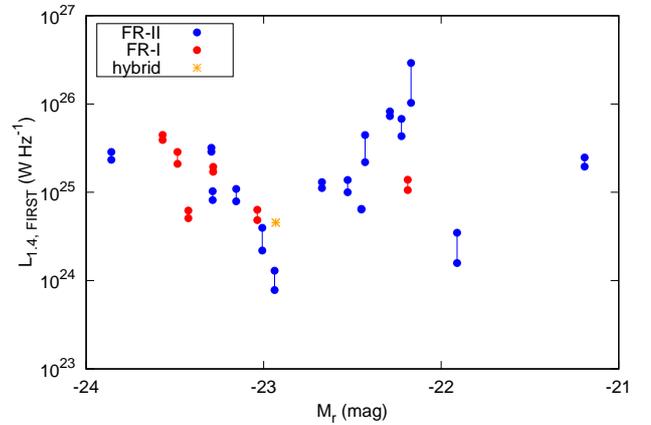}
\caption{The host galaxy optical absolute r-band magnitude versus lobe luminosity for each source, known as the Ledlow-Owen relation, taken after \citet{OwenLedlow94}. Each lobe is represented by a single point, with lobe pairs corresponding to a single source connected via a solid line. No clear separation of the classes is seen. Colours are as in Figure~\ref{fig:asymm_dist}.}
\label{fig:LedlowOwen}
\end{figure}

The 16 FR-IIs and 6 FR-Is in our sample have similar redshift distributions (indistinguishable at the $p=0.05$ level, shown in Figure~\ref{fig:z_dist}), and as such we can confidently compare the typical environments inhabited by the two morphological classes. Figure~\ref{fig:LedlowOwen}, taken after \citet{OwenLedlow94}, shows the relationship between lobe luminosity and host galaxy optical magnitude. We recover the previously found result that FR-Is are hosted by more massive galaxies (i.e. optically brighter) than FR-IIs, albeit at only at a 2$\sigma$ confidence level, most likely due to the small FR-I sample size. There is however no clear separation between FR-I and FR-II sources in this plot, consistent with the results of \citet{Best09} who similarly used a redshift-matched sample.

\begin{figure}
\centering
\includegraphics[angle=-90,width=1\columnwidth]{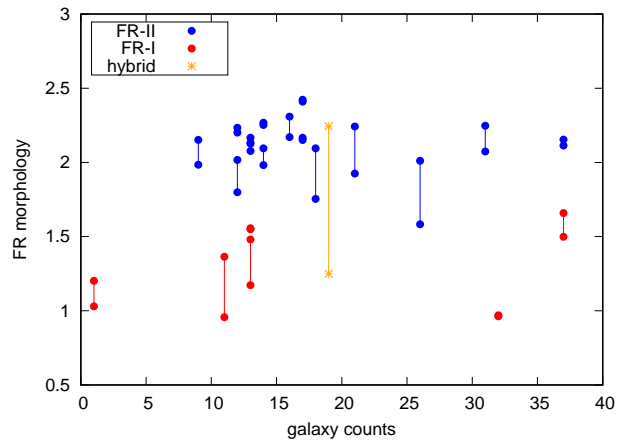}
\caption{The FR index as a function of complete ($m_z>20.5$~mag) SDSS galaxy counts within 1~Mpc of the active nucleus. Consideration of the large-scale environment does not provide a clear distinction between the morphologies in this sample. Colours are as in Figure~\ref{fig:asymm_dist}.}
\label{fig:FRindex_vs_clustering}
\end{figure}

Figure~\ref{fig:FRindex_vs_clustering} probes the large-scale environment (out to 1~Mpc) of the radio sources in our sample, quantified via counts of galaxies with $M_z < -20.47$~mag (corresponding to $m_z=20.5$~mag at $z=0.3$, our highest redshift). Perhaps surprisingly, FR-Is do not appear to inhabit denser environments as has been reported by several authors previously \citep[e.g.][]{SabaterEA13,MiraghaeiBest17}. This is again likely due to our selection effect against compact radio sources, which dominate number counts in dense environments \citep{ShabalaEA08}. Conversely, this result shows that the large-scale environment is not a crucial factor in the determination of the radio lobe morphology; the FR-I morphology is likely determined much closer to the active nucleus.

\subsection{FR-II sources}
\label{sec:FRII_analysis}

\subsubsection{Theoretical considerations}
\label{sec:theory}

Dynamical models of double-lobed radio sources make clear predictions for the relationship between FR-II size, radio continuum luminosity, and environment properties. For a classical FR-II expanding into an atmosphere with a power-law density of the form $\rho(r)=\rho_0 \left( r/r_0 \right)^{-\beta}$, Equations 1 and 4 of \citet{ShabalaGodfrey13} give the scalings between source linear size $D$, continuum luminosity $L_{\nu}$ and atmosphere gas density $\rho_0$ as,

\begin{eqnarray}
D \propto \rho_0^{-\left( \frac{1}{5-\beta} \right)}\\
L_\nu \propto \rho_0^{\frac{5+s}{12}} D^{3 - \left( \frac{4+\beta}{3} \right) \left( \frac{5+s}{4}\right)}
\label{eqn:model_scalings}
\end{eqnarray}
where $s\approx2-2.5$ is the power-law index of the electron energy distribution at the hotspot \citep{KaiserEA97,WillottEA99,TurnerEA18b}.

Assuming typical $\beta$ values between 0 and 2, the expected scalings are, 
\begin{eqnarray}
D \propto \rho_0^{[-0.33,-0.2]}\\
L_\nu \propto D^{[-2.63,-2.25]}\propto \rho_0^{[-0.13,+1.25]} .
\end{eqnarray}
The luminosity calculation above assumes no electron ageing, and a direct mapping between radio lobe pressure and continuum luminosity. In practice, factors such as spectral ageing \citep{Murgia03,Turner18}, non-equipartition magnetic fields \citep{CrostonHardcastle14}, magnetic field inhomogeneities \citep{Hardcastle13} and particle re-acceleration \citep{JonesEA99} will introduce significant scatter in the luminosity relations, and we therefore expect the size--density relation to be the tightest. 

\subsubsection{Dependence on the large-scale environment}

\begin{figure*}
\centering
\includegraphics[angle=-90,width=1\columnwidth]{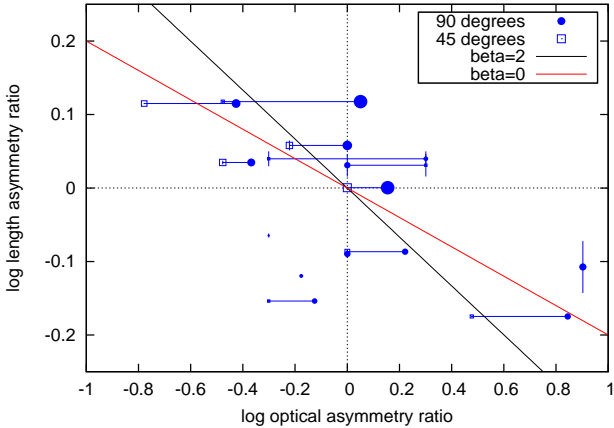}\hspace{0.03\textwidth}
\includegraphics[angle=-90,width=1\columnwidth]{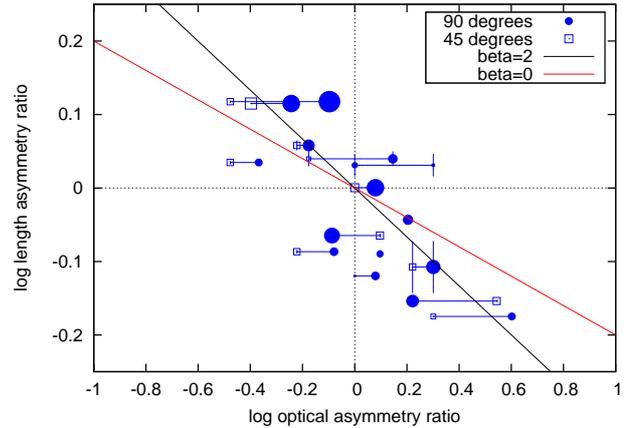}
\caption{Asymmetry in FR-II lobe length and environment. Symbol size is proportional to the total number of galaxies used in the environment asymmetry calculation, i.e. larger points correspond to higher galaxy counts. Environment asymmetry is quantified using galaxies with photometric redshift consistent with the AGN host, within either 500~kpc ({\it left}) or 1~Mpc ({\it right}) from the host galaxy. Model predictions correspond to a flat ($\beta=0$) density profile typical of inner ($\sim 30$~kpc) regions of clusters; and steep ($\beta=2$) environments more representative of large ($\gtrsim$100 kpc) clustercentric radii.}
\label{fig:FRII_length_env}
\end{figure*}

Figure~\ref{fig:FRII_length_env} shows the relationship between lobe length asymmetry and galaxy clustering. Both quantities are calculated by taking the ratio of the relevant quantity (length, or galaxy counts above the completeness limit) of the eastern lobe divided by the western lobe. Environment asymmetry is calculated for both 500~kpc and 1~Mpc radii from the AGN host, and using galaxies within 45 and 90 degree wedges centred on the lobe axis. Uncertainty in lobe length ratio is estimated by calculation of this quantity for two cuts in the signal-to-noise ratio, at 5$\sigma$ and 3$\sigma$ respectively; length measurements of prominent FR-II lobes are not expected to vary substantially with the adopted noise cut. Regardless of the environment metric, there is a consistent, clear anti-correlation between these two quantities as indicated by a deficit of sources in the top right and bottom left corners of the plot. Moreover, the slope of the observed relation ($D \propto \rho_0^{-0.29\pm0.07}$, a non-zero slope at the 4$\sigma$ level for the 500 kpc, 90 degree case) is consistent with expectation from analytical models that the galaxy clustering traces the underlying gas density profile, shown as solid lines for steep (black) and flat (red) gas density profiles. The best-fit relationship suggests the typical radio source in the sample expands into a host environment with a pressure profile falling off with radius as $\rho_0 \propto r^{-1.6}$, representative of gas distribution at large cluster-centric radii.


\begin{figure}
\centering
\includegraphics[angle=-90,width=1\columnwidth]{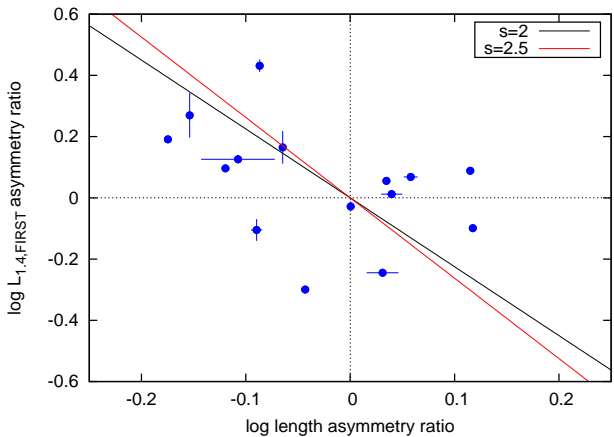}
\caption{Lobe luminosity and length asymmetry for FR-II sources. Symbols are as in Figure~\ref{fig:FRII_length_env}.}
\label{fig:FRII_length_lumin}
\end{figure}

The simplest analytical models of the kind described in Section~\ref{sec:theory} assume that radio lobes evolve self-similarly over their lifetime. This assumption is inconsistent with the observed increase in the radio source aspect ratio with length \citep{HardcastleEA98,TurnerEA18b}, which is also predicted by more realistic analytical models \citep{TurnerShabala15} and numerical simulations \citep{HardcastleKrause13}. Recent numerical simulation work (Vandorou et al., in prep.) shows that relaxing the self-similarity assumption can explain much of the observed scatter in the length--environment asymmetry relation shown in Figure~\ref{fig:FRII_length_env}.

\begin{figure*}
\centering
\includegraphics[angle=-90,width=1\columnwidth]{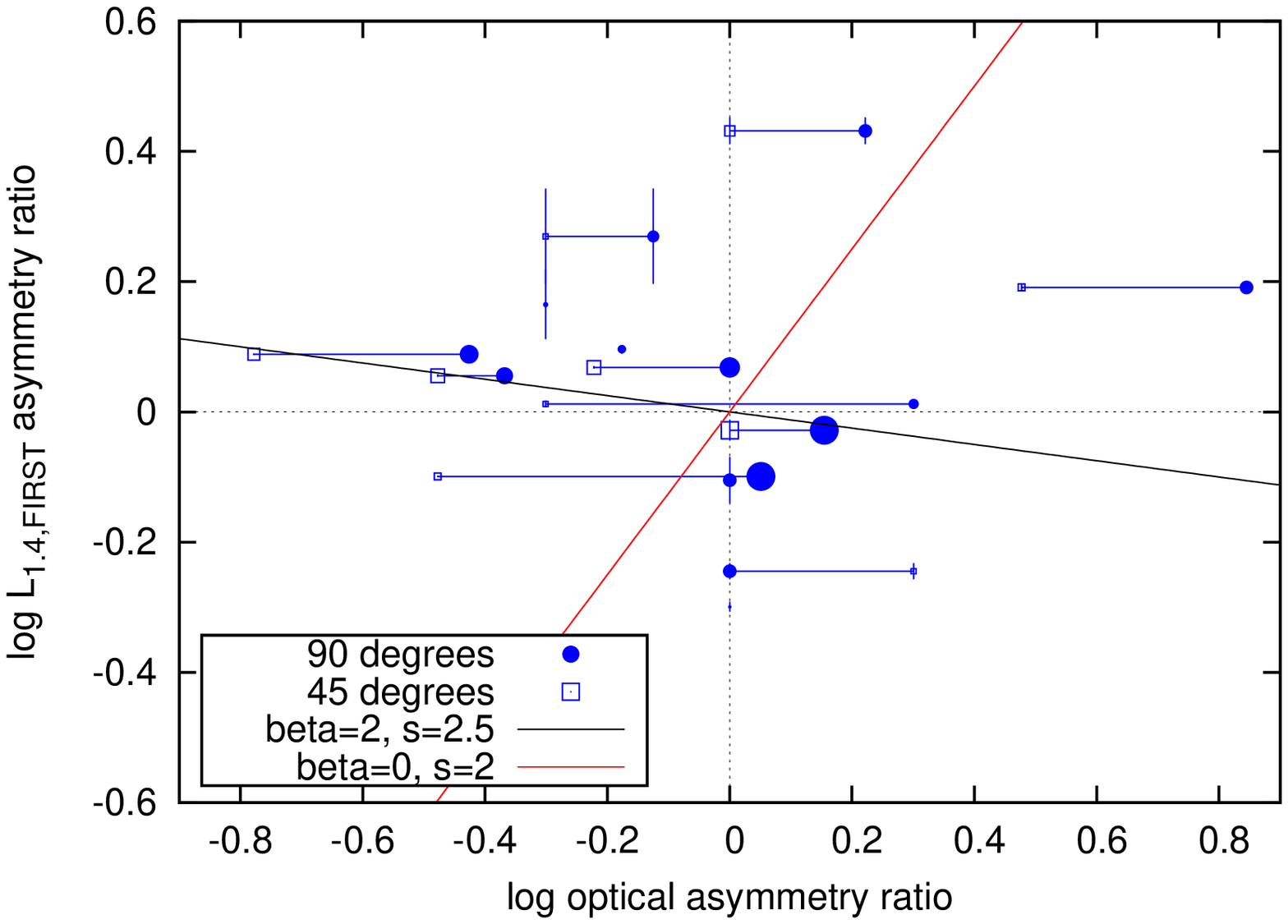}\hspace{0.03\textwidth}
\includegraphics[angle=-90,width=1\columnwidth]{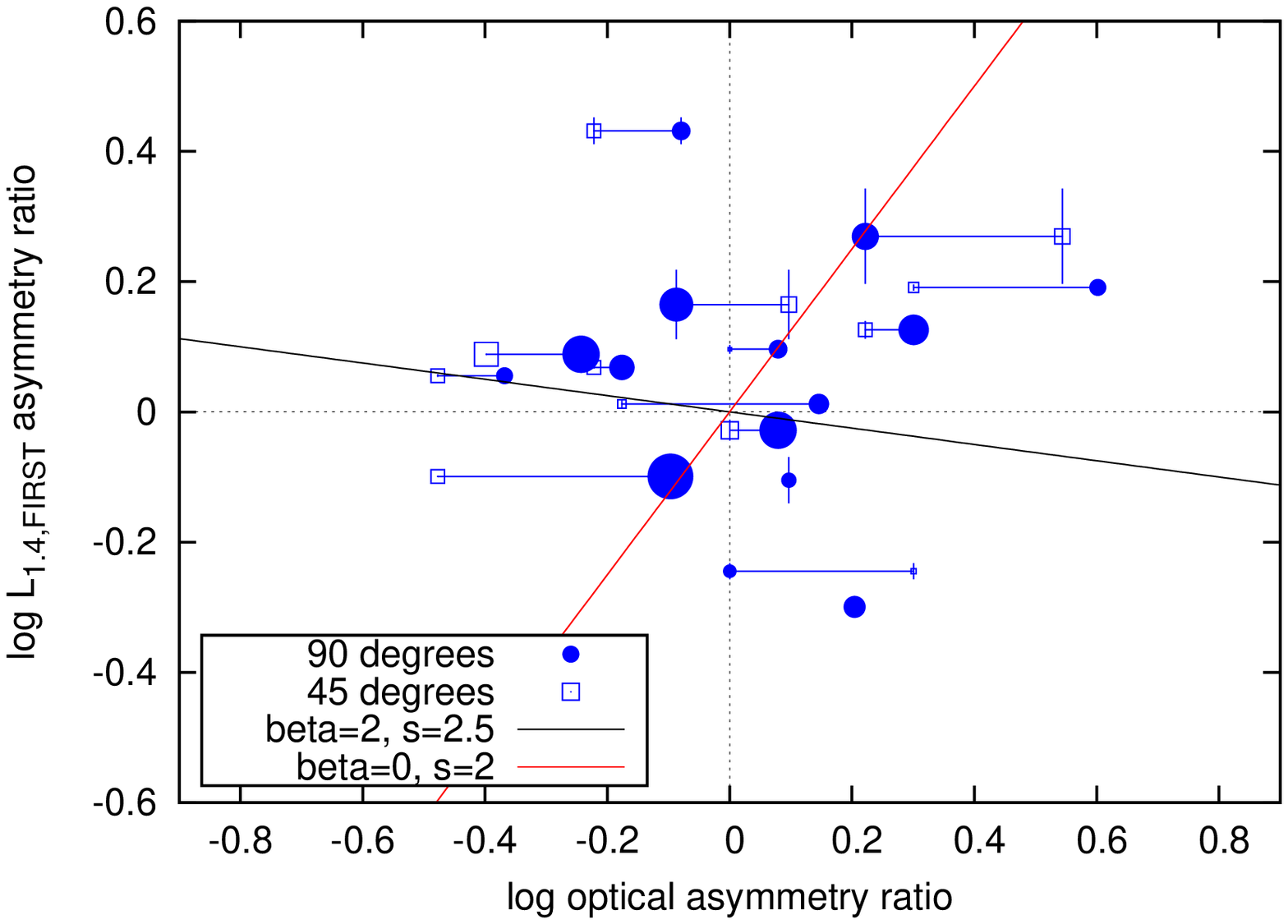}
\caption{Asymmetry in FR-II lobe luminosity and environment. Panels and symbols are as in Figure~\ref{fig:FRII_length_env}.}
\label{fig:FRII_lumin_env}
\end{figure*}

Figure~\ref{fig:FRII_length_lumin} shows the observed relationship between asymmetry in lobe length and luminosity. Definition of which emission belongs to the lobe is important: taking a higher threshold in the signal-to-noise ratio will systematically affect the lobe luminosity ratio by preferentially resolving out emission from the lower surface brightness lobe. Following our approach for lobe length measurement, the uncertainty in lobe luminosity ratio is again calculated by considering the difference in this quantity for signal-to-noise ratios of 3 and $5\sigma$. Both the approximate anti-correlation of $L_\nu \propto D^{-3.9\pm1.9}$ (non-zero slope at the 2$\sigma$ level) and large scatter in Figure~\ref{fig:FRII_length_lumin} are qualitatively consistent with the expectation from models. Idealized model predictions in Figure~\ref{fig:FRII_length_lumin} are only indicative as they do not take into account electron ageing, which will be significant for the large (and hence old) radio galaxies in our sample. The complex luminosity evolution of individual radio sources \citep[by over an order of magnitude, and strongly environment-dependent; ][]{HardcastleKrause13} will significantly contribute to the scatter in these relations; for example, using FR-II samples carefully matched in jet power and source size, \citet{ShabalaGodfrey13} showed that environment can contribute approximately 0.3~dex to scatter in radio luminosity. We defer a detailed discussion of these points to future work.

Finally, we plot the relationship between lobe luminosity asymmetry ratio and environment in Figure~\ref{fig:FRII_lumin_env}. Unlike lobe length, which increases monotonically with age, lobe luminosity is predicted to vary non-monotonically over the lifetime of a radio source; hence the lobe luminosity asymmetry ratio will depend on source age in addition to jet and environment properties \citep{HardcastleKrause13,Shabala18}. The lack of any correlation at above the $2\sigma$ level in Figure~\ref{fig:FRII_lumin_env} is expected from the wide range of factors which influence the lobe luminosity, as demonstrated by the broad range of predictions from analytical models (solid lines).

\subsection{FR-I and hybrid sources}
\label{sec:FR-I sources}

There are only six FR-I sources in our sample, hence we are unable to perform a similar analysis for these objects. From Table~\ref{tab:all_sources} there appears to be no obvious relationship between length or luminosity asymmetry and environment for FR-I sources, in broad agreement with complex expectations from the models. Jets are theoretically expected to propagate slower in denser environments \citep{KaiserAlexander97}; however, such environments may also be more conducive to keeping the jets collimated \citep{KrauseEA12}, allowing them to propagate further. Environmental boosting \citep{ArnaudEA10} is also likely to increase the jet surface brightness (and luminosity), enabling those jets propagating into denser gas to be visible to larger distances from the host galaxy \citep{ShabalaEA17,TurnerEA18a}. Given these competing effects and the small size of our sample, we are unable to draw any meaningful conclusions about the FR-I population.

A single hybrid radio source (RGZ J082835.2+322825) is present in our sample, and is shown in Figure~\ref{fig:hybrid}. The SW lobe shows a clear FR-II morphology, while the northern lobe appears to have a bright flare point followed by a gradual decrease in surface brightness. The FR indices are 2.24 and 1.25 respectively, consistent with the visual classification of this source as a hybrid. Small number statistics make environmental analysis difficult: there are 4 galaxies in a 90-degree wedge within 1~Mpc of the AGN host on the FR-I side, and 7 galaxies on the FR-II side, reducing to only 2 galaxies on each side for a 45-degree wedge; we refrain from speculating on the causes of this source's morphology.

\section{Conclusions}
\label{sec:conclusions}

We have presented a sample of extended radio AGN identified as part of the Radio Galaxy Zoo citizen science project. Our sample consists of 16 FR-II objects, 6 FR-Is and one hybrid morphology radio source. The environments into which these objects expand have been quantified using optical photometry from SDSS, to investigate the effect that the large-scale environment has on the size and luminosity evolution of radio sources. Small sample numbers preclude us from drawing any meaningful conclusions about the FR-I and hybrid populations.

For the FR-II sources, we find that:

\begin{itemize}
\item The length of an FR-II radio lobe is strongly negatively correlated (statistically significant at the $4\sigma$ level) with the number density of galaxies in the environment into which it expands; this relationship is consistent with analytical models \citep[e.g.][]{KaiserAlexander97,TurnerShabala15}.

\item Luminosity ratio of FR-II lobes is moderately (statistically significant at the $2\sigma$ level) negatively correlated with the lobe length ratio, again consistent with analytical models.

\item There is no clear correlation between the asymmetry in luminosity of FR-II lobes and the number density of galaxies in their vicinity; the large observed scatter is consistent with the sensitivity of this relation to changes in the environment density profile and lobe electron energy distribution.

\end{itemize}

The excellent agreement between data and predictions from analytical models suggests that galaxy clustering provides a useful measure of the radio lobe environment. In the coming years, a combination of radio source models \citep[e.g. ][]{TurnerShabala15,Hardcastle18} and large-area radio continuum surveys supplemented by galaxy clustering information, such as the GAMA Legacy ATCA Southern Survey (GLASS), the Evolutionary Map of the Universe \citep[EMU; ][]{NorrisEA11}, and the LoFAR Two-metre Sky Survey \citep[LoTSS; ][]{ShimwellEA16}, will enable a detailed census of the physical properties of radio AGN populations.

\section*{Acknowledgements}

P.R. thanks the University of Tasmania for a Dean's Summer Reseach Studentship. R.T. thanks the University of Tasmania for an Elite Research Scholarship and the CSIRO for a CASS studentship. S.S. thanks the Australian Research Council for an Early Career Fellowship, DE130101399. H.A. has benefitted from grant DAIP \#066/2018 of Universidad de Guanajuato. Partial support for the work of A.F.G. and L.R. comes from grant AST-1714205 to the University of Minnesota from the U.S. National Science Foundation.

This research has made use of the NASA/IPAC Extragalactic Database (NED) which is operated by the Jet Propulsion Laboratory, California Institute of Technology, under contract with the National Aeronautics and Space Administration.

The National Radio Astronomy Observatory is a facility of the National Science Foundation operated under cooperative agreement by Associated Universities, Inc.

Funding for the Sloan Digital Sky Survey (SDSS) has been provided by the Alfred P. Sloan Foundation, the Participating Institutions, the National Aeronautics and Space Administration, the National Science Foundation, the U.S. Department of Energy, the Japanese Monbukagakusho, and the Max Planck Society. The SDSS Web site is http://www.sdss.org/.

The SDSS is managed by the Astrophysical Research Consortium (ARC) for the Participating Institutions. The Participating Institutions are The University of Chicago, Fermilab, the Institute for Advanced Study, the Japan Participation Group, The Johns Hopkins University, Los Alamos National Laboratory, the Max-Planck-Institute for Astronomy (MPIA), the Max-Planck-Institute for Astrophysics (MPA), New Mexico State University, University of Pittsburgh, Princeton University, the United States Naval Observatory, and the University of Washington.

This publication has been made possible by the participation of more than 11,000 volunteers in the Radio Galaxy Zoo project. The data in this paper are the results of the efforts of the Radio Galaxy Zoo volunteers, without whom none of this work would be possible. Their efforts are individually acknowledged at http://rgzauthors.galaxyzoo.org.

\bibliography{RGZasymmetry.bib} 
\bibliographystyle{mnras}

\appendix

\begin{figure*}
\section{FR-II radio and optical images}
\label{sec:appendixa}
\end{figure*}

\begin{figure*}
	\centering
	\includegraphics[width=0.8\columnwidth]{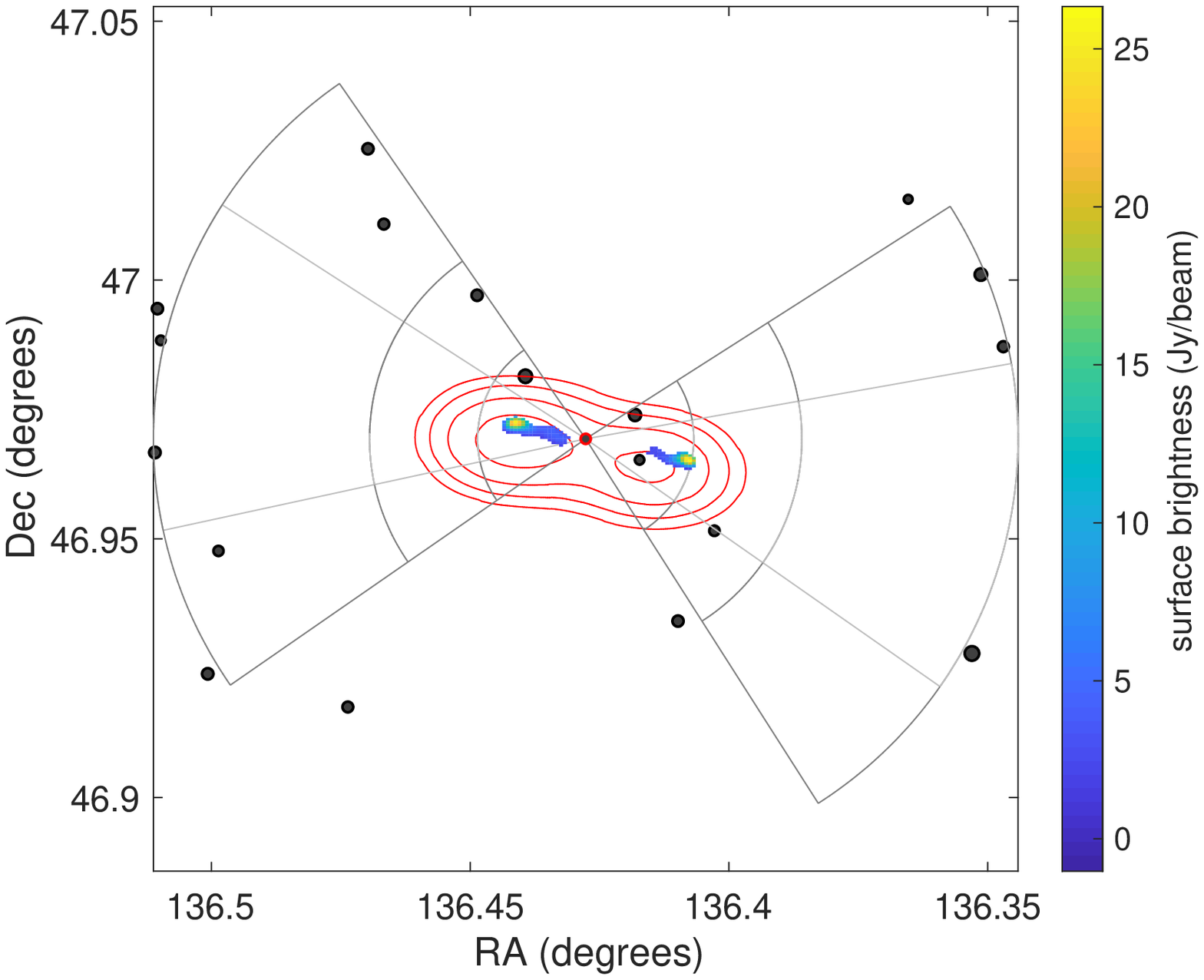}\hspace{0.15\textwidth}
	\includegraphics[width=0.8\columnwidth,clip]{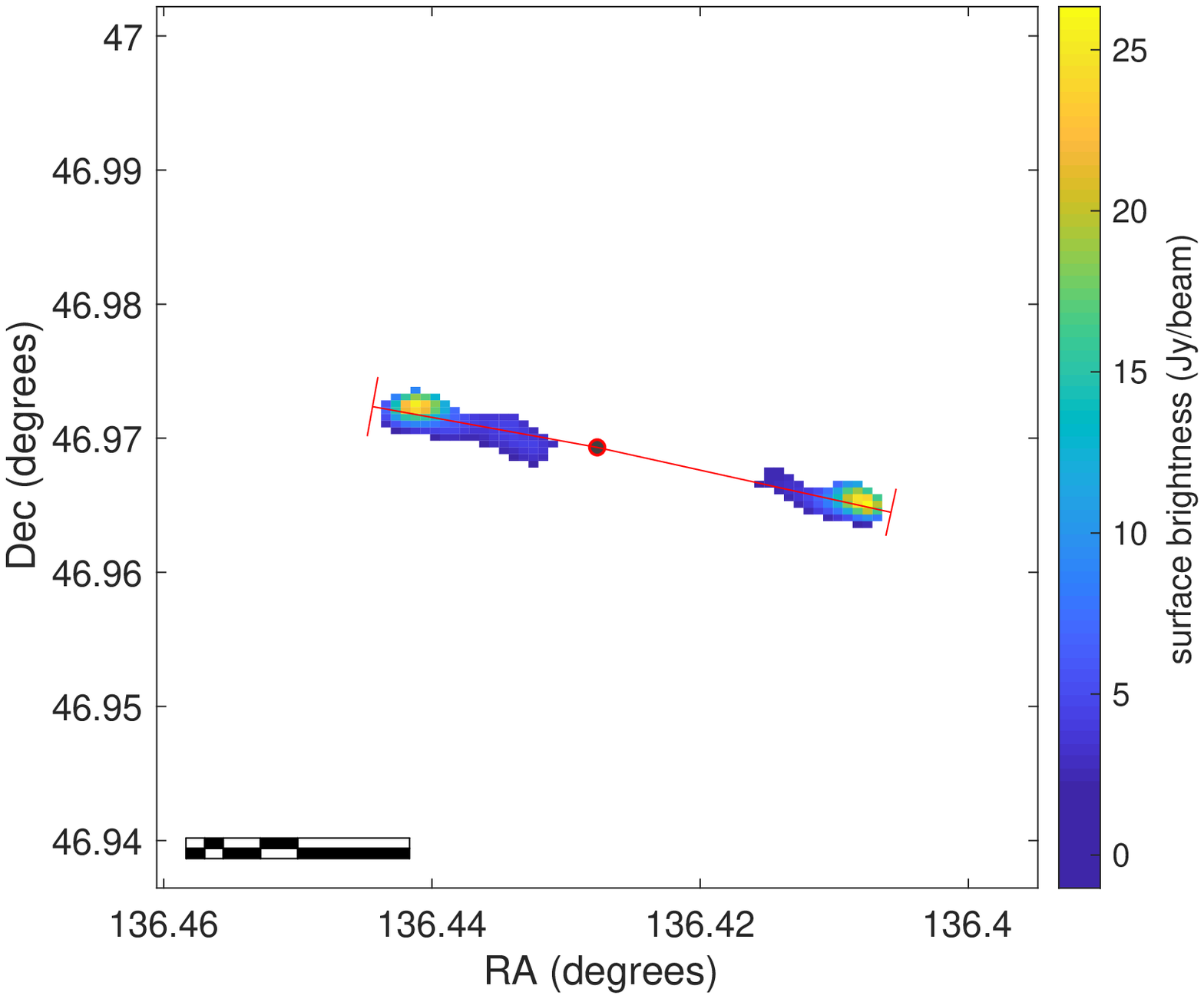}
	\caption{FR-II source RGZ J090542.6+465809; symbols are as in Figure~\ref{fig:FRII_alpha_11}.}
	\label{fig:FRII_alpha_1}
\end{figure*}

\begin{figure*}
	\centering
	\includegraphics[width=0.8\columnwidth]{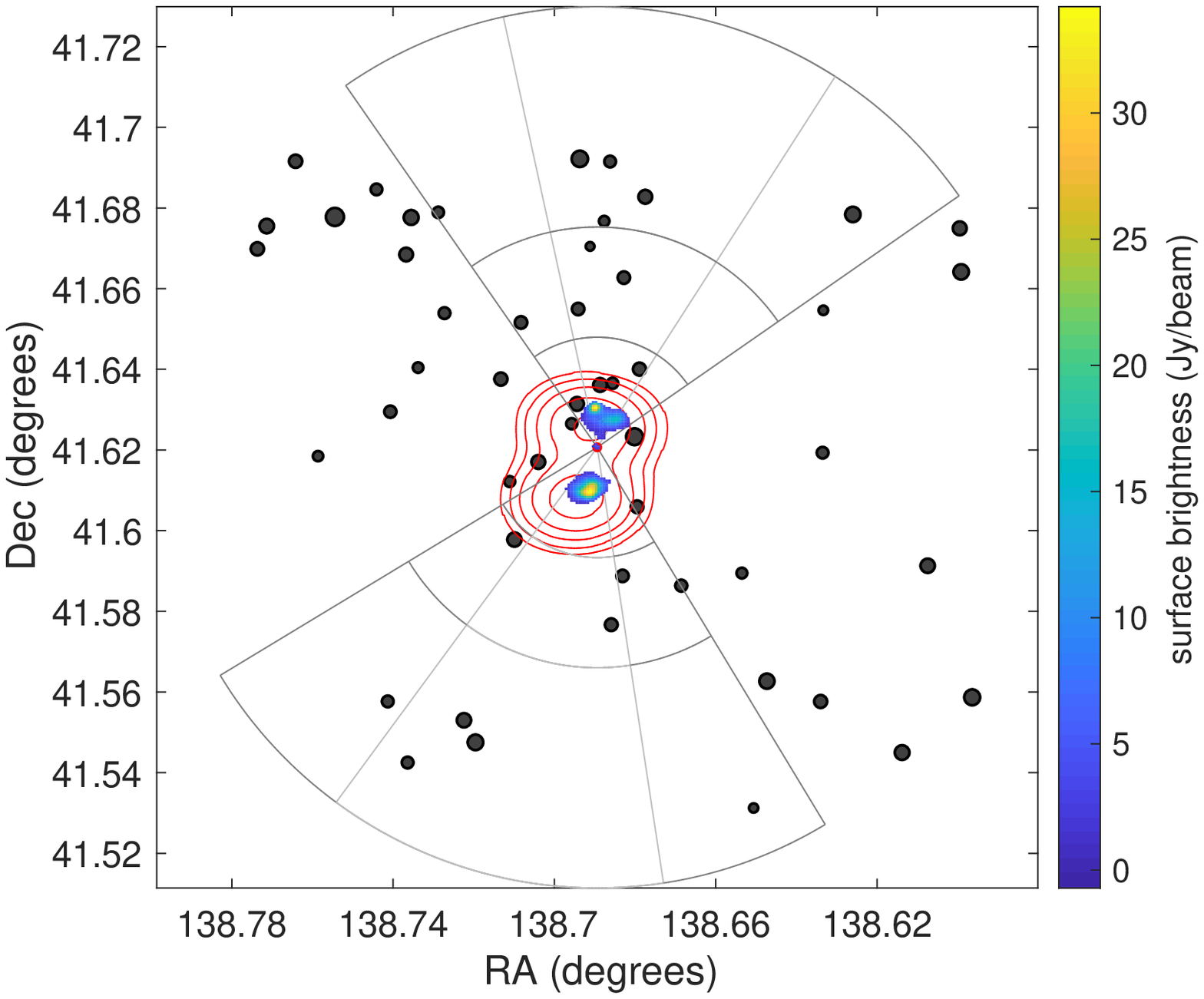}\hspace{0.15\textwidth}
	\includegraphics[width=0.8\columnwidth,clip]{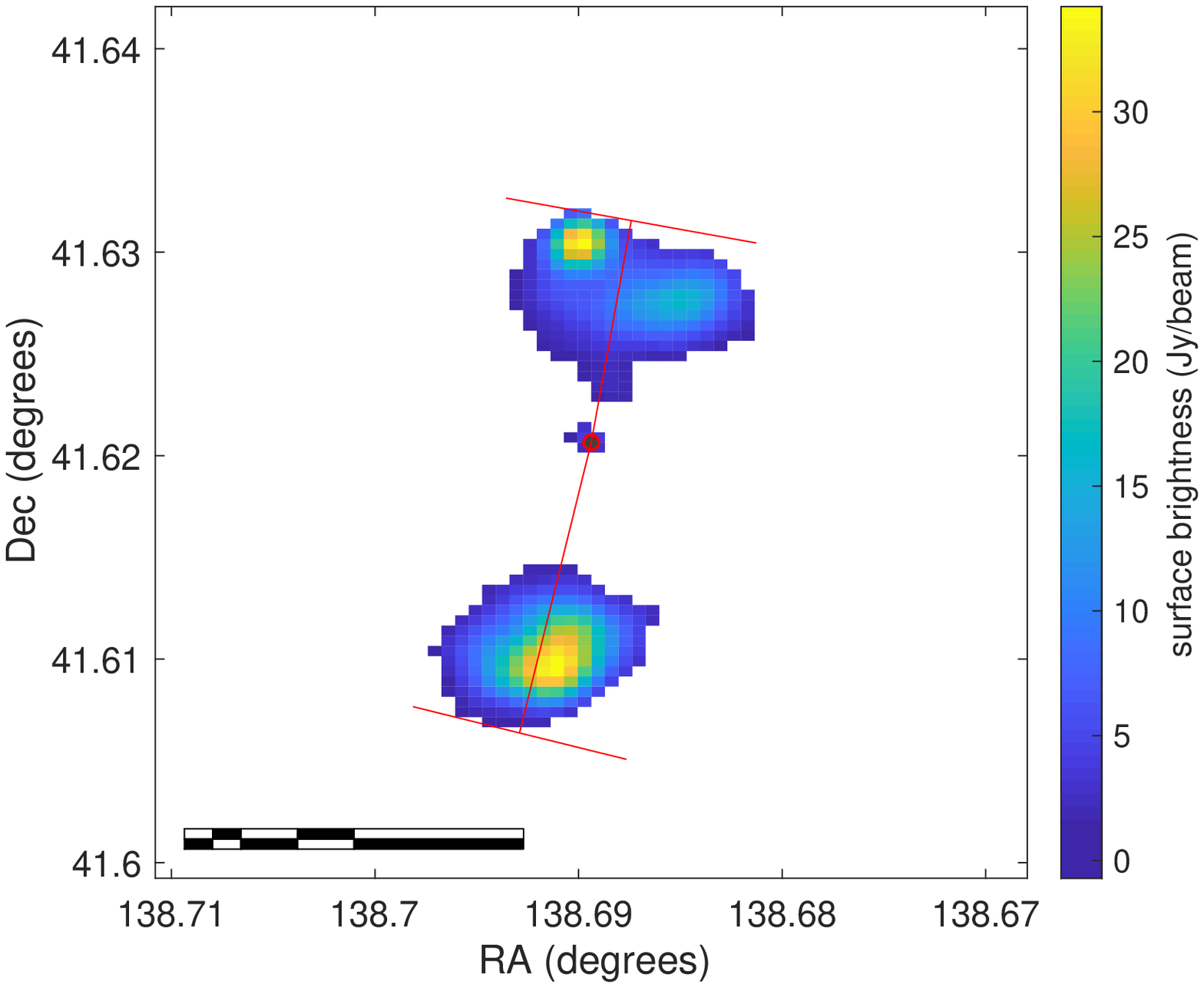}
	\caption{FR-II source RGZ J091445.5+413714; symbols are as in Figure~\ref{fig:FRII_alpha_11}.}
	\label{fig:FRII_alpha_2}
\end{figure*}

\begin{figure*}
	\centering
	\includegraphics[width=0.8\columnwidth]{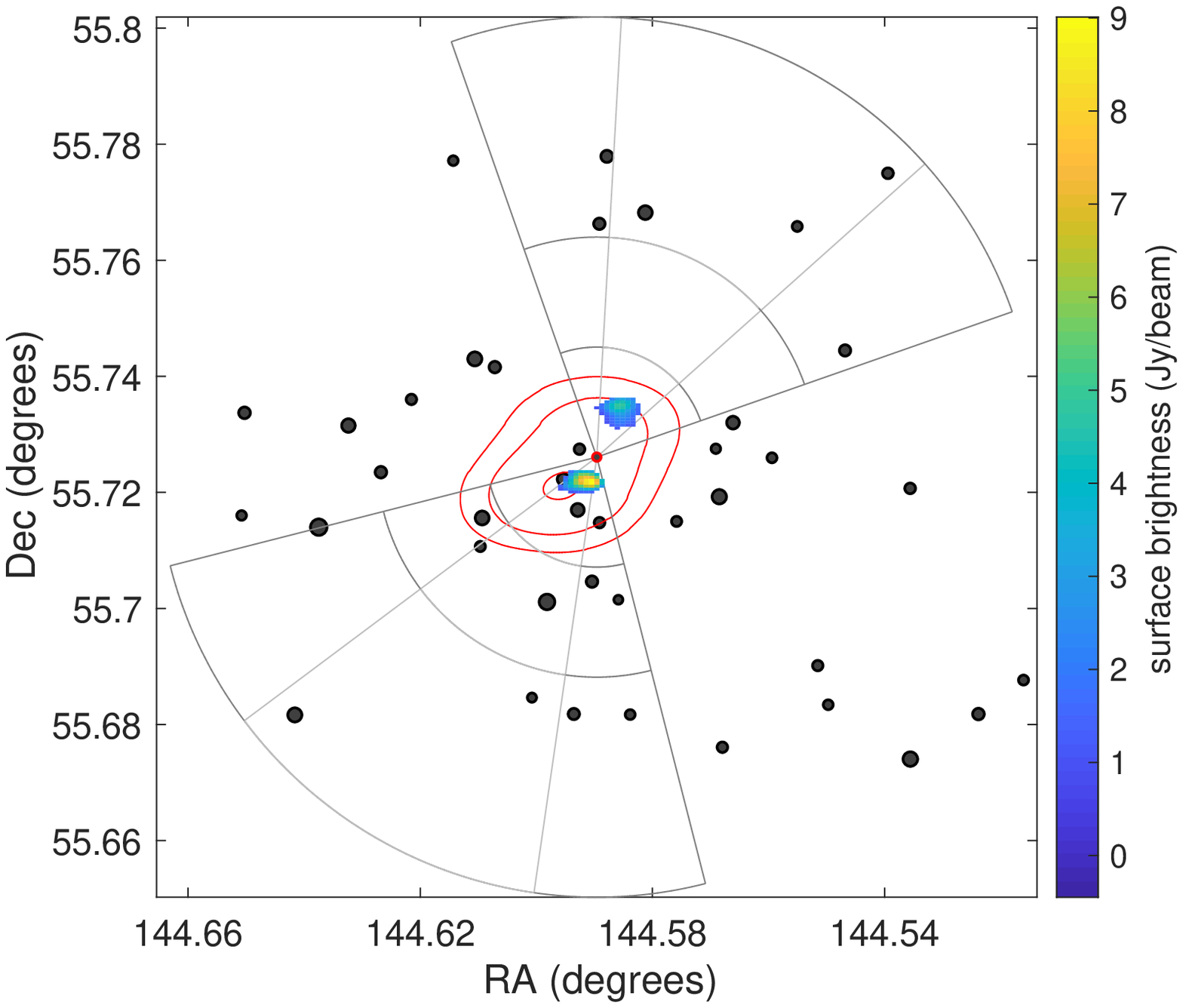}\hspace{0.15\textwidth}
	\includegraphics[width=0.8\columnwidth]{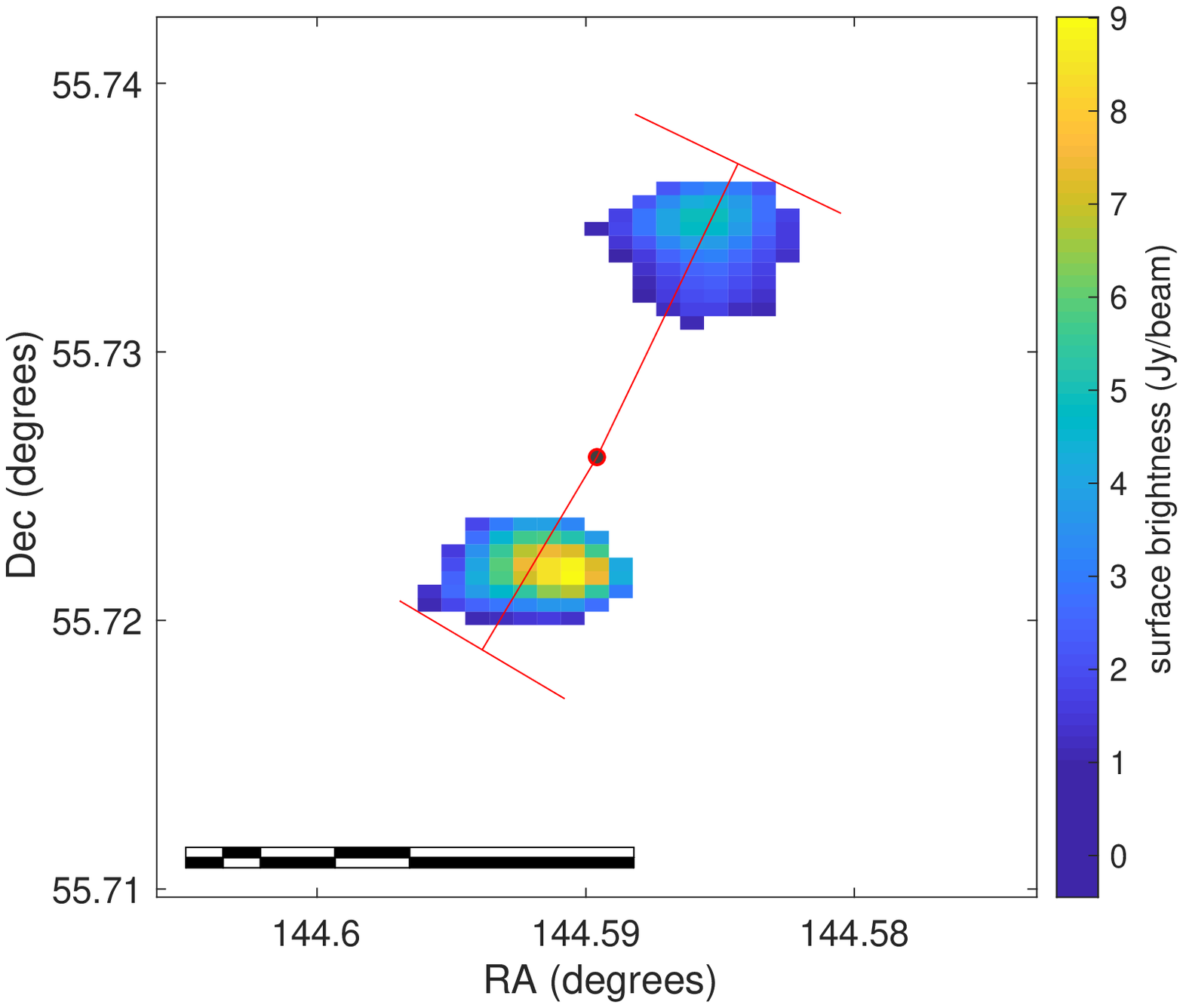}
	\caption{FR-II source RGZ J093821.5+554333; symbols are as in Figure~\ref{fig:FRII_alpha_11}.}
	\label{fig:FRII_alpha_3}
\end{figure*}

\begin{figure*}
	\centering
	\includegraphics[width=0.8\columnwidth]{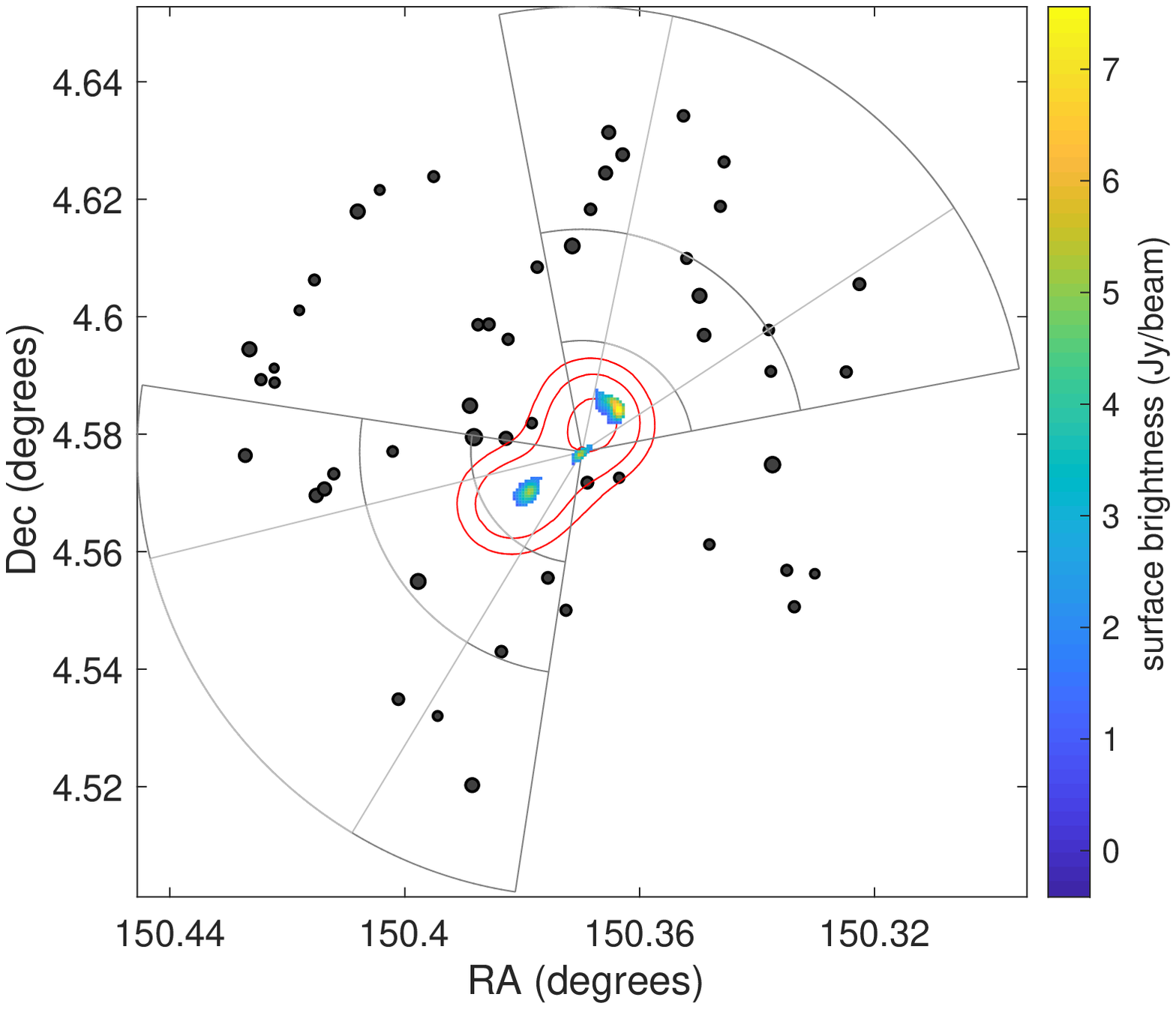}\hspace{0.15\textwidth}
	\includegraphics[width=0.8\columnwidth]{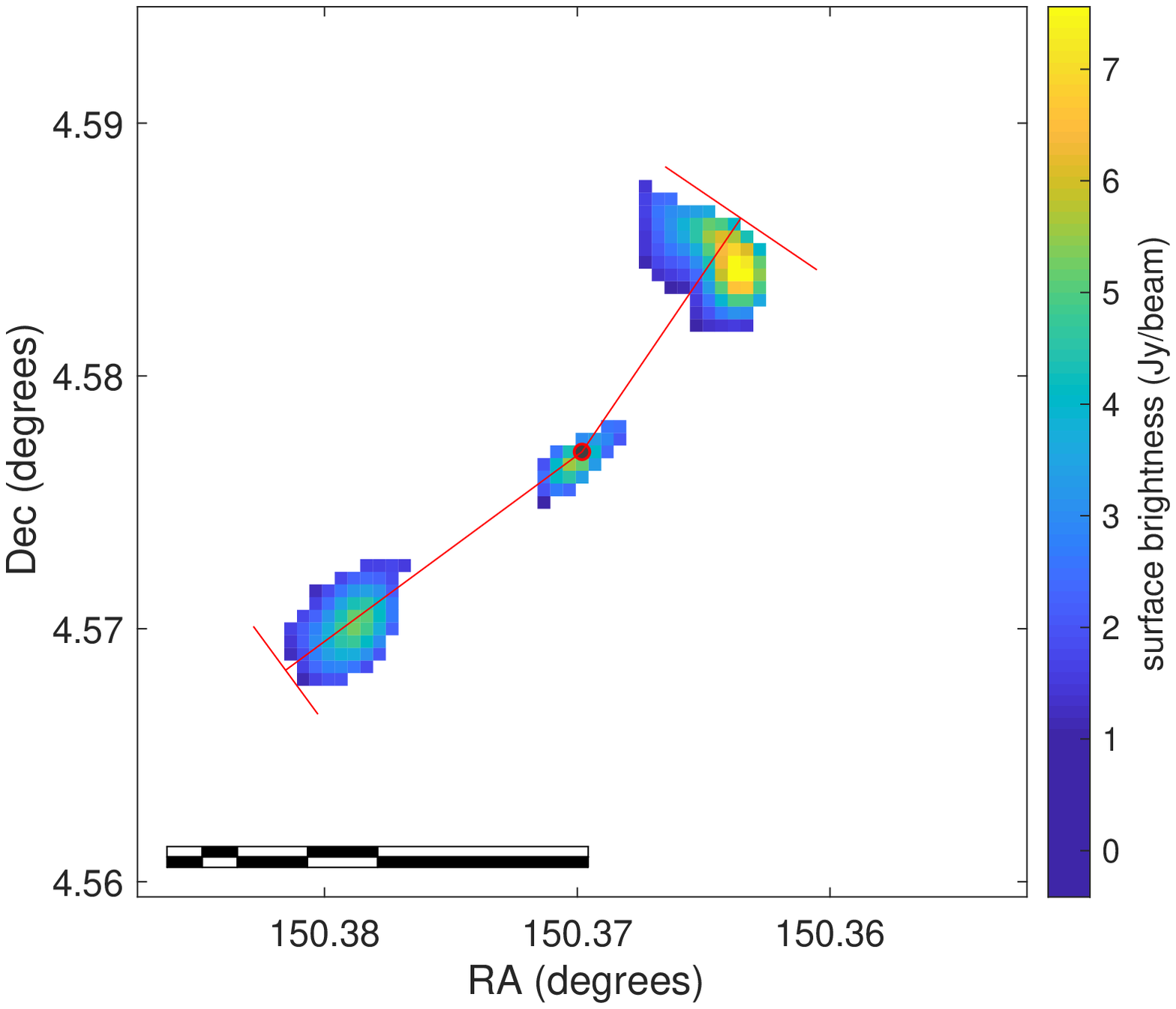}
	\caption{FR-II source RGZ J100128.8+043437; symbols are as in Figure~\ref{fig:FRII_alpha_11}.}
	\label{fig:FRII_beta_1}
\end{figure*}

\begin{figure*}
	\centering
	\includegraphics[width=0.8\columnwidth]{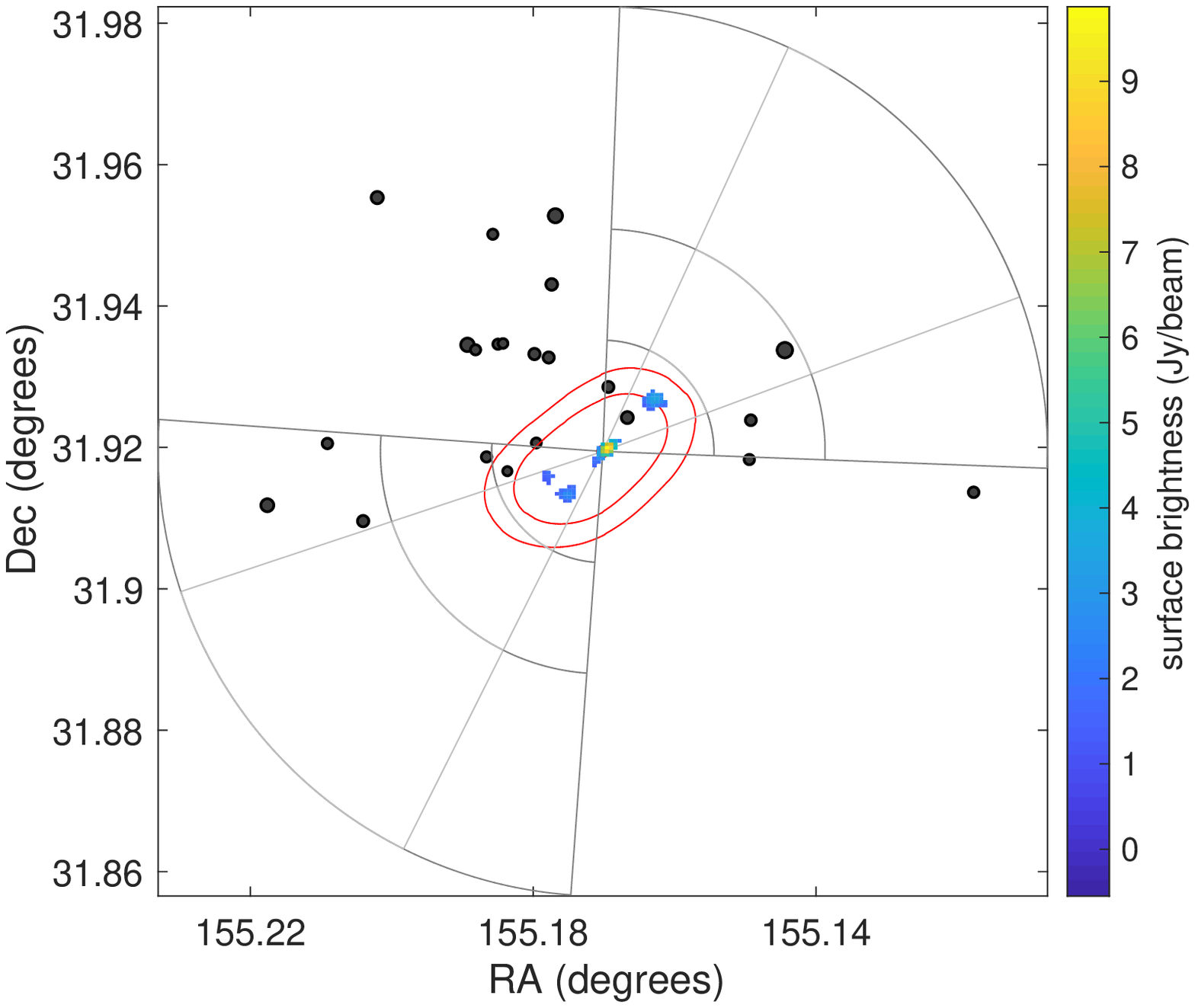}\hspace{0.15\textwidth}
	\includegraphics[width=0.8\columnwidth]{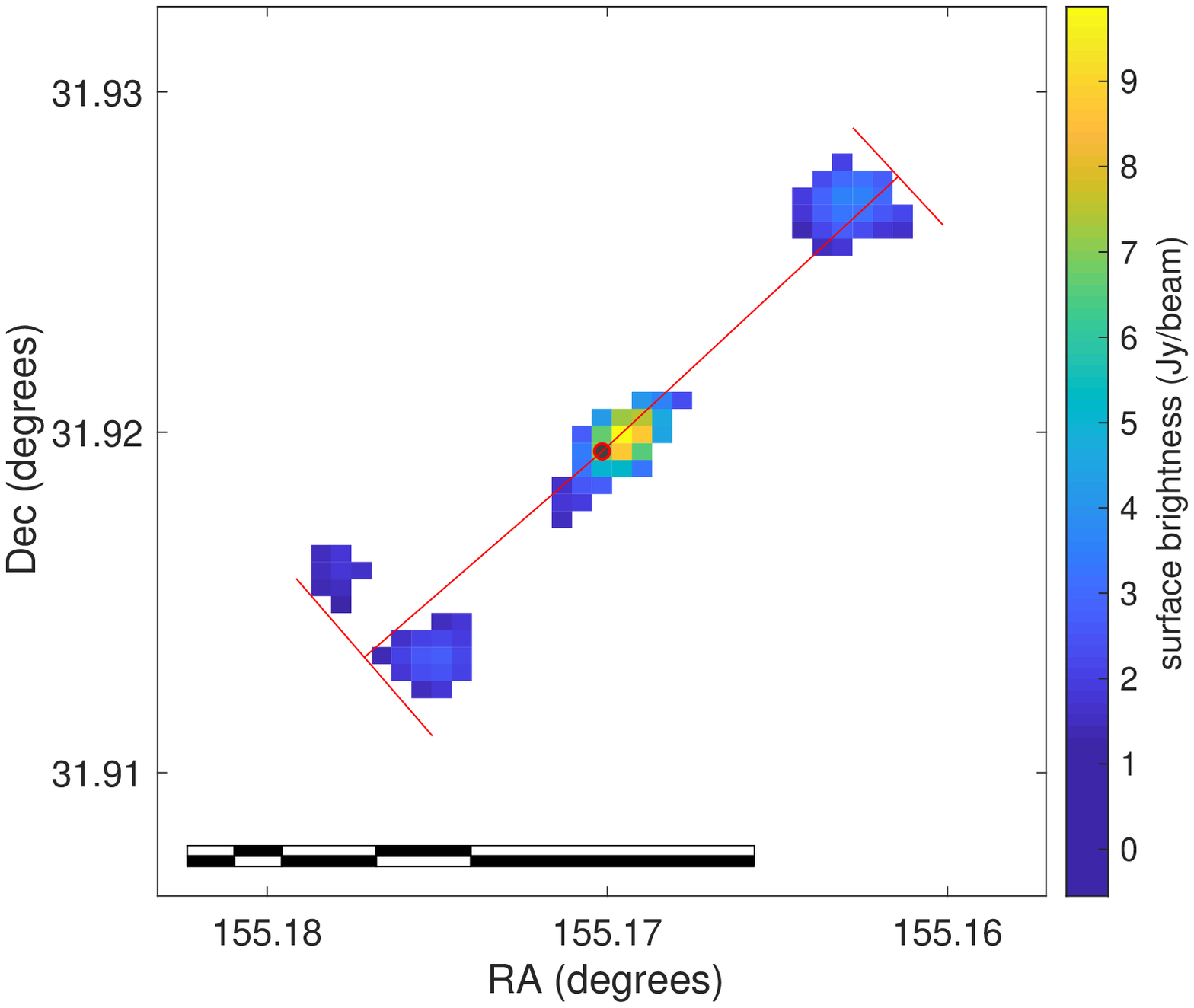}
	\label{fig:FRII_beta_2}
	\caption{FR-II source RGZ J102040.8+315509; symbols are as in Figure~\ref{fig:FRII_alpha_11}.}
\end{figure*}

\begin{figure*}
	\centering
	\includegraphics[width=0.8\columnwidth]{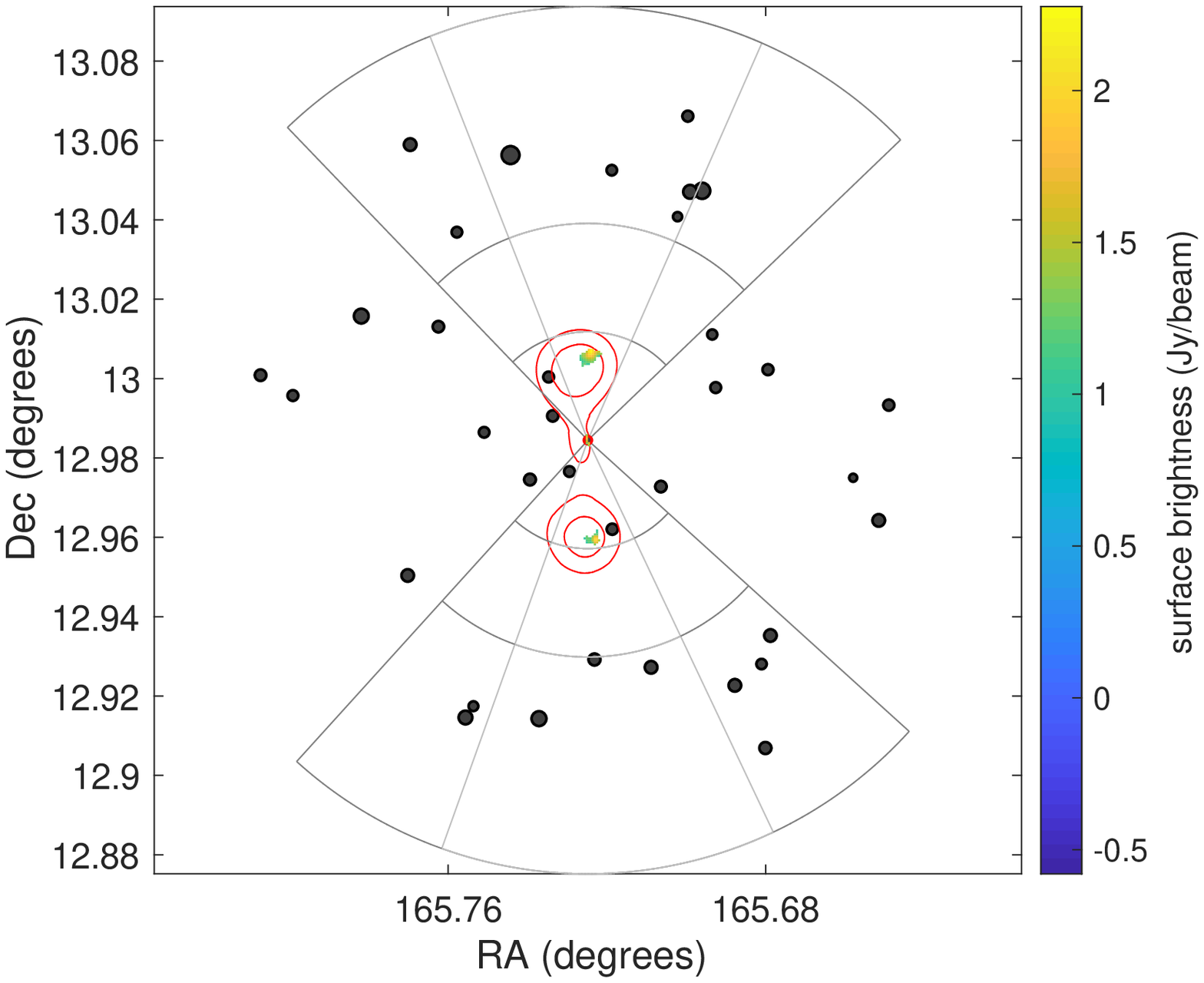}\hspace{0.15\textwidth}
	\includegraphics[width=0.8\columnwidth]{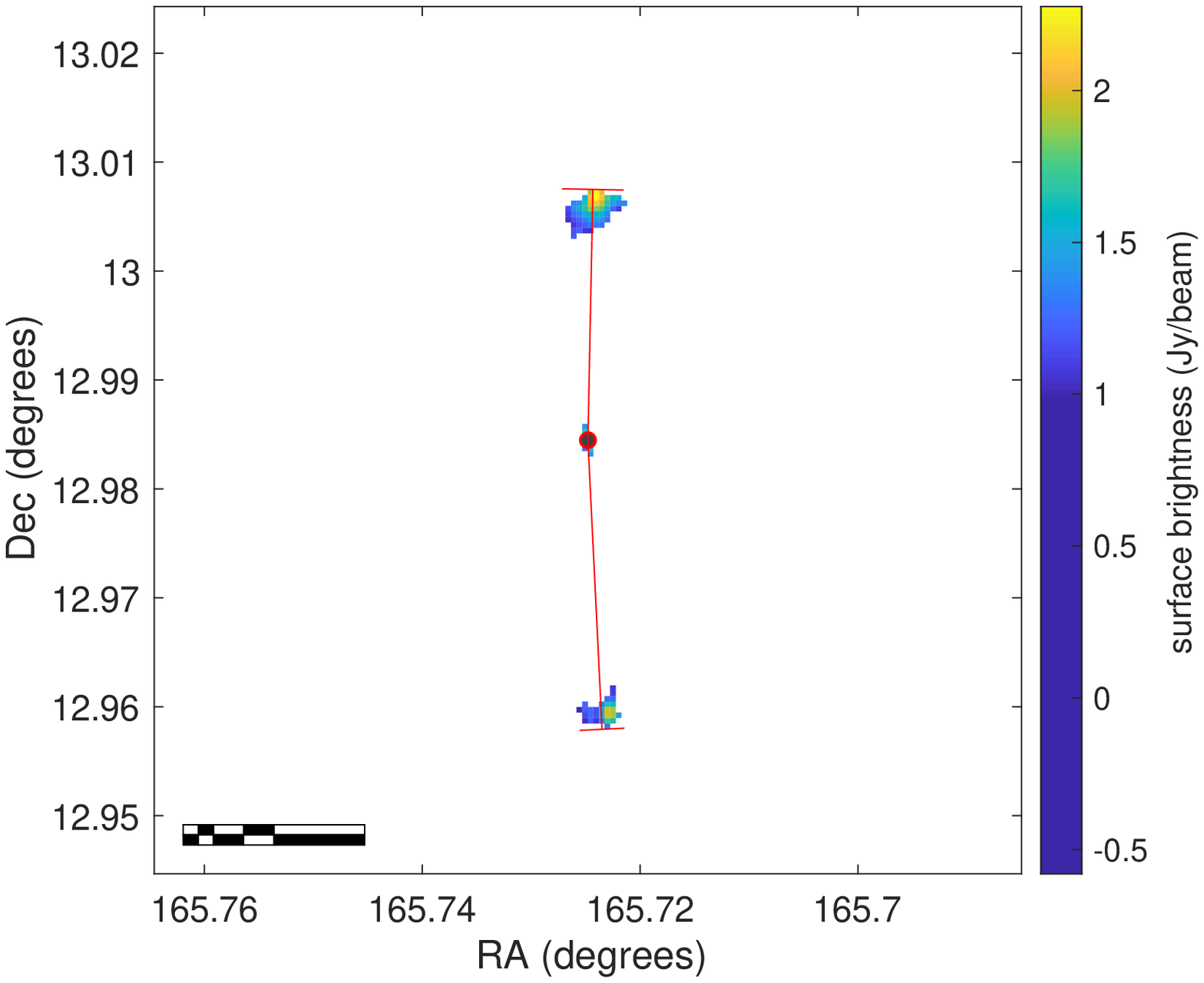}
	\caption{FR-II source RGZ J110253.0+125904; symbols are as in Figure~\ref{fig:FRII_alpha_11}.}
	\label{fig:FRII_alpha_4}
\end{figure*}

\begin{figure*}
	\centering
	\includegraphics[width=0.8\columnwidth]{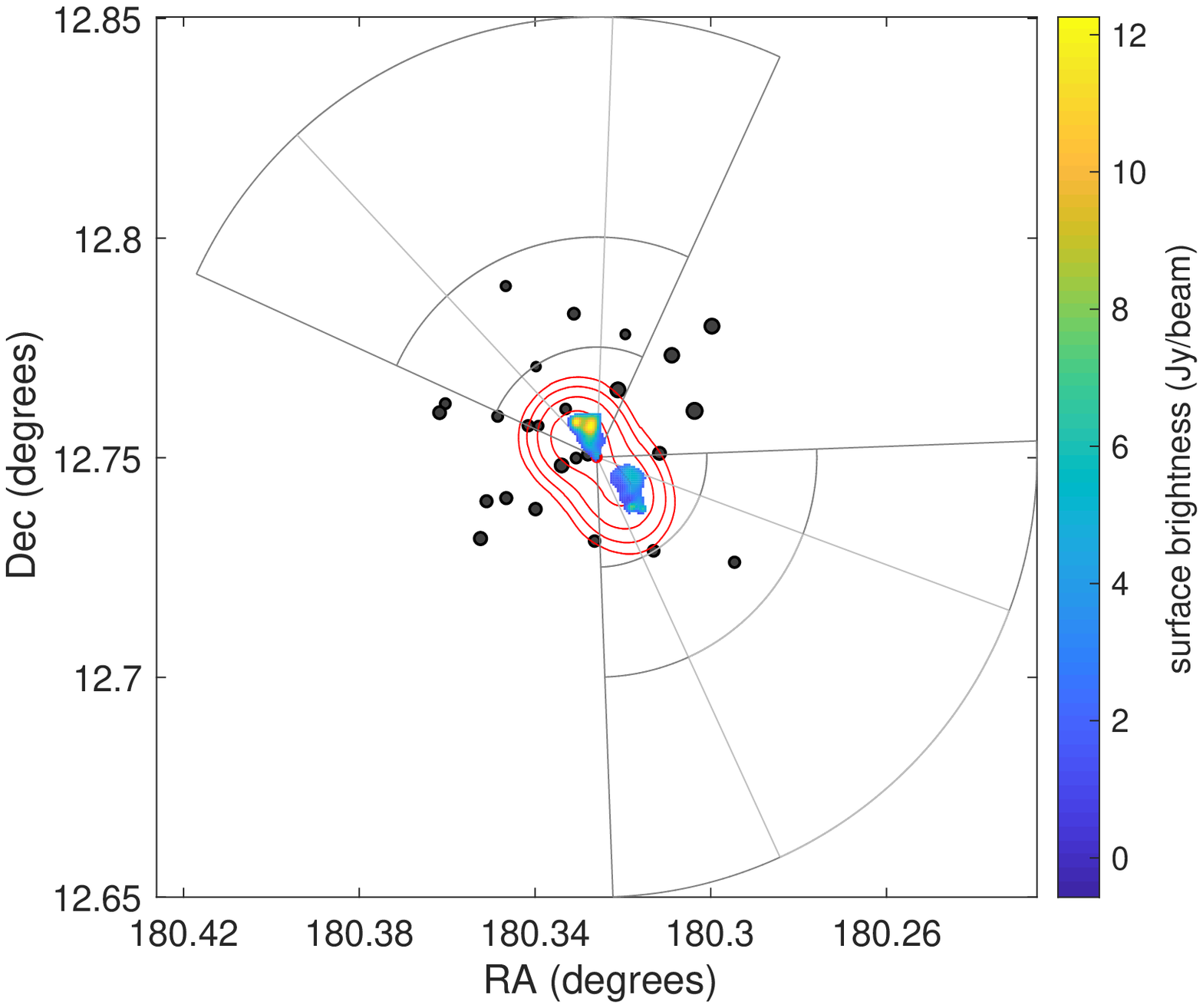}\hspace{0.15\textwidth}
	\includegraphics[width=0.8\columnwidth]{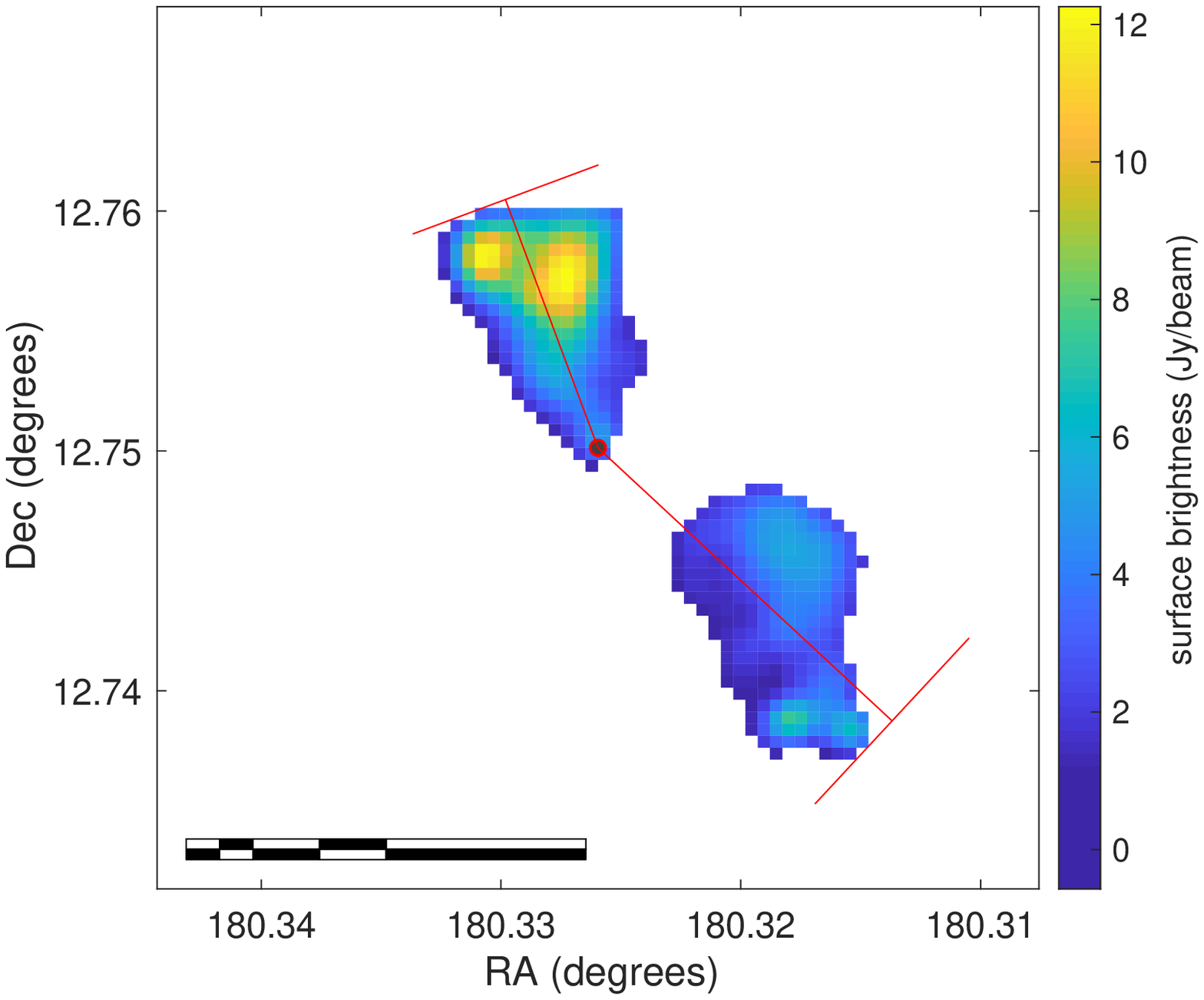}
	\caption{FR-II source RGZ J120118.2+124500; symbols are as in Figure~\ref{fig:FRII_alpha_11}.}
	\label{fig:FRII_alpha_5}
\end{figure*}

\begin{figure*}
	\centering
	\includegraphics[width=0.8\columnwidth]{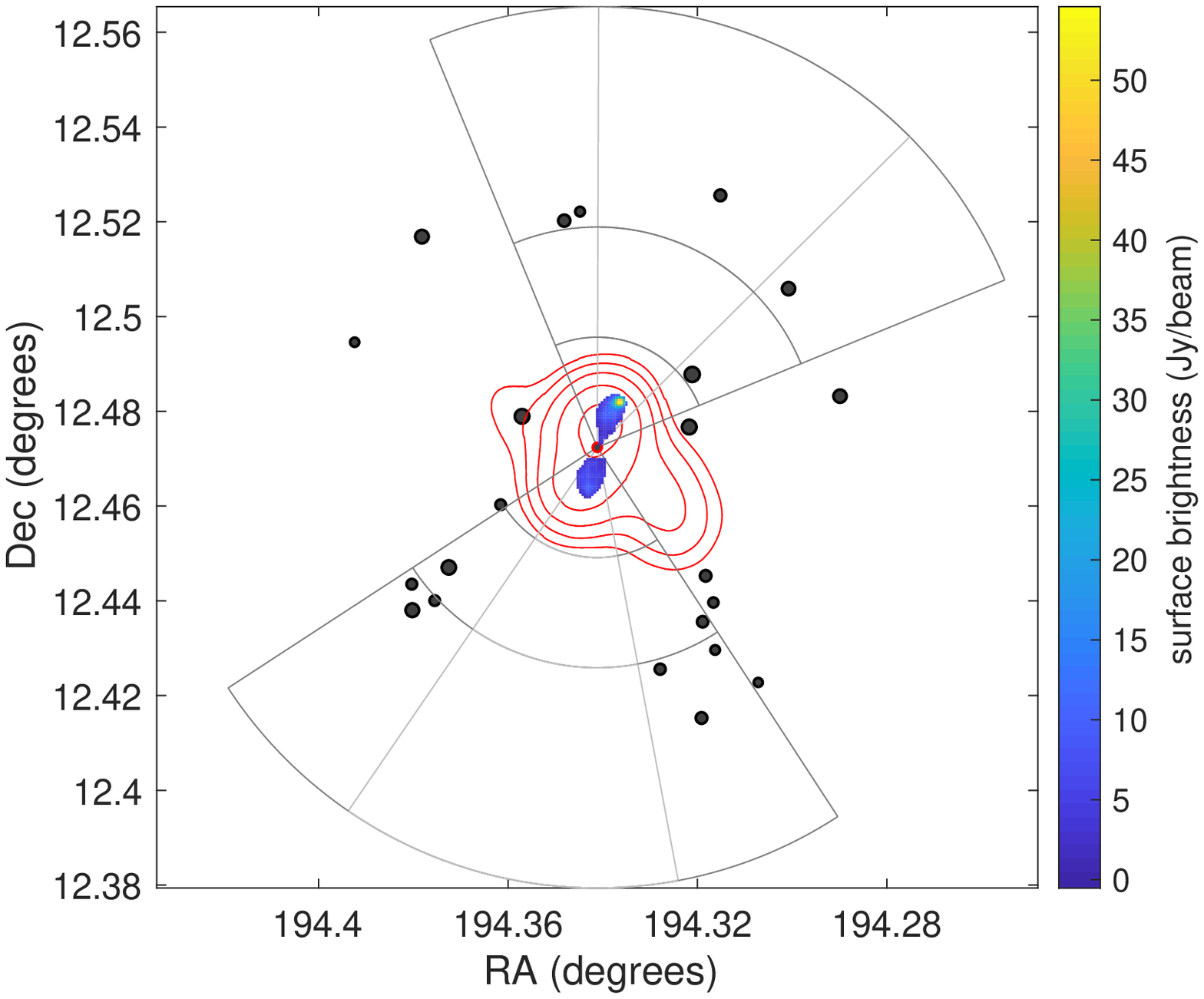}\hspace{0.15\textwidth}
	\includegraphics[width=0.8\columnwidth]{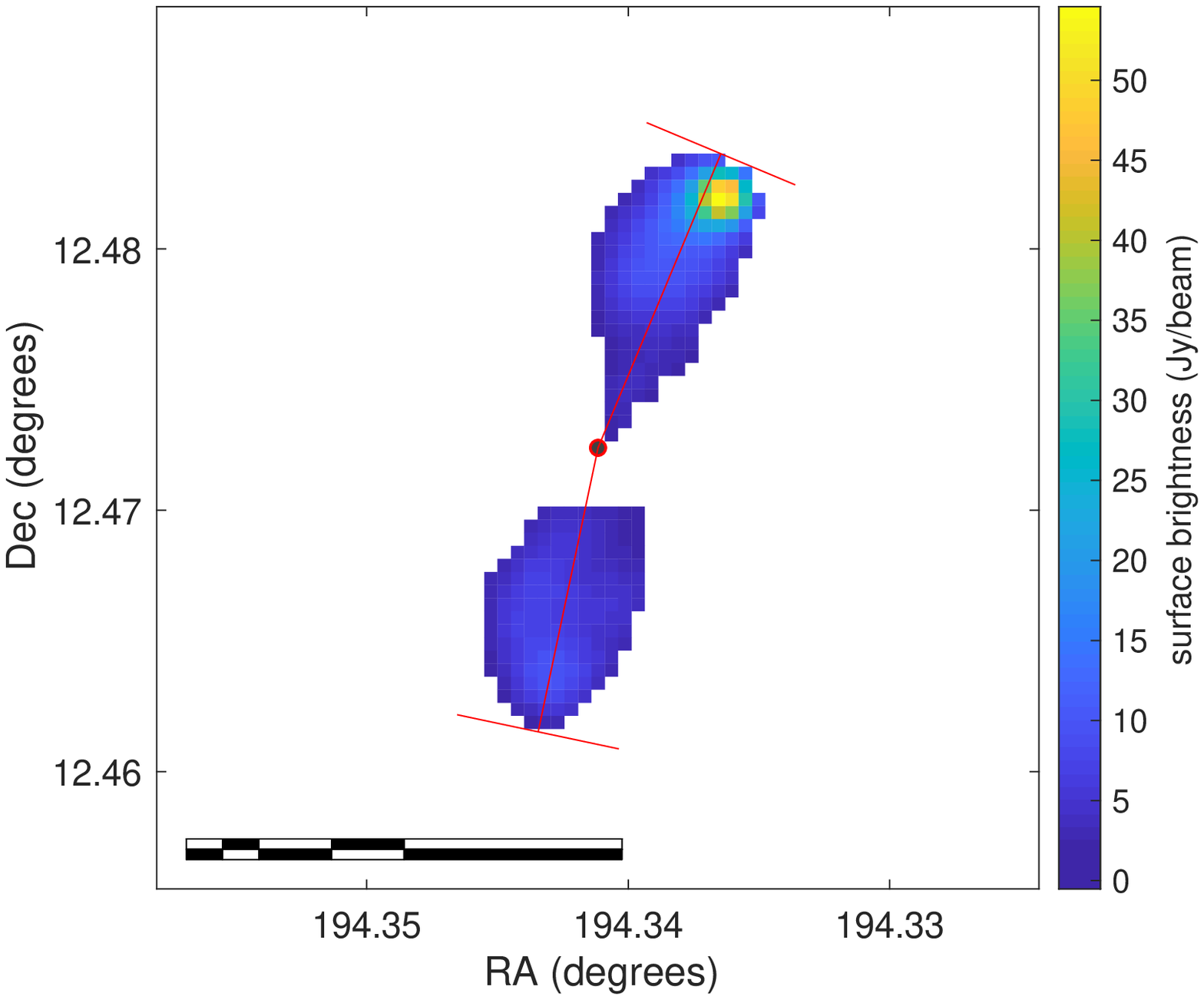}
	\caption{FR-II source RGZ J125721.9+122820; symbols are as in Figure~\ref{fig:FRII_alpha_11}.}
	\label{fig:FRII_alpha_6}
\end{figure*}

\begin{figure*}
	\centering
	\includegraphics[width=0.8\columnwidth]{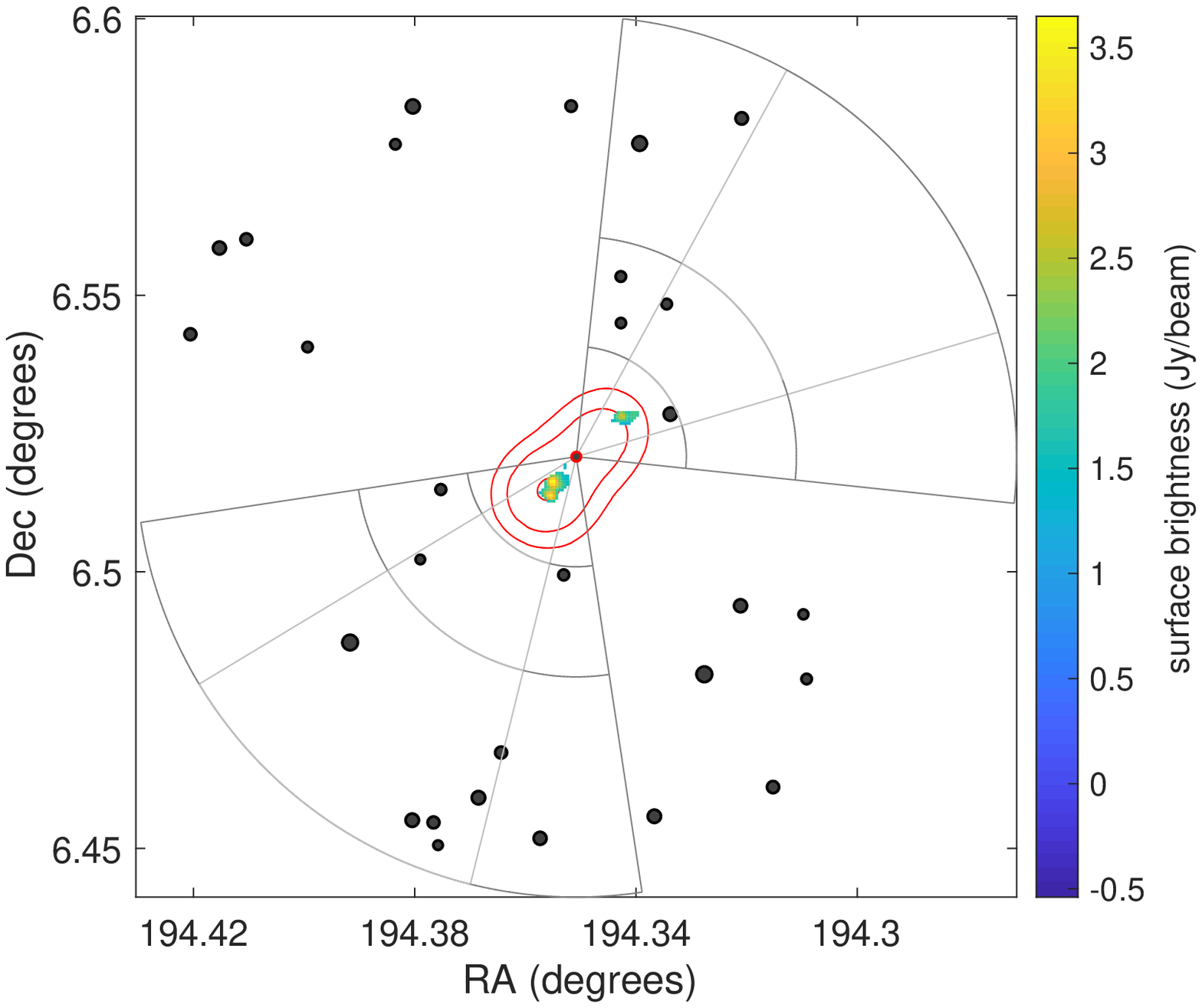}\hspace{0.15\textwidth}
	\includegraphics[width=0.8\columnwidth]{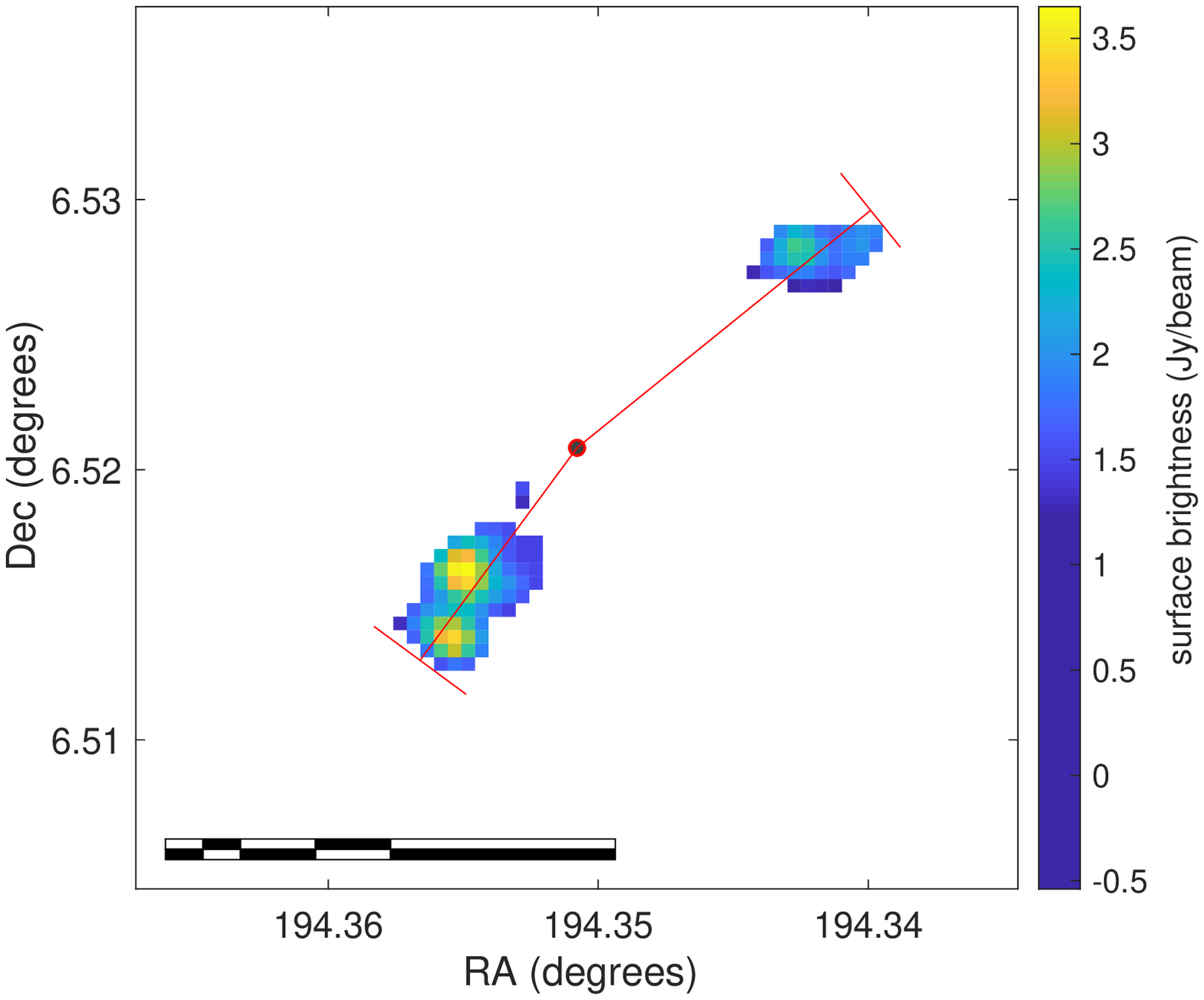}
	\caption{FR-II source RGZ J125724.2+063114; symbols are as in Figure~\ref{fig:FRII_alpha_11}.}
	\label{fig:FRII_beta_3}
\end{figure*}

\begin{figure*}
	\centering
	\includegraphics[width=0.8\columnwidth]{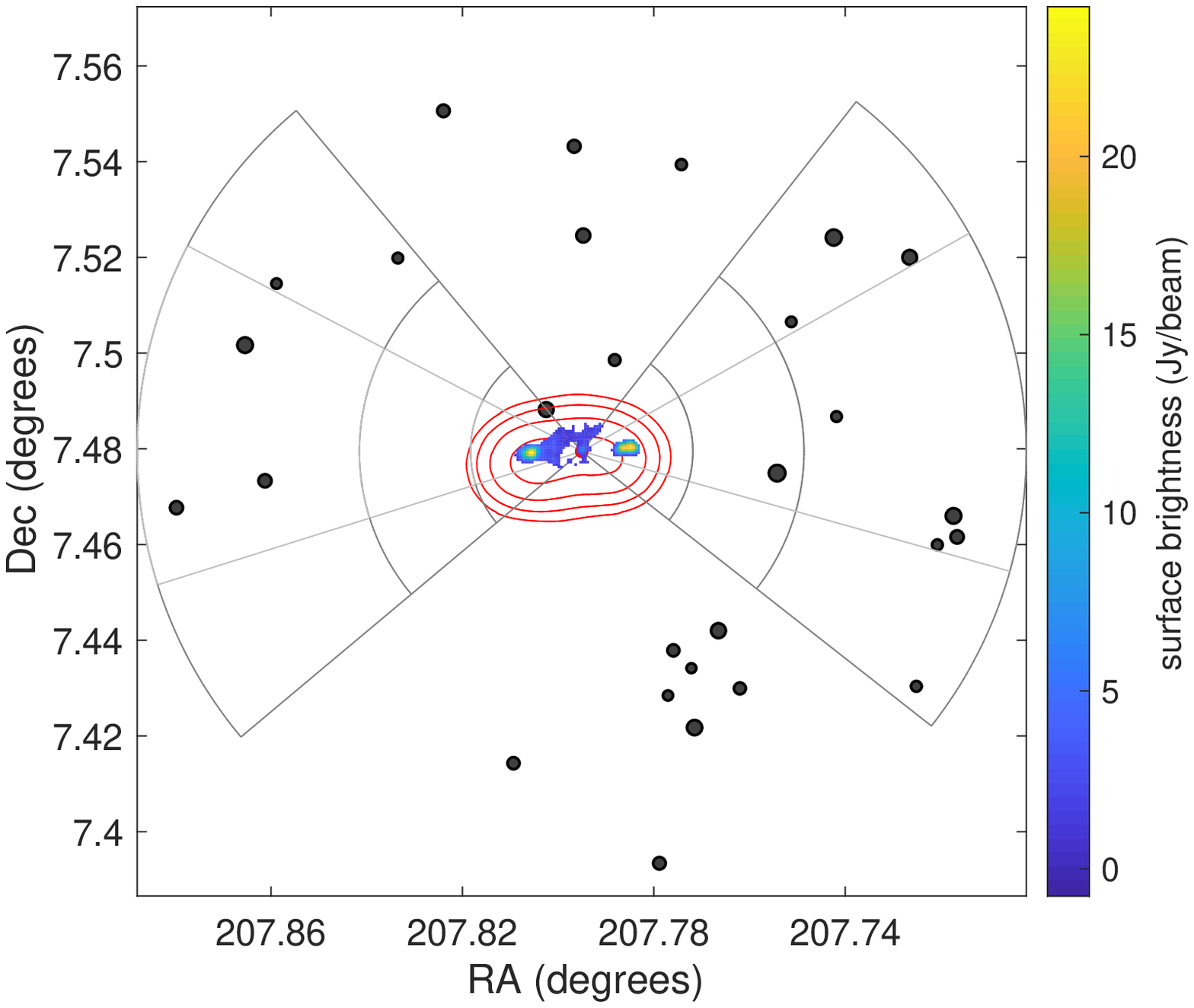}\hspace{0.15\textwidth}
	\includegraphics[width=0.8\columnwidth]{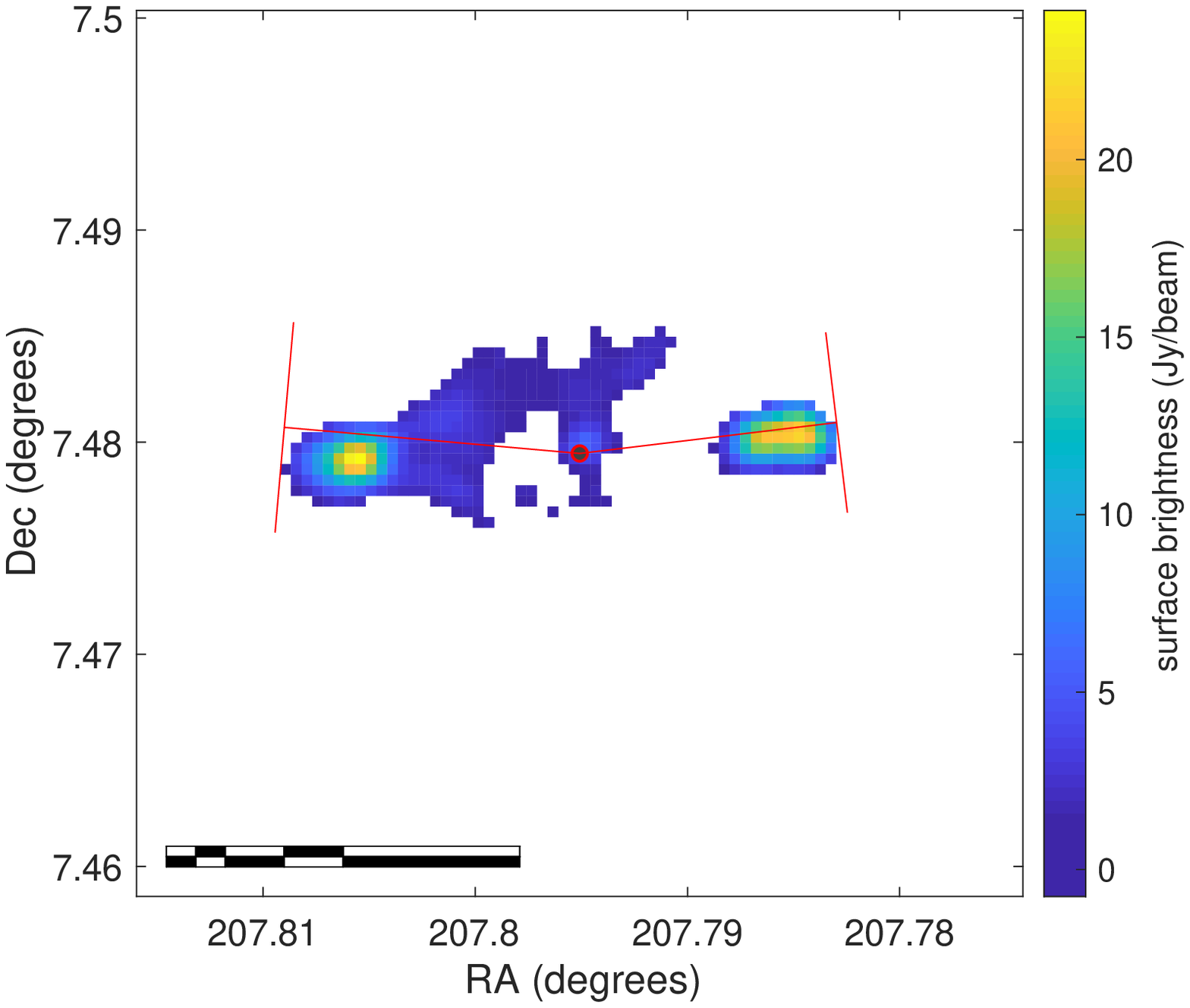}
	\caption{FR-II source RGZ J135110.8+072846; symbols are as in Figure~\ref{fig:FRII_alpha_11}.}
	\label{fig:FRII_alpha_7}
\end{figure*}

\begin{figure*}
	\centering
	\includegraphics[width=0.8\columnwidth]{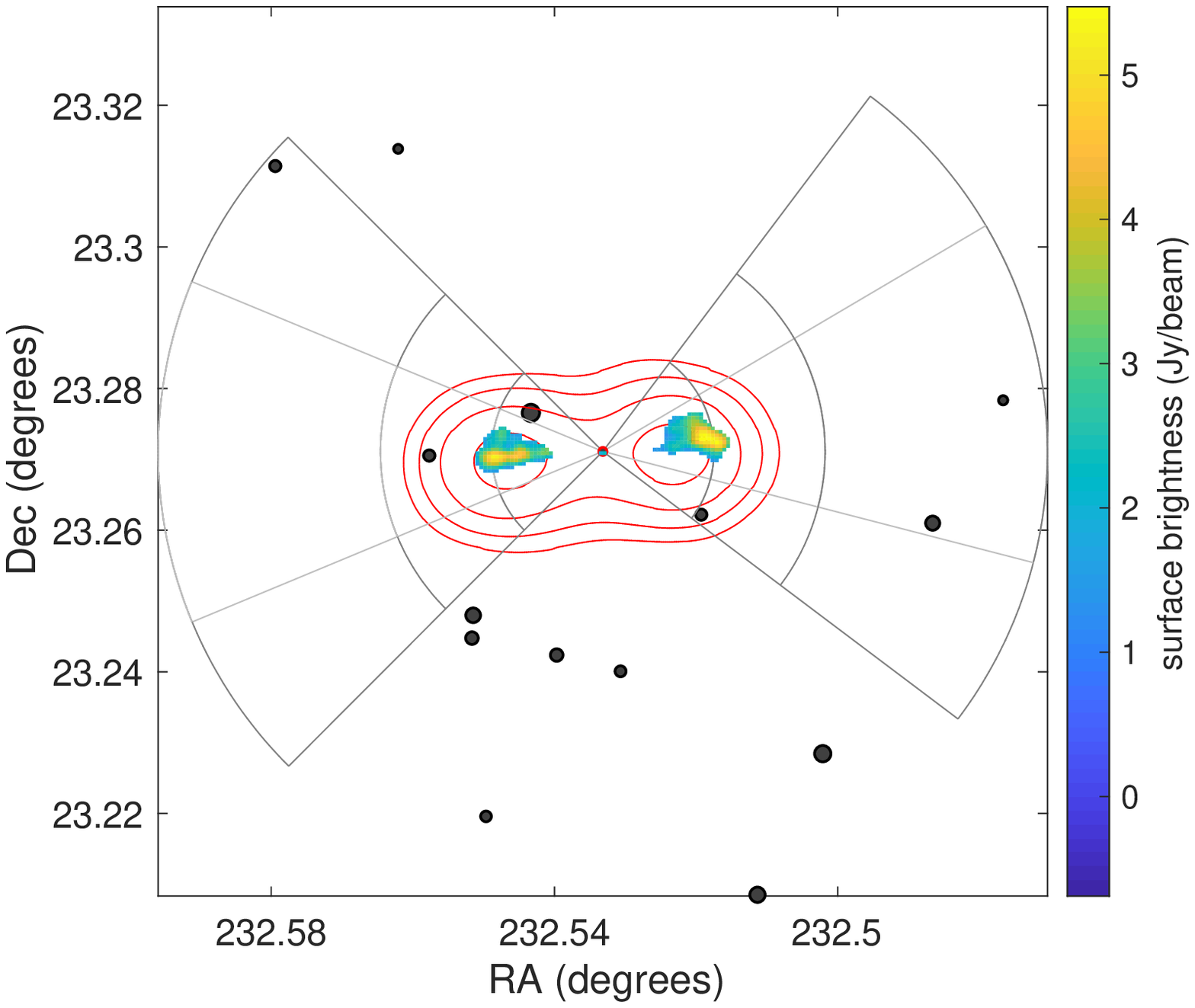}\hspace{0.15\textwidth}
	\includegraphics[width=0.8\columnwidth]{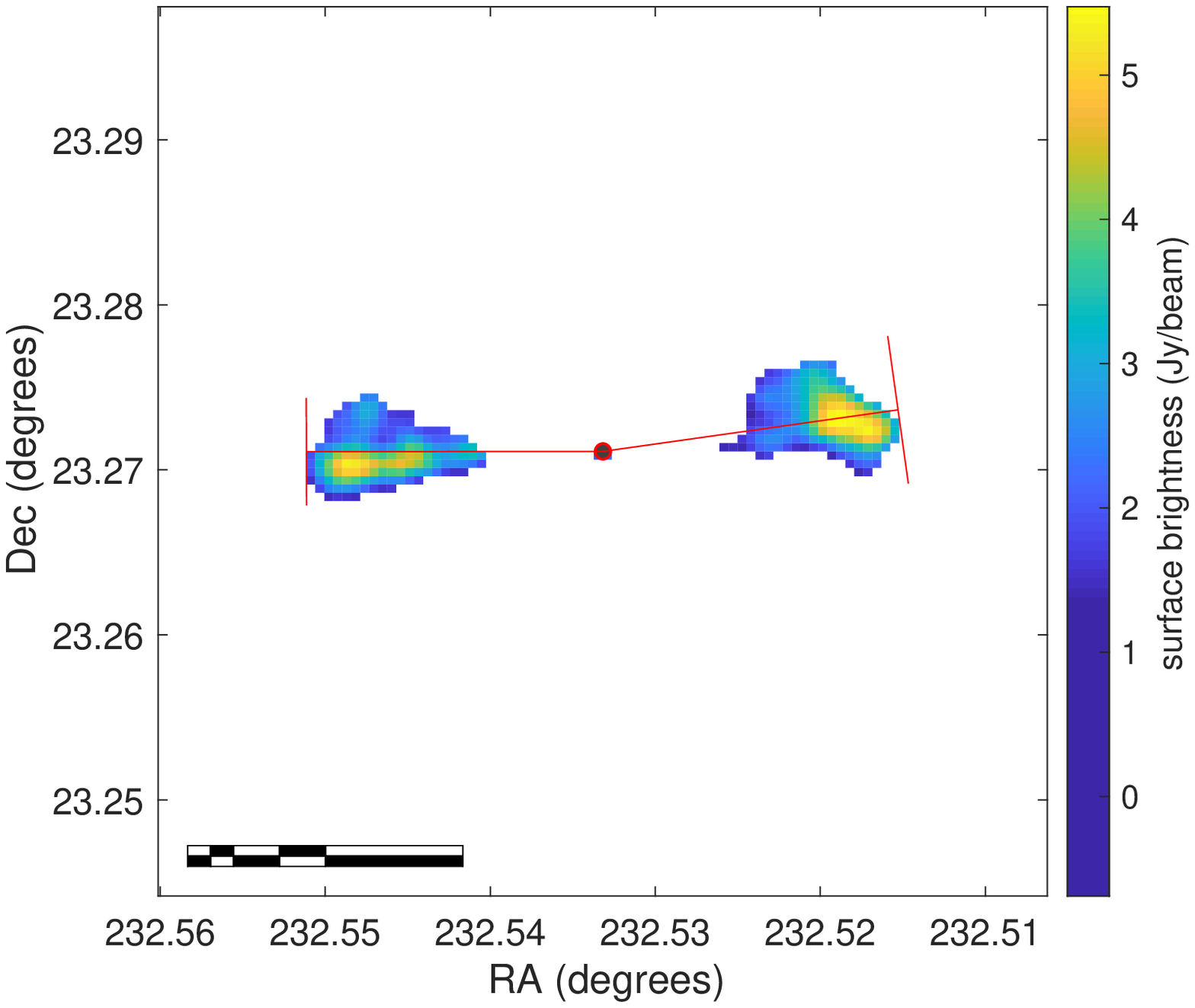}
	\caption{FR-II source RGZ J153008.0+231616; symbols are as in Figure~\ref{fig:FRII_alpha_11}.}
	\label{fig:FRII_alpha_8}
\end{figure*}

\begin{figure*}
	\centering
	\includegraphics[width=0.8\columnwidth]{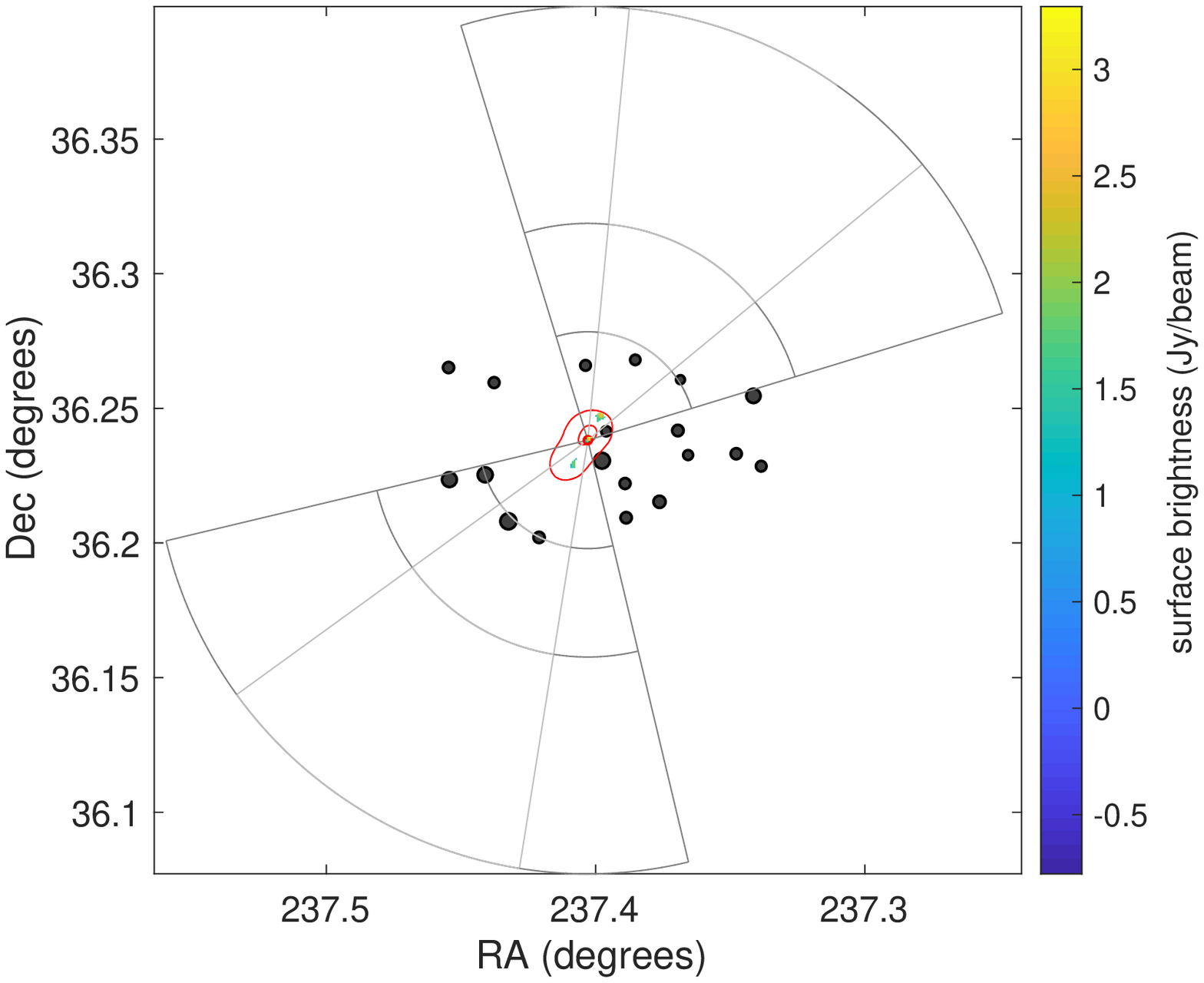}\hspace{0.15\textwidth}
	\includegraphics[width=0.8\columnwidth]{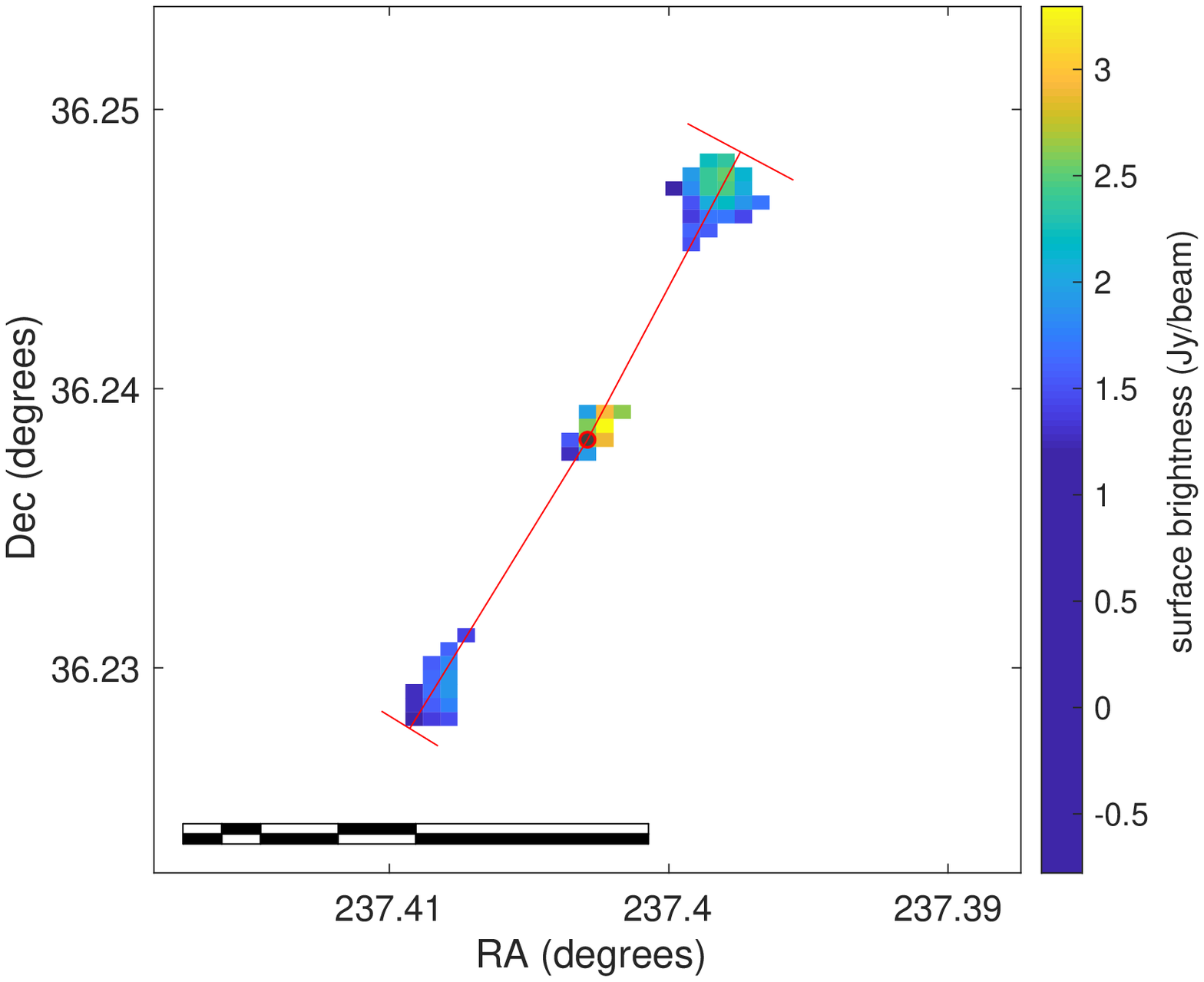}
	\caption{FR-II source RGZ J154936.7+361417; symbols are as in Figure~\ref{fig:FRII_alpha_11}.}
	\label{fig:FRII_alpha_9}
\end{figure*}

\begin{figure*}
	\centering
	\includegraphics[width=0.8\columnwidth]{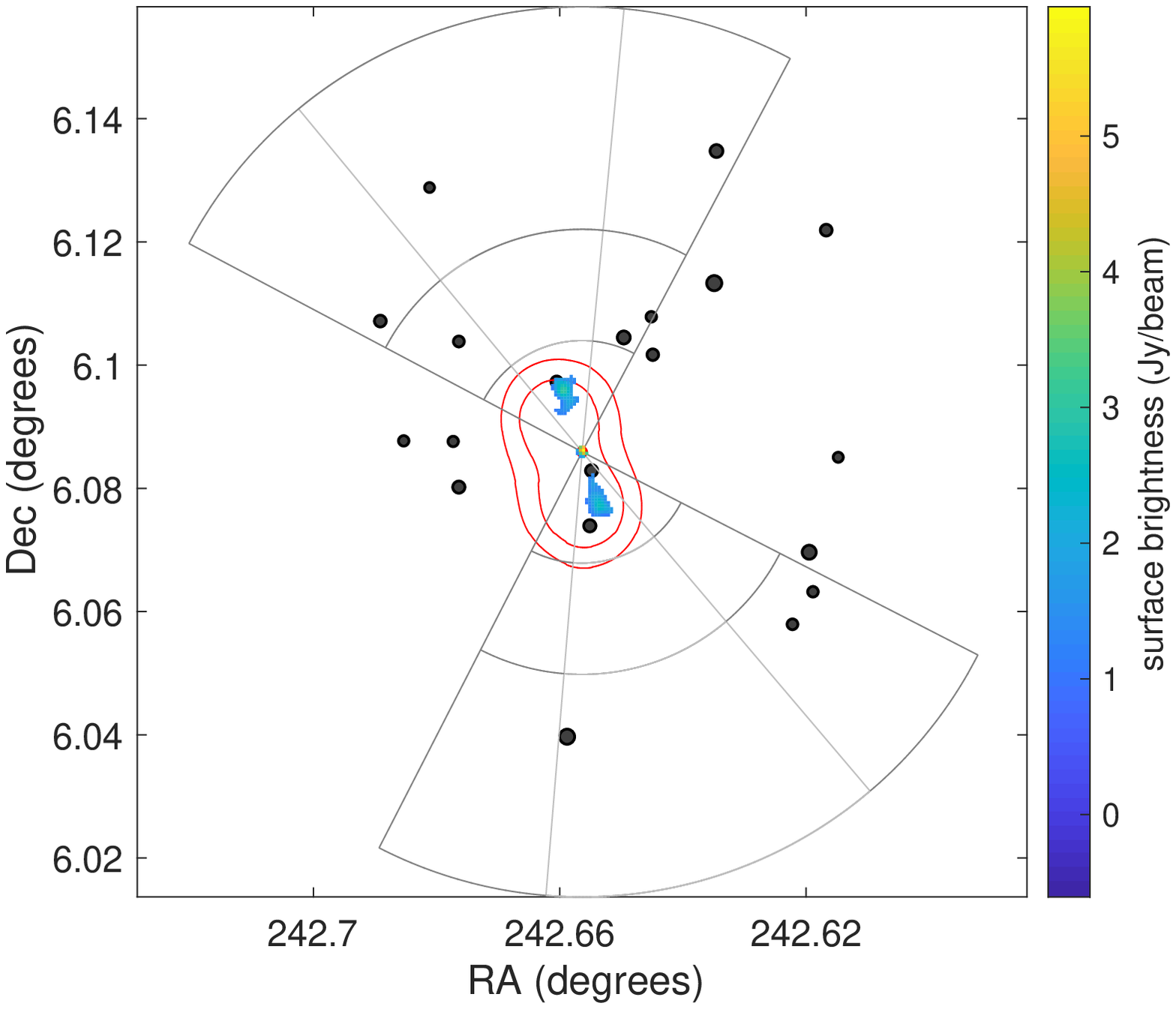}\hspace{0.15\textwidth}
	\includegraphics[width=0.8\columnwidth]{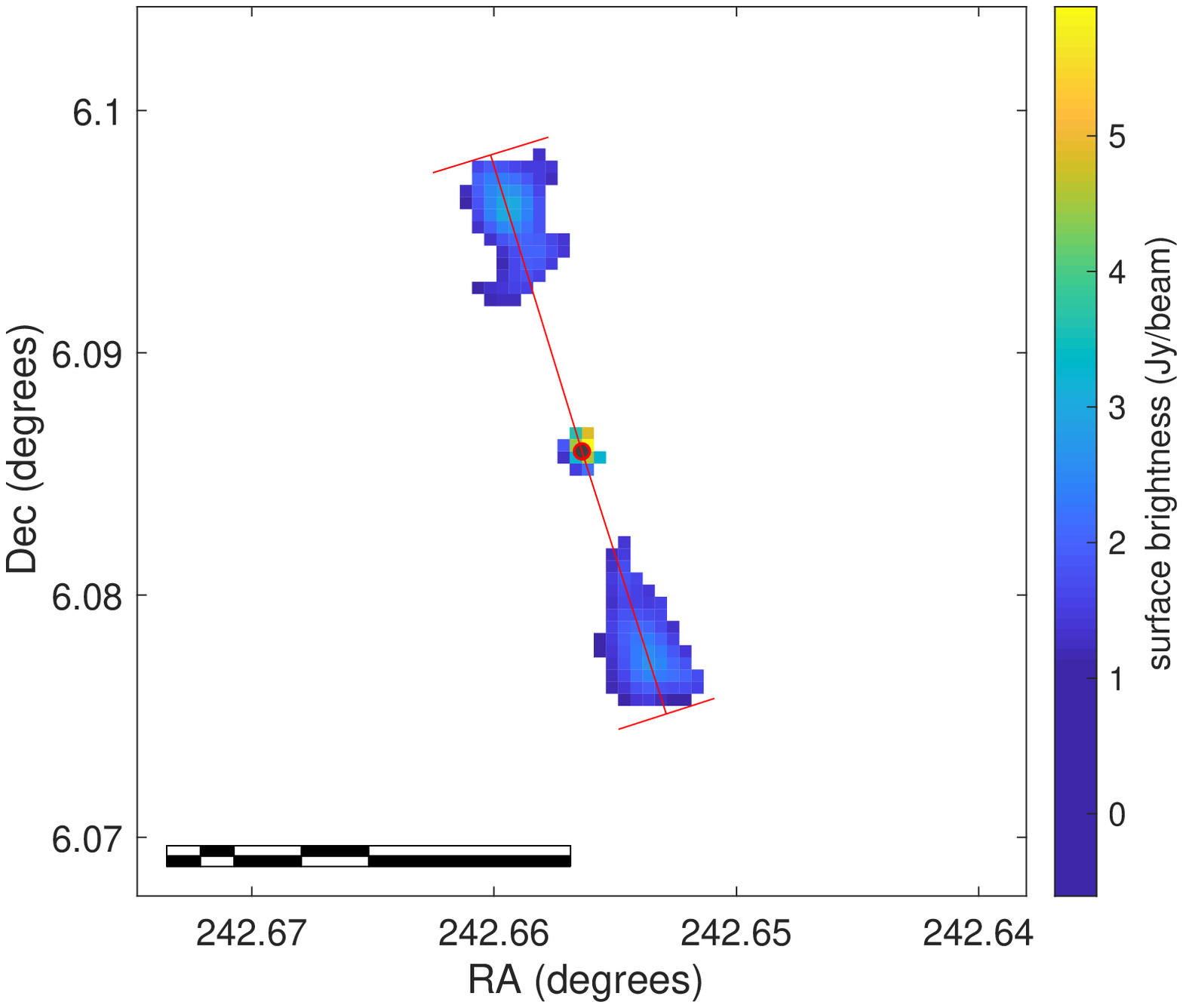}
	\caption{FR-II source RGZ J161037.5+060509; symbols are as in Figure~\ref{fig:FRII_alpha_11}.}
	\label{fig:FRII_alpha_10}
\end{figure*}

\begin{figure*}
	\centering
	\includegraphics[width=0.8\columnwidth]{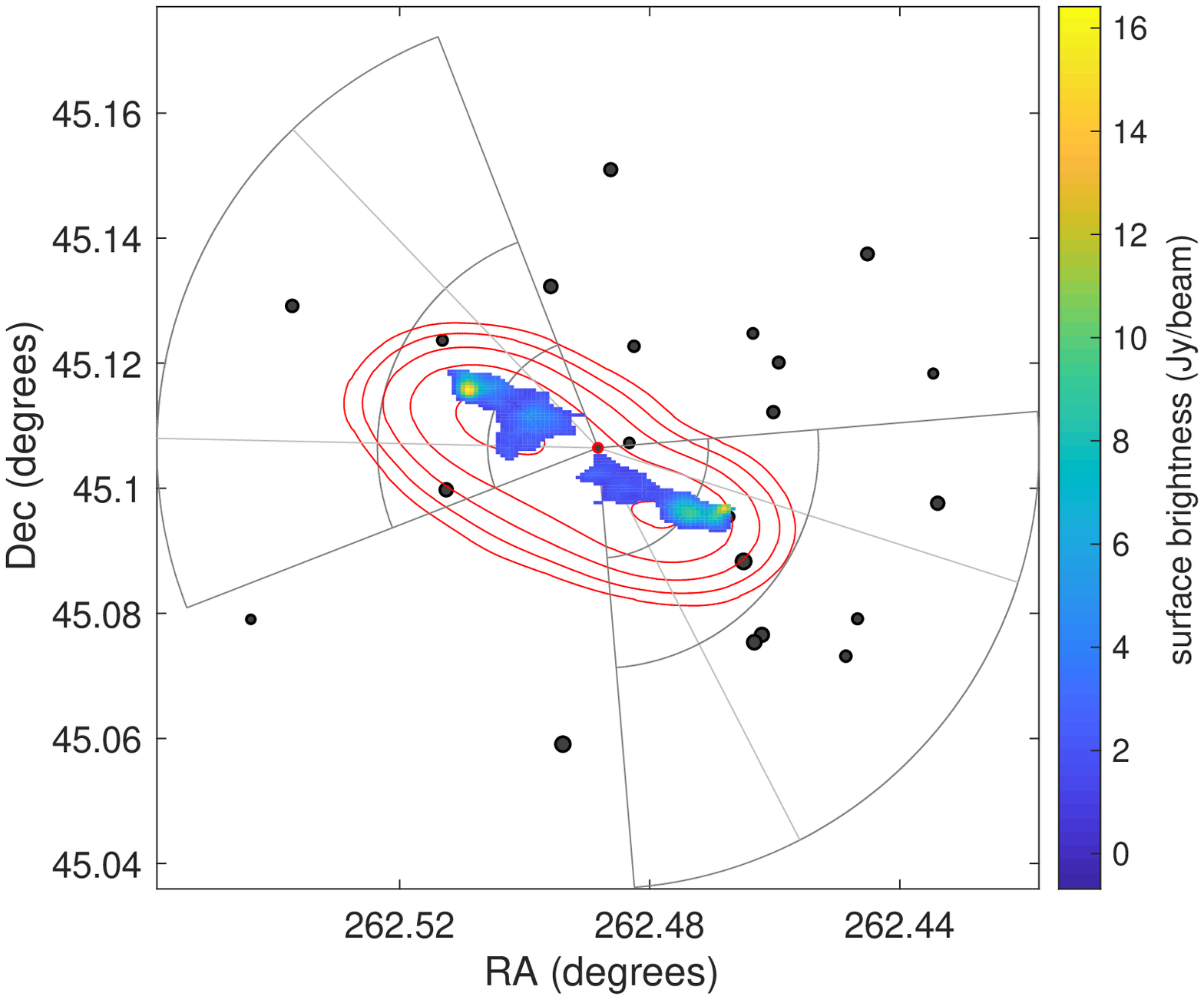}\hspace{0.15\textwidth}
	\includegraphics[width=0.8\columnwidth]{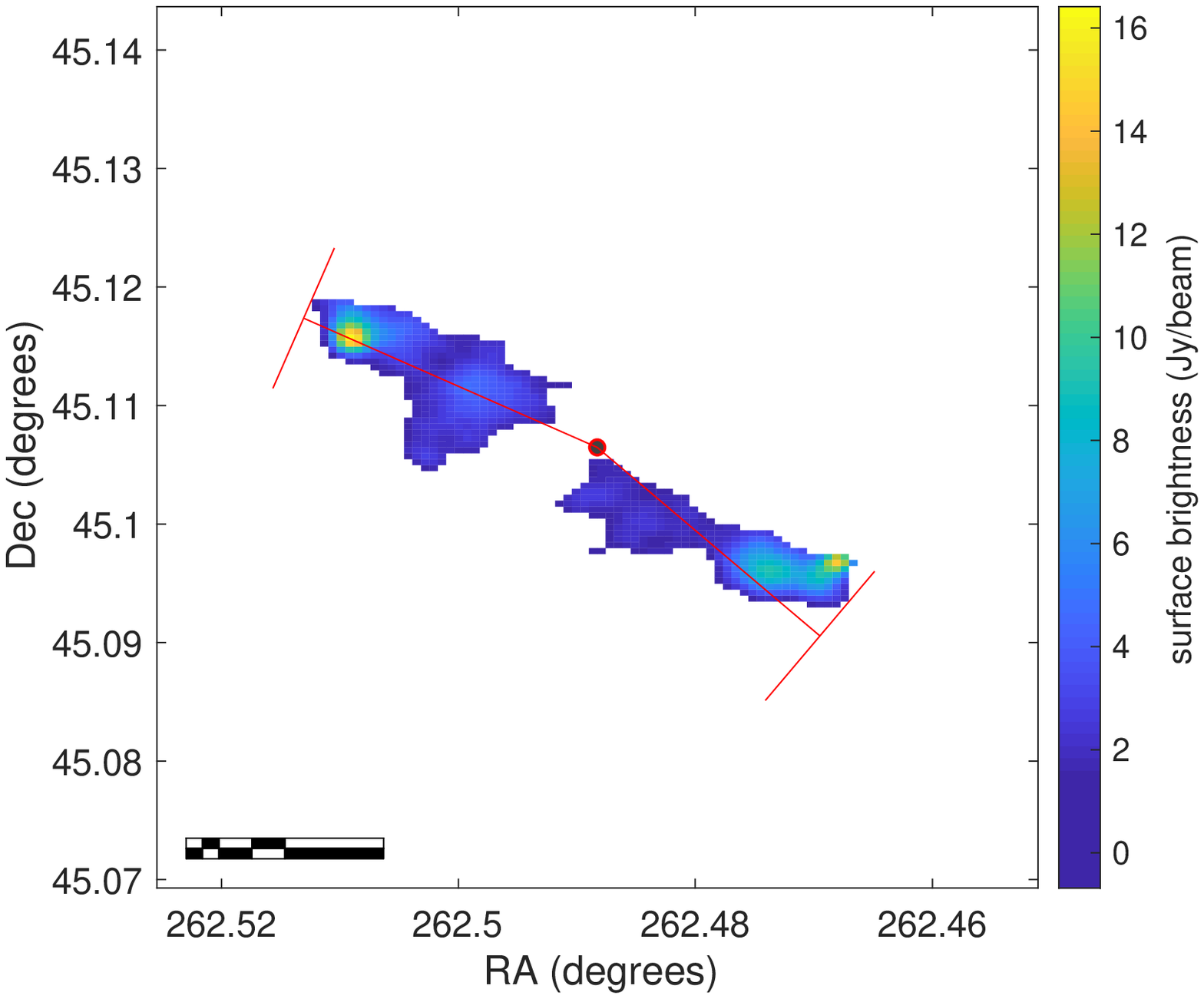}
	\caption{FR-II source RGZ J172957.2+450623; symbols are as in Figure~\ref{fig:FRII_alpha_11}.}
	\label{fig:FRII_alpha_12}
\end{figure*}

\begin{figure*}
	\centering
	\includegraphics[width=0.8\columnwidth]{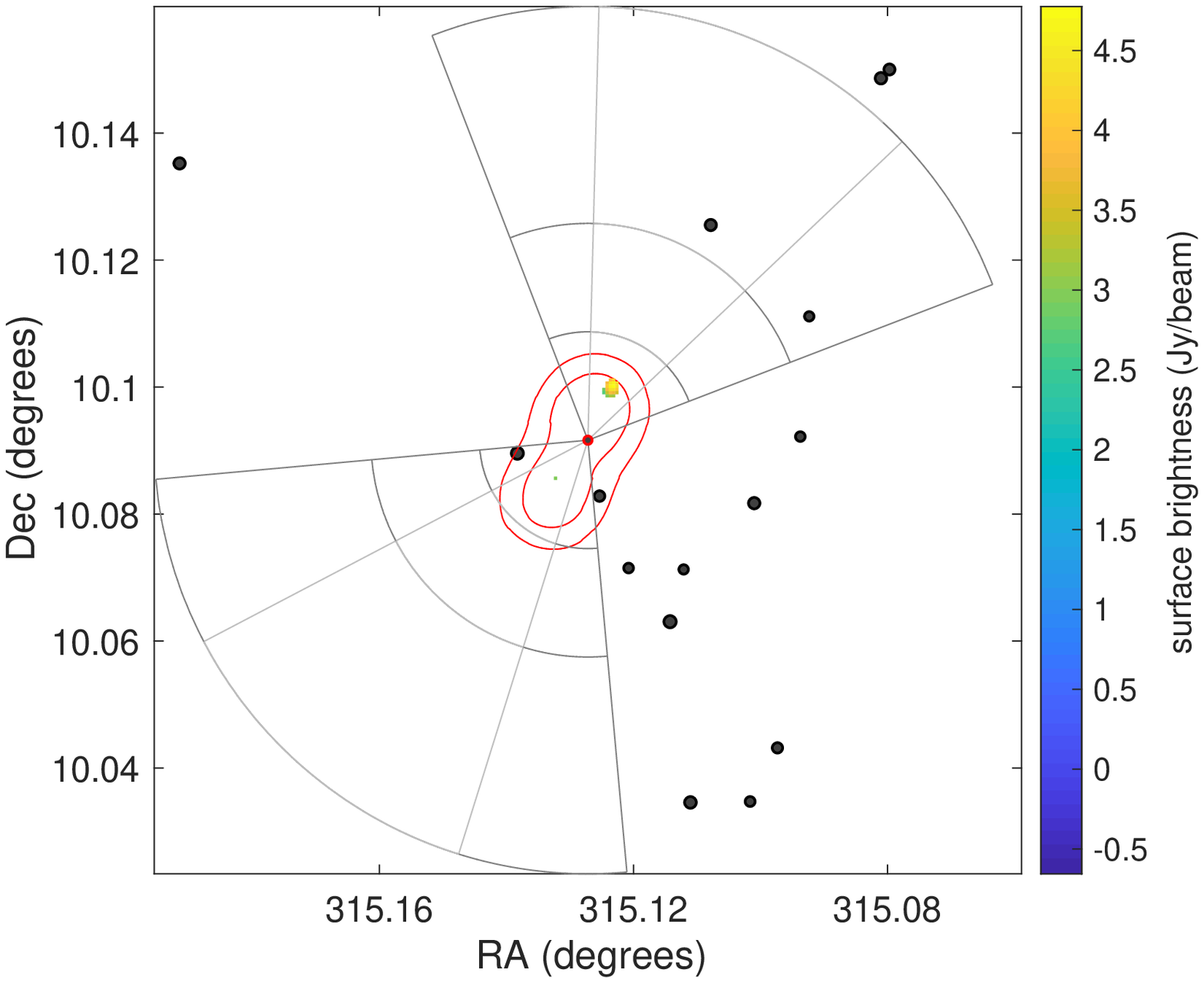}\hspace{0.15\textwidth}
	\includegraphics[width=0.8\columnwidth]{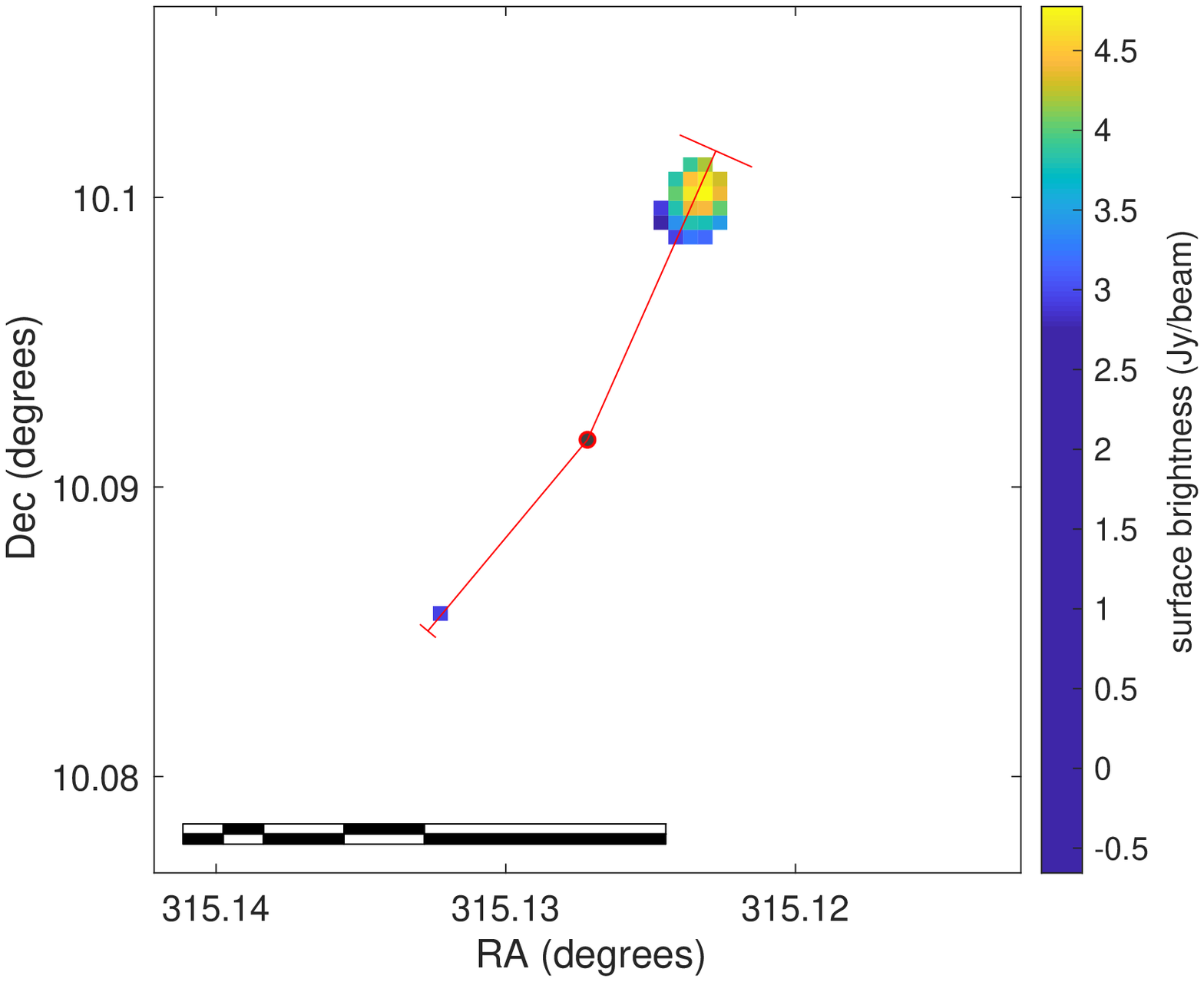}
	\caption{FR-II source RGZ J210030.5+100529; symbols are as in Figure~\ref{fig:FRII_alpha_11}. This is the only FR-II source excluded from analysis, due to its hotspot dominance of integrated lobe luminosity.}
	\label{fig:FRII_beta_4}
\end{figure*}

\clearpage
\begin{figure*}
\section{FR-I radio and optical images}
\label{sec:appendixb}
\end{figure*}

\begin{figure*}
	\centering
	\includegraphics[width=0.8\columnwidth]{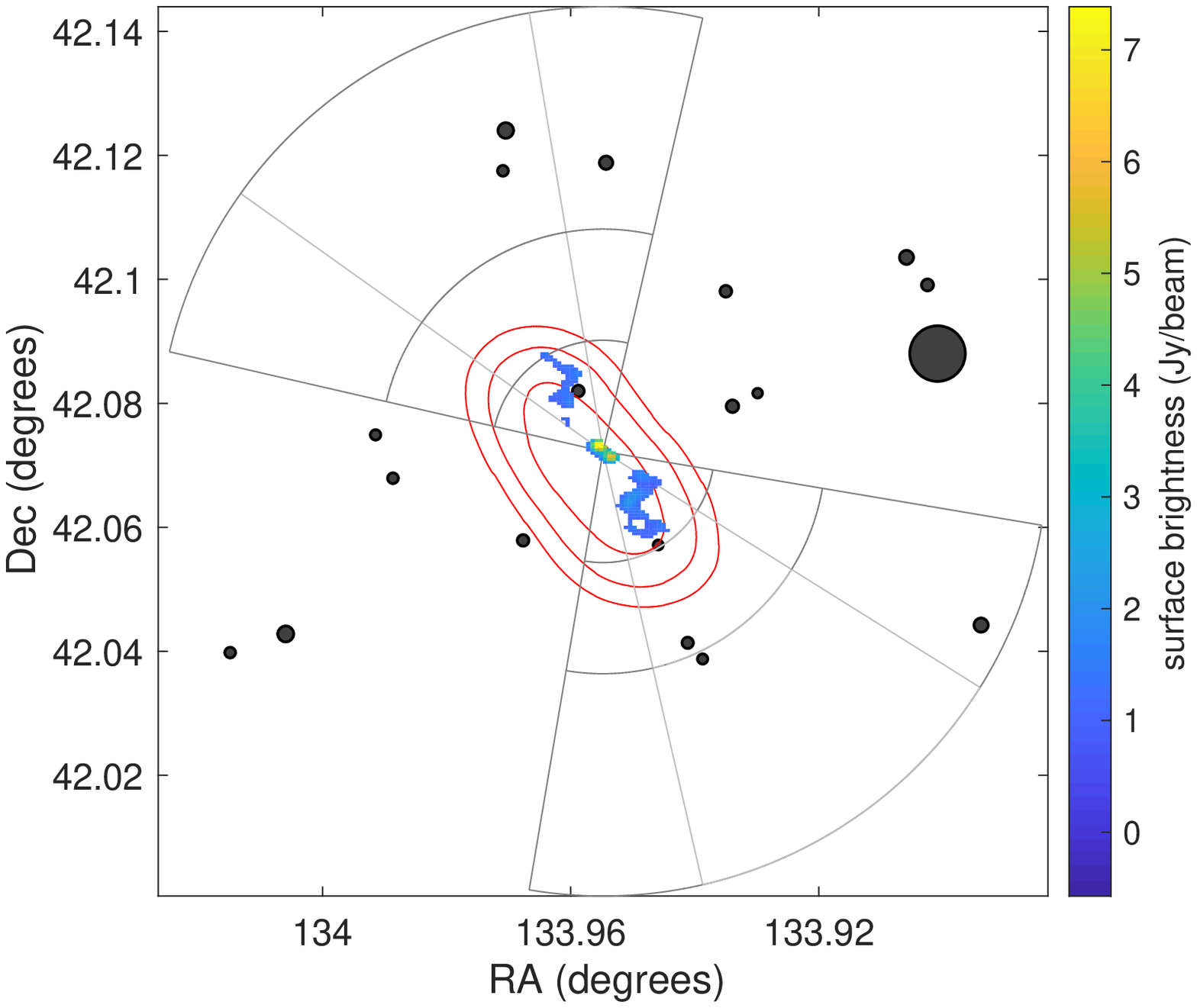}\hspace{0.15\textwidth}
	\includegraphics[width=0.8\columnwidth]{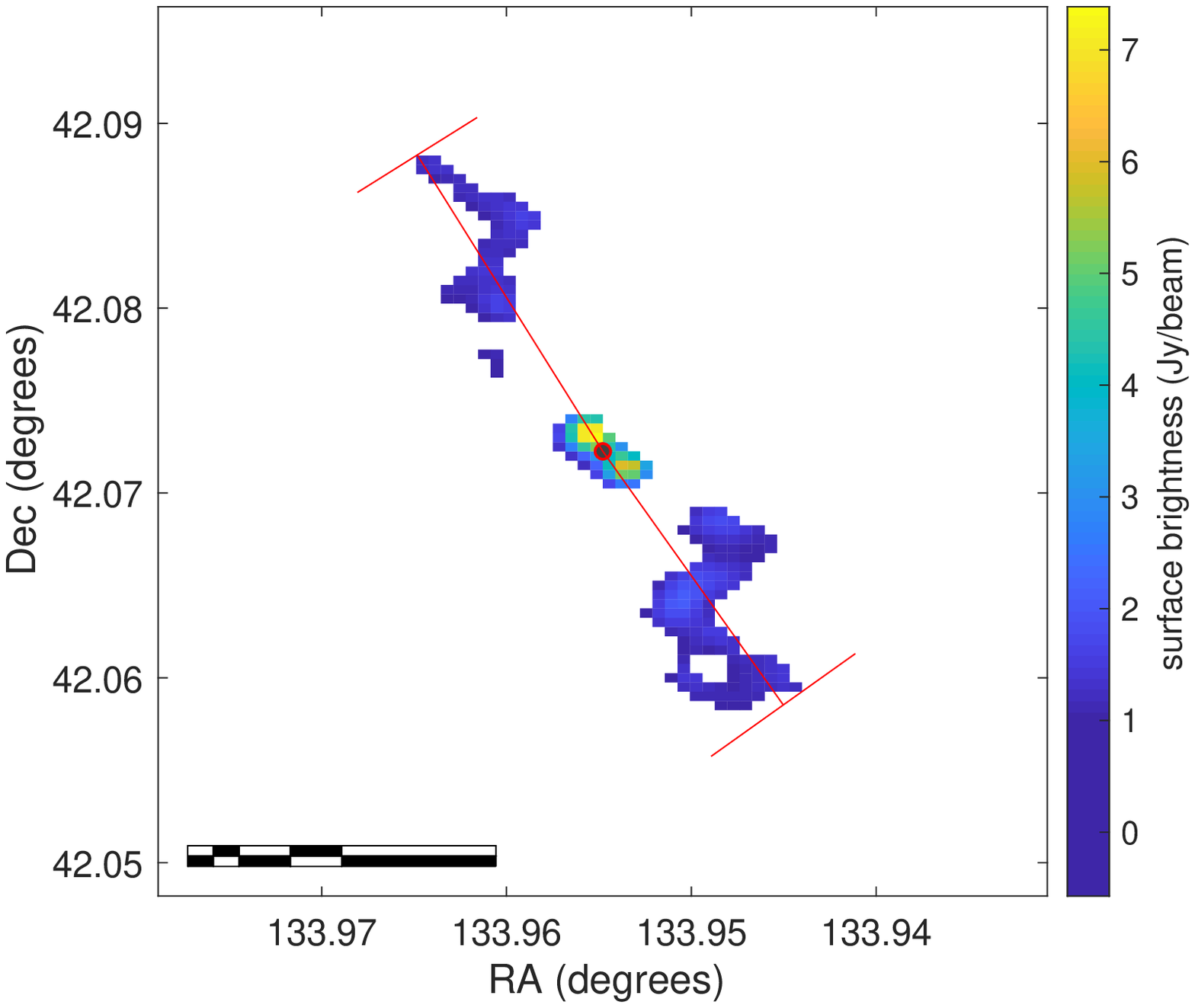}
	\caption{FR-I source RGZ J085549.1+420420; symbols are as in Figure~\ref{fig:FRII_alpha_11}.}
	\label{fig:FRI_1}
\end{figure*}

\begin{figure*}
	\centering
	\includegraphics[width=0.8\columnwidth]{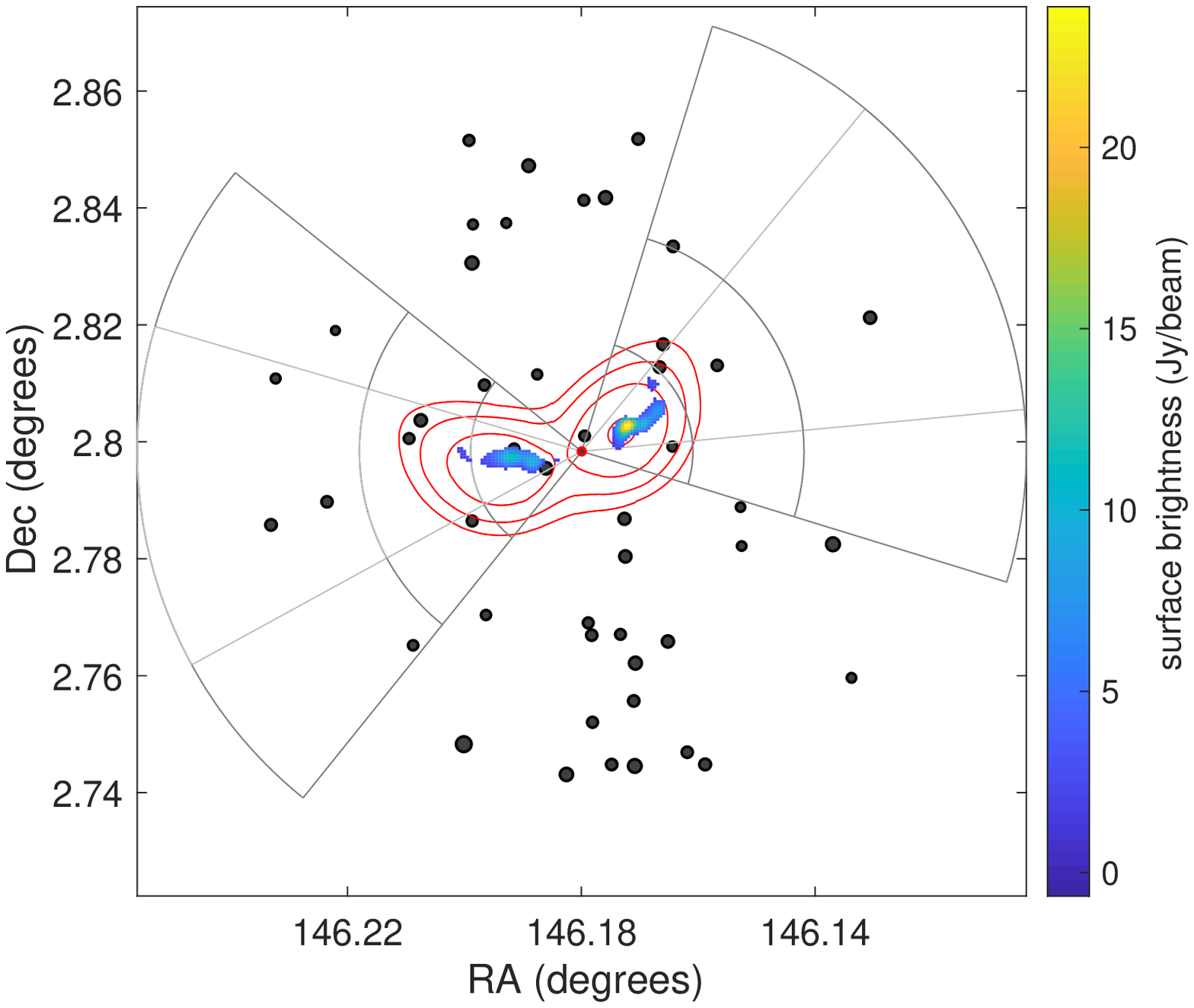}\hspace{0.15\textwidth}
	\includegraphics[width=0.8\columnwidth]{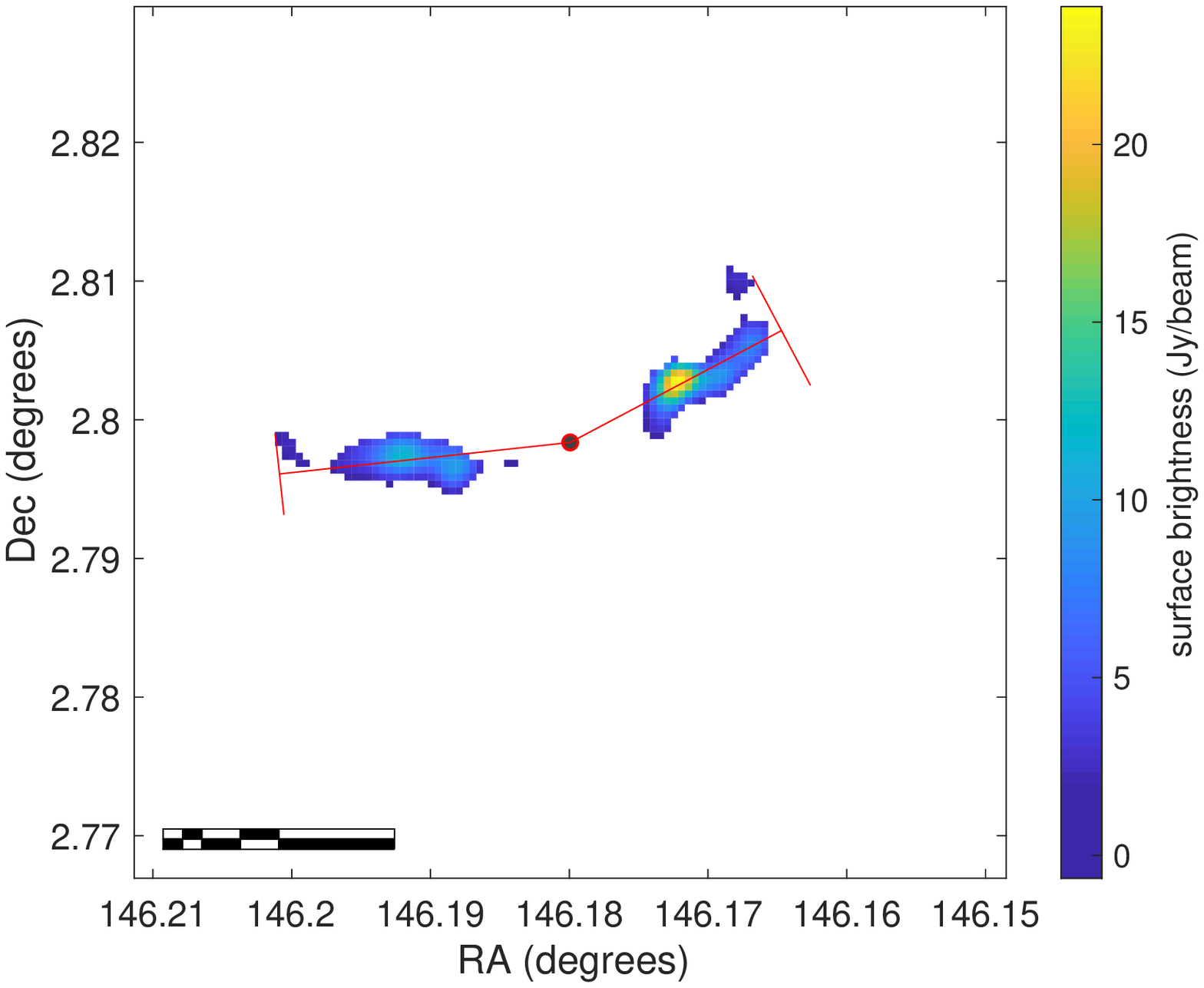}
	\caption{FR-I source RGZ J094443.2+024754; symbols are as in Figure~\ref{fig:FRII_alpha_11}.}
	\label{fig:FRI_2}
\end{figure*}

\begin{figure*}
	\centering
	\includegraphics[width=0.8\columnwidth]{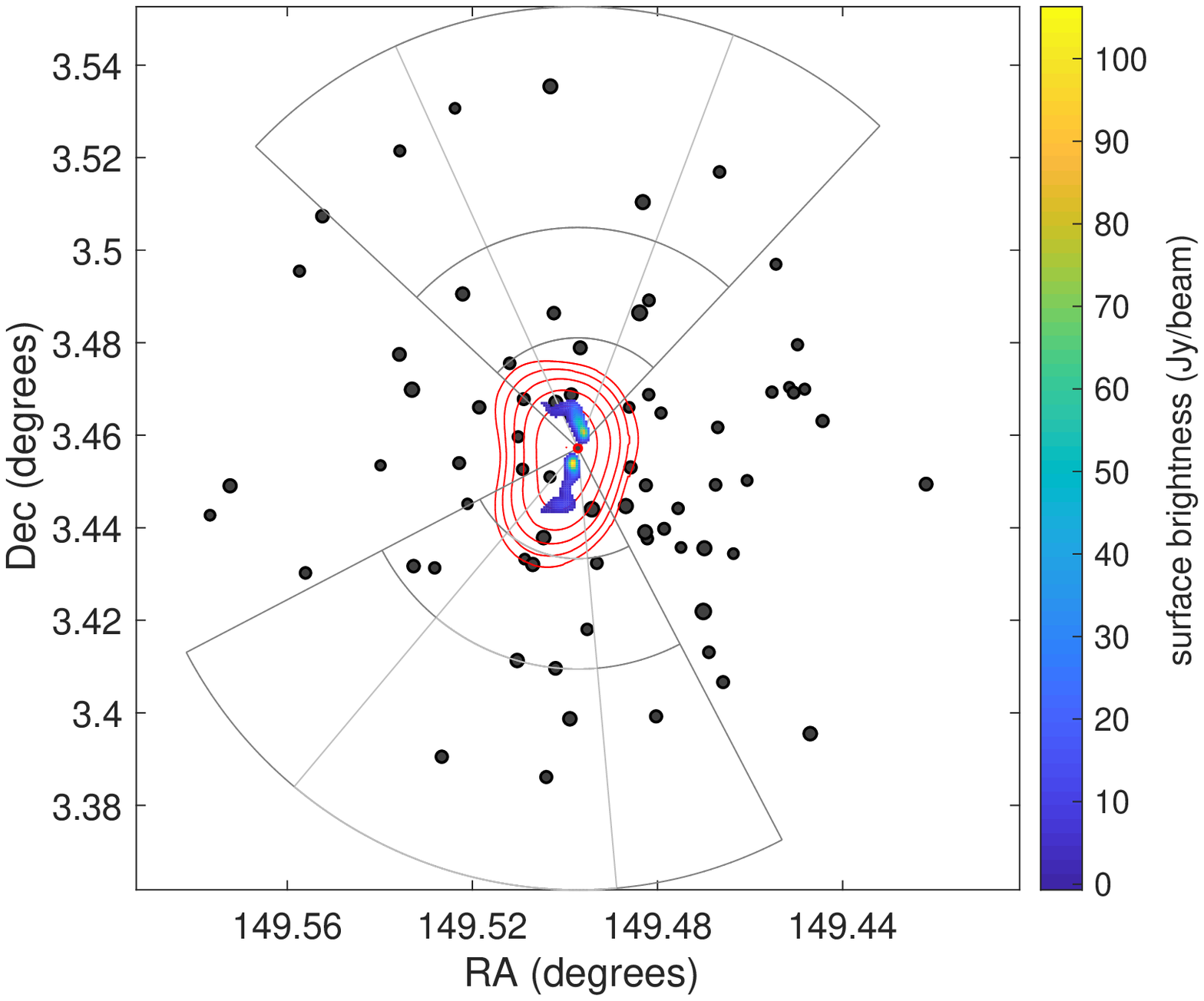}\hspace{0.15\textwidth}
	\includegraphics[width=0.8\columnwidth]{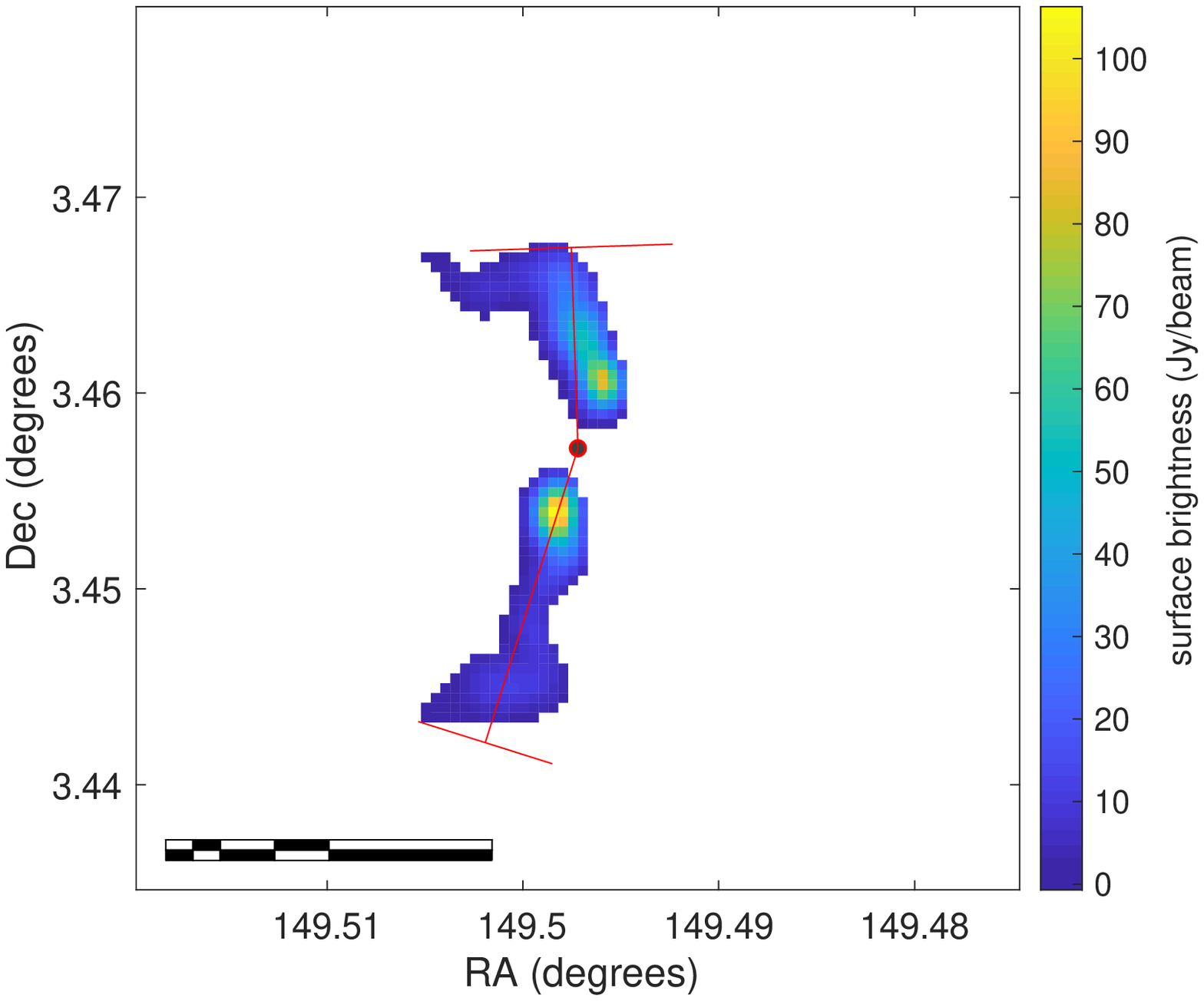}
	\caption{FR-I source RGZ J095759.3+032725; symbols are as in Figure~\ref{fig:FRII_alpha_11}.}
	\label{fig:FRI_3}
\end{figure*}

\begin{figure*}
	\centering
	\includegraphics[width=0.8\columnwidth]{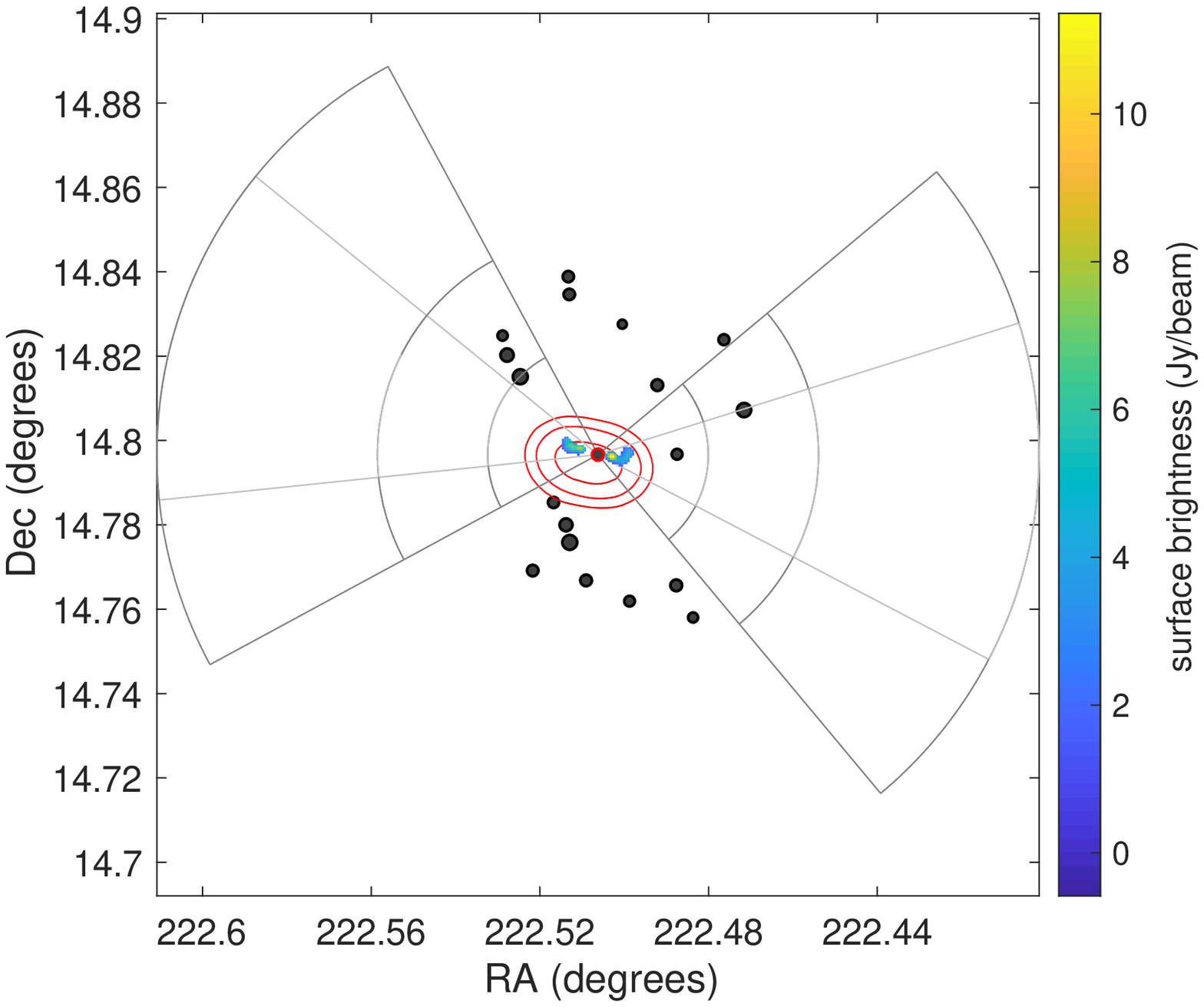}\hspace{0.15\textwidth}
	\includegraphics[width=0.8\columnwidth]{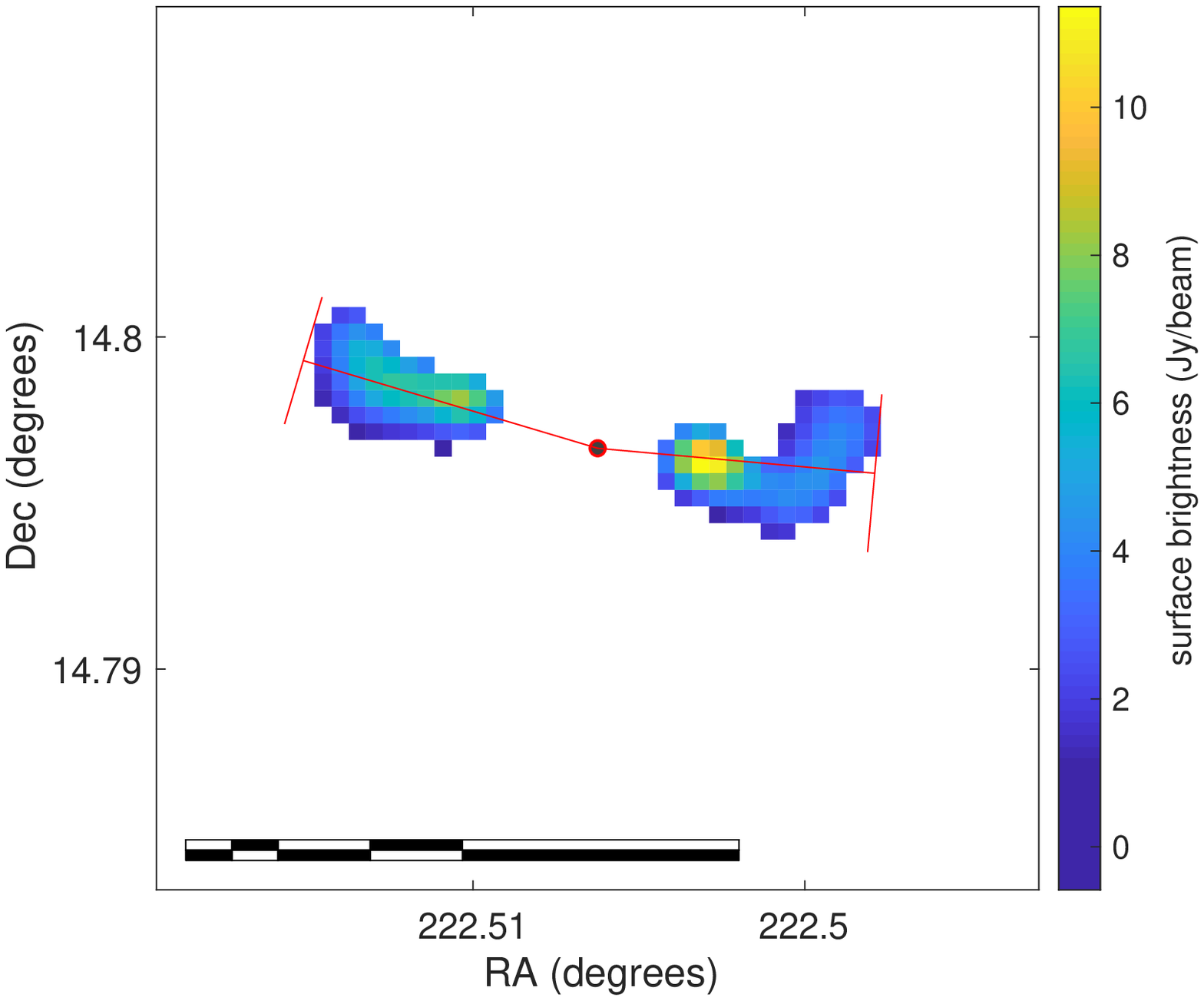}
	\caption{FR-I source RGZ J145001.5+144747; symbols are as in Figure~\ref{fig:FRII_alpha_11}.}
	\label{fig:FRI_5}
\end{figure*}

\begin{figure*}
	\centering
	\includegraphics[width=0.8\columnwidth]{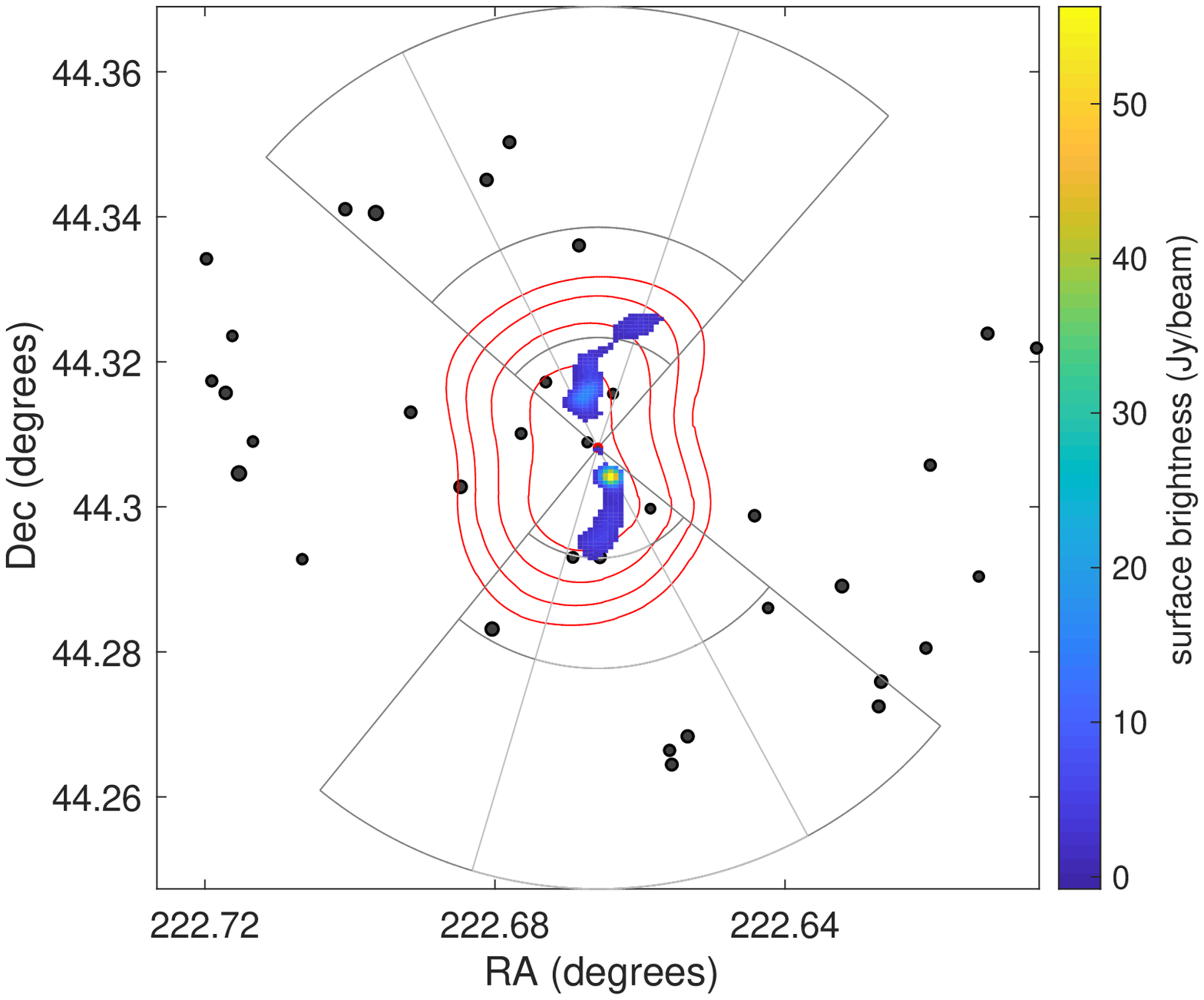}\hspace{0.15\textwidth}
	\includegraphics[width=0.8\columnwidth]{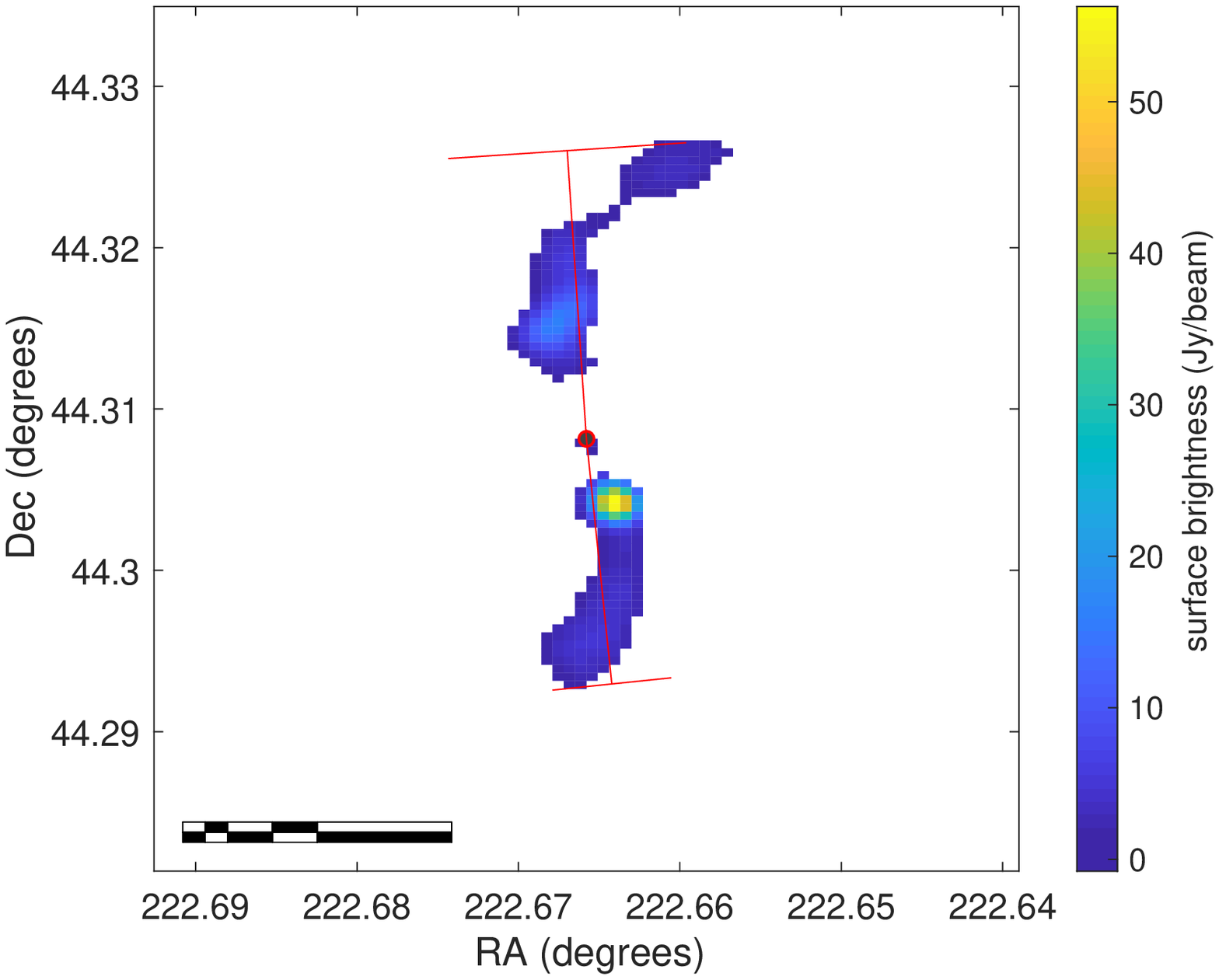}
	\caption{FR-I source RGZ J145039.8+441829; symbols are as in Figure~\ref{fig:FRII_alpha_11}.}
	\label{fig:FRI_6}
\end{figure*}

\label{lastpage}
\end{document}